\documentclass[%
12pt,
aip,
amsmath,amsfonts,amssymb,
author-numerical,
reprint,
floatfix,
]{revtex4-2}

\usepackage[utf8]{inputenc}
\usepackage[T1]{fontenc}
\usepackage{graphicx,animate}
\usepackage[colorlinks,citecolor=green,linkcolor=blue,bookmarks=false,hypertexnames=true]{hyperref}
\usepackage{mathrsfs}

\newtheorem{remark}{Remark}

\renewcommand{\d}[1]{\ensuremath{\operatorname{d}\!{#1}}}
\newcommand{\D}[1]{\ensuremath{\operatorname{D}\!{#1}}}
\newcommand{\dt}[1]{\frac{\d{#1}}{\d{t}}}
\newcommand{\Dt}[1]{\frac{\D{#1}}{\D{t}}}
\newcommand{\del}[2]{\frac{\delta{#1}}{\delta{#2}}}
\def\const{\ensuremath{\operatorname{const}}}

\def\i{\ensuremath{\operatorname{i}}}

\def\cp{\ensuremath{\operatorname{\circlearrowleft}}}

\def\Id{\ensuremath{\operatorname{Id}}}
\def\curl{\nabla\times}

\def\lie{\ensuremath{\operatorname{\pounds}}}

\def\Diff{\ensuremath{\operatorname{Diff}}}
\def\Diffa{\ensuremath{\Diff_\mathrm{area}}}

\makeatletter
\newcommand*{\bdot}{}
\DeclareRobustCommand*{\bdot}{%
  \mathbin{\mathpalette\bdot@{}}%
}
\newcommand*{\bdot@scalefactor}{.5}
\newcommand*{\bdot@widthfactor}{1.15}
\newcommand*{\bdot@}[2]{%
  \sbox0{$#1\vcenter{}$}
  \sbox2{$#1\cdot\m@th$}%
  \hbox to \bdot@widthfactor\wd2{%
    \hfil
    \raise\ht0\hbox{%
      \scalebox{\bdot@scalefactor}{%
        \lower\ht0\hbox{$#1\bullet\m@th$}%
      }%
    }%
    \hfil
  }%
}
\makeatother

\DeclareFontFamily{U}{mathx}{\hyphenchar\font45}
\DeclareFontShape{U}{mathx}{m}{n}{
      <5> <6> <7> <8> <9> <10>
      <10.95> <12> <14.4> <17.28> <20.74> <24.88>
      mathx10
      }{}
\DeclareSymbolFont{mathx}{U}{mathx}{m}{n}
\DeclareMathSymbol{\btimes}{1}{mathx}{"91}

\usepackage{pict2e}
\makeatletter
\DeclareRobustCommand{\intprod}{%
  \mathbin{\mathpalette\int@prod{(0.1,0)(0.9,0)(0.9,0.8)}}%
}
\DeclareRobustCommand{\intprodr}{%
  \mathbin{\mathpalette\int@prod{(0.1,0.8)(0.1,0)(0.9,0)}}}
\newcommand{\int@prod}[2]{%
  \begingroup
  \sbox\z@{$\m@th#1+$}%
  \setlength\unitlength{\wd\z@}%
  \begin{picture}(1,1)
  \roundcap
  \polyline#2
  \end{picture}%
  \endgroup
}
\makeatother

\def\a{\mathbf a}
\def\b{\mathbf b}
\def\c{\mathbf c}
\def\x{\mathbf x}
\def\f{\mathbf f}
\def\l{\mathbf l}
\def\q{\mathbf q}
\def\r{\mathbf r}
\def\u{\bar{\mathbf u}}
\def\v{\boldsymbol\eta}
\def\m{\bar{\mathbf m}}

\def\t{\bar\vartheta}

\def\ts{\vartheta_\sigma}
\def\p{\bar \psi}
\def\ps{\psi_\sigma}
\def\pss{\psi_{\sigma^2}}
\def\xx{\bar\xi}

\def\z{\hat{\mathbf z}}
\def\n{\hat{\mathbf n}}

\def\J{\mathbb J}
\def\L{\mathscr L}
\def\H{\mathscr H}
\def\M{\mathscr M}
\def\C{\mathscr C}
\def\U{\mathscr U}
\def\V{\mathscr V}
\def\W{\mathscr W}
\def\E{\mathscr E}
\def\G{\mathscr G}

\def\Fm{\U_{\m}}
\def\Gm{\V_{\m}}
\def\Hm{\W_{\m}}
\def\Fs{\U_{\rho_{(n_\alpha)}}}
\def\Gs{\V_{\rho_{(n_\alpha)}}}
\def\Hs{\W_{\rho_{(n_\alpha)}}}

\def\Hr{H_\mathrm{r}}
\def\gb{g_\mathrm{b}}
\def\Nr{N_\mathrm{r}}

\def\IL{IL$\smash{^{(0,\alpha)+}}$}

\begin{document}

\title{Extended shallow-water theories with thermodynamics and geometry}

\author{F.J.\ Beron-Vera} \email{fberon@miami.edu}
\affiliation{Department of Atmospheric Sciences, Rosenstiel School
of Marine and Atmospheric Science, University of Miami, Miami,
Florida 33149, USA}

\date{\today}
\begin{abstract}
  Driven by growing momentum in two-dimensional geophysical flow
  modeling, this paper introduces a general family of ``thermal''
  rotating shallow-water models.  The models are capable of
  accommodating thermodynamic processes, such as those acting in
  the ocean mixed layer, by allowing buoyancy to vary in horizontal
  position and time as well as with depth, in a polynomial fashion
  up to an arbitrary degree.  Moreover, the models admit Euler--Poincare
  variational formulation and possess Lie--Poisson Hamiltonian
  structure. Such a geometric property provides solid fundamental
  support to the theories described with consequences for numerical
  implementation and the construction of unresolved motion
  parametrizations. In particular, it is found that stratification
  halts the development of small-scale filament rollups recently
  observed in a popular model, which, having vertically homogeneous
  density, represents a special case of the models presented here.
\end{abstract}


\maketitle

\section{Introduction}

\begin{figure}[t!]
  \centering%
  \includegraphics[width=\linewidth]{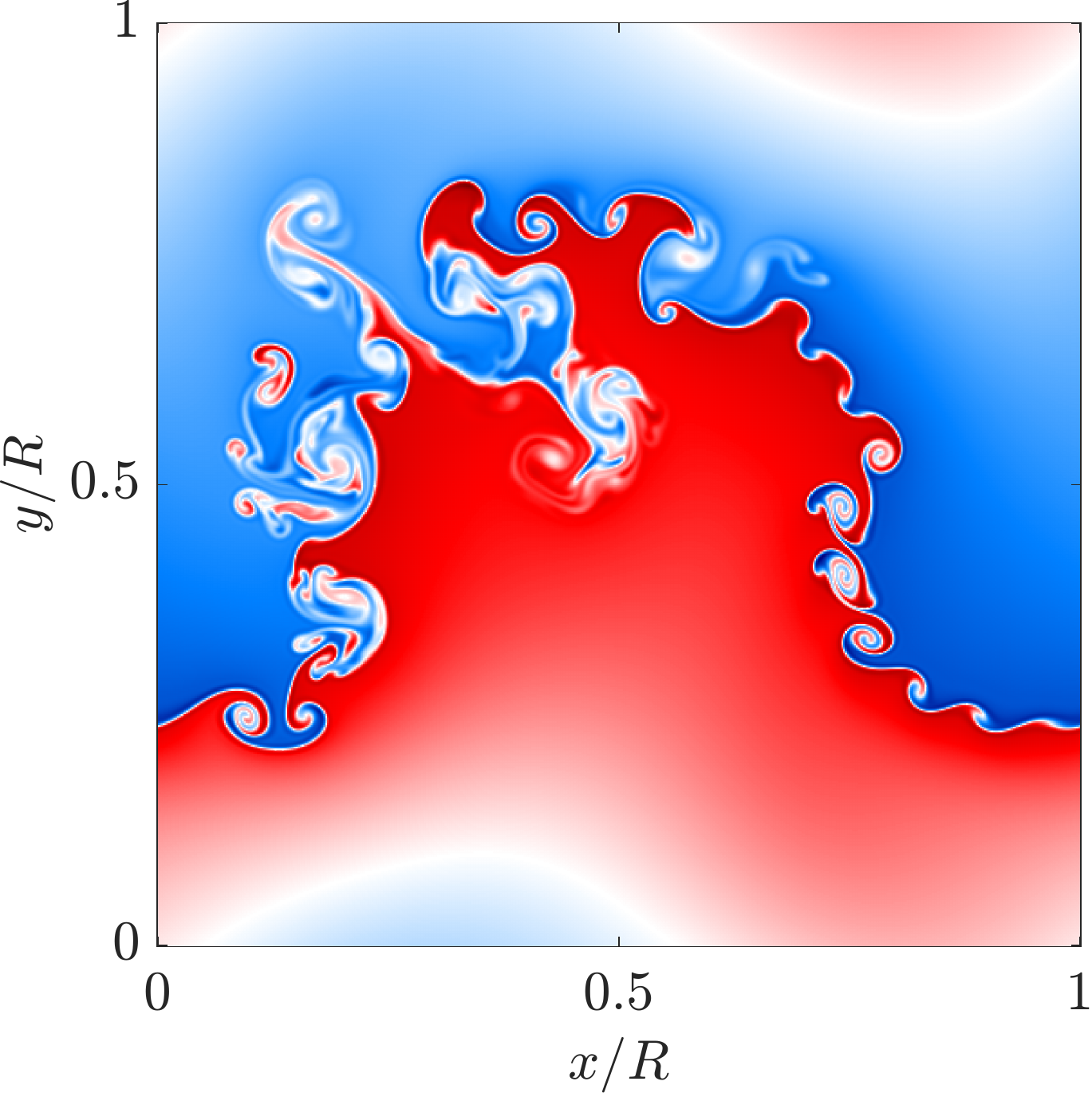}%
  \caption{Buoyancy snapshot from a numerical solution of a QG
  approximation to the IL$^0$PE model in a doubly periodic domain.
  Note the Kelvin--Helmholtz-like rollup filaments (length is scaled
  by the Rossby radius of deformation of the system).}
  \label{fig:il0}%
\end{figure}

This paper is motivated by renewed interest \citep{Warneford-Dellar-13,
Warneford-Dellar-14, Warneford-Dellar-17, Gouzien-etal-17, Zeitlin-18,
Eldred-etal-19, Lahaye-etal-20, Holm-etal-21, Beron-21-POFa,
Beron-21-RMF, Moreles-etal-21} in ``thermal shallow-water modelling''
of geophysical flows. Due to Pedro Ripa's investigation of its
geometric structure and stability properties,\citep{Ripa-GAFD-93,
Ripa-JGR-96, Ripa-RMF-96, Ripa-JFM-95, Ripa-DAO-99} the core model
has been known to many \citep{Dellar-03,
Sanchez-etal-16,Mungkasi-Roberts-16, Rehman-etal-18} as \emph{Ripa's
model}. Its origins can be traced back to the mid 1960s and early
1970s, when various authors \citep{Obrien-Reid-67, Dronkers-69,
Lavoie-72} independently introduced the idea of incorporating
thermodynamics to the rotating shallow-water (or, more fairly,
Laplace's tidal) equations, i.e., the two-dimensional Euler equations
for homogeneous fluid with Coriolis force in the horizontal and the
hydrostatic approximation, by vertically averaging the pressure
gradient force of the full \emph{primitive equations} (or \emph{PE}),
i.e., the three-dimensional Euler equations for arbitrary stratified
incompressible fluid with Coriolis force in the horizontal and the
hydrostatic and Boussinesq approximations.  The result is a model
with four prognostic variables: the two components of the velocity,
layer thickness, and buoyancy, all varying in the horizontal and
time.  The main advantage of the resulting model over the rotating
shallow-water model is its ability to incorporate heat and freshwater
fluxes across the air--sea interface.  This feature was extensively
used to simulate mixed-layer (i.e., the topmost part of the ocean,
above the thermocline) dynamics, particularly through the 1980s and
1990s (e.g., \citet{Schopf-Cane-83, McCreary-etal-93, Beier-97})
realistically at a relatively low computational cost.  Moreover,
recent high-resolution numerical simulations \citep{Beron-21-POFa,
Holm-etal-21} of a \emph{quasigeostrophic} (\emph{QG}) approximation
of the system (i.e., with the pressure gradient force nearly balancing
the Coriolis force \citep{Pedlosky-87}) revealed the formation of
Kelvin--Helmholtz-like rollup filaments (Fig.\ \ref{fig:il0}; for
additional snapshots, cf.\@~Fig.\@~\ref{fig:il0-il01}, left panel,
multimedia view) that resemble quite well submesoscale (1--10 km)
features identified in satellite-derived chlorophyll distributions
on the surface of the ocean.  Such submesoscale circulations are
believed to play an important role in the exchange of gases through
the ocean surface and thus global climate, and are the subject of
intense research.\citep{McWilliams-16}

In this paper I present a family of rotating shallow-water models
with thermodynamics that extend Ripa's model by allowing buoyancy
to vary in the vertical. The dependence on the vertical coordinate
is polynomial, up to an arbitrary degree.  This extension enables
representation of important mixed-layer dynamics processes such as
restratification by baroclinic instability.\citep{Haine-Marshall-98}
I will spend a substantial part of the paper discussing its geometric
properties (Euler--Poincare variational formulation \citep{Holm-etal-98a}
and generalized (noncanonical) Hamiltonian structure
\citep{Morrison-Greene-80}). This will enable making an explicit
connection, so far overlooked, with earlier work by
\citet{Morrison-Greene-80}.  \citet{Ripa-JFM-95} presents a rotating
shallow-water model with thermodynamics with velocity shear in the
vertical.  Such a model, however, does not possess the geometric
structure of the model presented here.  I emphasize this property
of the present model, which is shared with the PE from which it
derives.  Such a structure can be exploited to formulate integration
schemes to avoid spurious numerical effects.  Structure-preserving
algorithm development is a topic of active investigation
nowadays.\citep{Kraus-etal-16, Morrison-17, Eldred-etal-19,
Gawlik-Gay-21}  Additional motivation for preserving structure in
rotating shallow-water theories with thermodynamics is that this
enables the application of a flow-topology-preserving framework
\citep{Holm-15} for building parametrizations \citep{Cotter-etal-20}
of unresolvable motions and this way investigating their contribution
to transport at resolvable scales. Furthermore, the Hamiltonian
formulation of fluid dynamics provides a systematic manner to find
conservation laws, linked with symmetries via Noether's theorem.
These can be used to derived a-priori flow stability statements
\citep{Holm-etal-83} as well as to gain insight into the nature of
growing perturbations to unstable states \citep{Ripa-JFM-91} or
infer saturation bounds on their nonlinear growth.\citep{Shepherd-88a}

Following notation introduced in \citet{Ripa-JFM-95}, I will use
the acronym \emph{IL}, for \emph{inhomogeneous layer}, to refer to
the rotating shallow-water models with thermodynamics discussed in
this paper.  {Superscript(s) will be appended to IL
to indicate the amount of vertical variation permitted in the
horizontal velocity and buoyancy fields.}  With this notation,
Ripa's model will be referred to as the IL$^0$PE model, to mean
that neither velocity nor buoyancy change with depth. In turn, the
full PE will be referred to as the IL$^\infty$PE model, since
velocity and buoyancy vary with depth unrestrictedly.

The rest of the paper is organized as follows.  The general IL model
with thermodynamics in PE is introduced in Sec.\@~\ref{sec:IL0alpha+}.
Its geometric structure is discussed in extent in
Sec.\@~\ref{sec:geometry}.  The interpretation of the model as a
rotating shallow-water with thermodynamics is given in
Sec.\@~\ref{sec:thermo}, where several submodels are considered
along with their QG form, insightful in the study of processes such
as baroclinic instability leading to the rollup filaments in
Fig.\@~\ref{fig:il0}, and corresponding geometric structure.
{The effects of stratification on the development
of such rollup filaments are discussed in Section \ref{sec:strat}.}
Section \ref{sec:conclusions} summarizes the paper in addition to
offering a discussion and an outlook on avenues for future research.
Finally, three appendices are included with the proof of a result
that extends earlier results by other authors \citep{Marsden-etal-84}
and some abstract mathematics details that provide deeper geometrics
mechanics interpretations of the equations discussed.

\section{The \IL PE model}\label{sec:IL0alpha+}

Let $\x = (x,y) \in D \subseteq \mathbb R^2$ denote position in
some fixed domain of a $\beta$-plane.  Let $\u(\x,t)$ denote
instantaneous velocity at time $t$ inside a fluid layer limited
from above by a rigid lid.  The bottom of the layer is soft as the
layer floats atop an infinitely deep layer of quiescent fluid, i.e.,
a reduced gravity is assumed.  However, the model equations hold
for a system with a rigid bottom and a free surface. The overbar
notation is used to emphasize that independence on the vertical
coordinate ($z$) is achieved via (vertical) average.  

If $D \subset \mathbb R^2$ has a solid boundary, then no normal
flow through it is implied, i.e., $\u\cdot\n\vert_{\partial D} =
0$; conversely, if $D = \mathbb R^2$ then we will assume that $\u$
decays to zero as $|\x|\to\infty$. A physically relevant setting
is one with $D$ representing a zonal channel, infinitely long ($D
= \mathbb R\times [0,L_y]$) or reentrant ($D =\mathbb R/ L_x\mathbb
Z\times [0,L_y]$).  The boundary condition in the first case is a
combination of the aforementioned ones; in the second case, one
requires periodicity in $x$ i.e., $\u(x,y,t) = \u(x+L_x,y,t)$.
Doubly reentering domains, i.e., $D = \mathbb R/L_x\mathbb Z\times
\mathbb R/ L_y\mathbb Z$, and domains including islands, i.e., with
$\partial D$ being the union of multiply connected components, are
also plausible, with boundary conditions being periodicity of $\u$
in $x$ and $y$ and no flow through each connected component of the
boundary, respectively.\citep{Marsden-Weinstein-83}

Let 
\begin{equation}
  \rho_{n_\alpha}(\x,t),\quad n_\alpha = 1,2,\dotsc,\alpha+2, 
\end{equation}
be a finite set functions, each one satisfying
\begin{equation}
  \int_{D_{\u}} \d{^2}\x\, \rho_{n_\alpha} = \const,
  \label{eq:intrho}
\end{equation}
where $D_{\u} \subset D$ is a fluid region (i.e., carried along
with the flow of $\u$), and consider the {function
\begin{equation}
  \varphi_{\alpha+}(\rho_1,
  \tilde\rho_2,\dotsc,\tilde\rho_{\alpha+2}),  
\end{equation}
where
\begin{equation}
  \tilde\rho_{\tilde n_\alpha} := \rho_1^{-1}\rho_{\tilde
  n_\alpha},\quad
  \tilde n_\alpha = 2,\dotsc,\alpha+2,
\end{equation}
which is assumed to be differentiable in all of its arguments.}
When $D \subset \mathbb R^2$ has a solid boundary, an additional
boundary condition is
\begin{equation}
  \nabla\tilde\rho_{\tilde n_\alpha}\times\n\vert_{\partial D} = 0
  \label{eq:bc}
\end{equation}
(the above applies on each component of the boundary of a multiply
connected $D$). The nature of \eqref{eq:bc} will be clarified below.

Let
\begin{equation}
  \m := \rho_1(\u+\f),\quad \z\cdot\curl\f = f = f_0+\beta y,
  \label{eq:m}
\end{equation}
where $\f(\x)$ is the vector potential of the local angular velocity
of the Earth relative to a fixed frame, $\frac{1}{2}f\z$. Finally,
define
\begin{equation}
  \bar q := \frac{\z\cdot\curl\u + f}{\rho_1} \equiv
  \rho^{-1}_1\z\cdot\nabla\times\rho^{-1}_1\m. 
  \label{eq:q}
\end{equation}
If $\rho_1$ is taken as the thickness of the fluid layer, \emph{as
we will throughout this paper}, then \eqref{eq:q} is the vertical
average of Ertel's potential vorticity in the IL$^\infty$PE model;
\citep{Ripa-JFM-95} we will refer $\bar q$ to simply as \emph{potential
vorticity}.  

The \IL PE is given by:
\begin{equation}
\left.  
\begin{aligned}
  \partial_t\m + \lie_{\u}\m - \rho_1\nabla K + \nabla
  \rho_1^2
  \tilde\partial_{\rho_1}\varphi_{\alpha+}
  &= 0,\\
  \partial_t\rho_{n_\alpha} + \nabla \cdot \rho_{n_\alpha}\u &= 0,
\end{aligned}
\right\}
\label{eq:il1-}
\end{equation}
where
\begin{equation}
  K := \tfrac{1}{2}|\u|^2 + \u\cdot\f
  \label{eq:K}
\end{equation}
and
\begin{equation}
  \tilde\partial_{\rho_1} :=
  \partial_{\rho_1}\vert_{\tilde\rho_{\tilde n_\alpha}},
\end{equation}
and we have adopted the convenient \emph{notation}\footnote{This
object can be interpreted as a Lie derivative upon appropriate
interpretation of $\a$ and $\b$, which I intentionally omit as the
algebraic approach taken serves my purposes. Appendices
\ref{app:constraints} and \ref{app:lie} do discuss some differential
geometry aspects which are needed to provide a deeper geometric
interpretation of some of the results of the paper, but these can
be safely ignored by the reader if not interested.}
\begin{equation}
  \lie_{\a}\b := (\a\cdot\nabla)\b  + b_j\nabla a^j +
  (\nabla\cdot\a)\b = b_j\nabla a^j + \partial_j\b a^j
  \label{eq:lie}
\end{equation}
for any vectors $\a,\b(\x,t)$ (summation over repeated (lowered and
raised) indices is implied \emph{unless indicated in between
parentheses}). Note that $\lie_{\u}\m$ includes three terms.  The
first represents the advection of $\m$ by $\u$, and the second and
third terms respectively account for the stretching and expansion
of fluid elements as $\m$ is being transported by $\u$.

Dividing the first equation in \eqref{eq:il1-} by $\rho_1$, and
using the second equation for $n_\alpha = 1$, it takes more familiar
form
\begin{equation}
  \partial_t\u + (\u\cdot\nabla)\u + f\z\times\u = - \rho_1^{-1}
  \nabla \rho_1^2 \tilde\partial_{\rho_1}\varphi_{\alpha+}.
  \label{eq:mom1}
\end{equation}
This facilitates the interpretation of the term $- \rho_1^{-1}
\nabla \rho_1^2 \tilde\partial_{\rho_1}\varphi_{\alpha+}$ as a
generalized pressure gradient force.  The second equation in
\eqref{eq:il1-} is a consequence of Reynolds' transport theorem
applied on \eqref{eq:intrho}, expressing continuity of the functions
$\rho_{n_\alpha}$. These functions, therefore, represent conserved
\emph{densities}, and thus are different than $\tilde\rho_{\tilde
n_\alpha}$, which represent quantities \emph{transported by the
flow of} $\u$, i.e., by continuity of $\rho_1$,
\begin{equation}
  \Dt{\tilde\rho_{n_\alpha}} = 0,
  \label{eq:rhotilde}
\end{equation}
where 
\begin{equation}
  \Dt{}:= \partial_t + (\u\cdot\nabla)
\end{equation}
is the material derivative. 

I will make explicit the interpretation of the \IL PE model
\eqref{eq:il1-} as a family of rotating shallow-water systems with
thermodynamics after discussing its geometric structure first.

\section{Geometry}\label{sec:geometry}

The \IL PE model both admits an \emph{Euler--Poincare variational
formulation} and possesses a \emph{generalized Hamiltonian structure},
as it is shown here.  By Euler--Poincare variational principle I
mean a Hamilton's principle for fluids that leads to motion equations
in Eulerian variables. This is accomplished by writing the Lagrangian
in terms of Eulerian variables and then extremizing the corresponding
action under constrained variations representing fluid particle
path variations at fixed Lagrangian labels and time.  This principle,
as just described, was apparently first discussed in \citet{Newcomb-62}.
Yet it belongs to a much general variational formulation of mechanics,
written in the abstract language of differential geometry by
\citet{Holm-etal-98a, Holm-etal-02}, who made connections with seminal
work by Henri Poincare, reviewed in \citet{Marle-13}.  By generalized
Hamiltonian structure I mean a noncanonical Hamiltonian representation
of the equations of motion in terms of Eulerian variables as discussed
by \citet{Morrison-Greene-80}.  The abstract formulation, following
earlier work by \citet{Arnold-66b}, is due to \citet{Marsden-Weinstein-82,
Marsden-Weinstein-83}.  In particular, \citet{Marsden-etal-84}
showed how to obtain the Euler equation for compressible fluid
motion as a generalized Hamiltonian system via reduction by symmetry
of the corresponding canonical Hamiltonian formulation, i.e., the
Euler equation written in Lagrangian variables.  

Fixing notation, calligraphic letters, e.g., $\U$, will be used to
denote real-valued \emph{functionals of fields}, e.g., $\mu(\x,t)
= (\mu^1(\x,t), \mu^2(\x,t), \dotsc)$.  We will restrict attention
to functionals of the specific form:
\begin{equation}
  \U[\mu] = \int U(\x; \mu, \nabla\mu, \dotsc)
 \label{eq:U}
\end{equation}
where $\int := \int_D\d{}^2\x$ operates on everything on its right.
Note that $U$ in \eqref{eq:U} does not depend explicitly on time.
Otherwise it has a finite number of arguments, and is a sufficiently
smooth function in each of them.  The latter makes $\U$ smooth
enough for the functional derivative wrt $\mu$, i.e., the unique
element $\del{}{\mu}\U$ satisfying
\begin{equation}
  \left.\frac{\d{}}{\d{\varepsilon}}\right|_{\varepsilon=0} \U[\mu
  + \varepsilon\delta\mu] = \int\del{\U}{\mu}\bdot\delta{\mu} :=
  \del{\U}{\mu^a}\delta{\mu^a},
\end{equation}
to be well defined. In particular, $\mu^a(\x',t) = \int
\delta(\x-\x')\mu^a(\x,t)$; then
\begin{equation}
  \del{\mu^a(\x',t)}{\mu^b(\x,t)} = \delta^a_b\delta(\x-\x').
  \label{eq:delta}
\end{equation}
The shorthand $\U_\mu$ for $\smash{\del{}{\mu}}\U$ will be used in
inline equation display or offline when notationally more convenient.

\subsection{Euler--Poincare variational formulation}\label{sec:ep}

Consider the functional
\begin{equation}
  \L[\u,\rho] := \int \rho_1 \big(K - \varphi_{\alpha+}\big)  
  \label{eq:L}
\end{equation}
where $\rho = (\rho_{n_\alpha})$. Its functional derivatives are:
\begin{align}
  \del{\L}{\u} &= \m,\\ \del{\L}{\rho_1} &=  K - \varphi_{\alpha+}
  - \rho_1\tilde\partial_{\rho_1}\varphi_{\alpha+} + \tilde\rho_{\tilde
  n_\alpha}\partial_{\tilde\rho_{\tilde n_\alpha}}\varphi_{\alpha+},\\
  \del{\L}{\rho_{\tilde n_\alpha}} &= - \partial_{\tilde\rho_{\tilde
  n_\alpha}}\varphi_{\alpha+}.
\end{align} 
Noting that
\begin{equation}
  \nabla\varphi_{\alpha+} = (\tilde\partial_{\rho_1}\varphi_{\alpha+})\nabla\rho_1 +
  \big(\partial_{\tilde\rho_{\tilde
  n_\alpha}}\varphi_{\alpha+}\big)\nabla\tilde\rho_{\tilde n_\alpha}
  \label{eq:grad}
\end{equation}
and
\begin{equation}
  (\tilde\partial_{\rho_1}\varphi_{\alpha+})\nabla\rho_1 +
  \nabla\big(\rho_1\tilde\partial_{\rho_1}\varphi_{\alpha+}\big) =
  \rho_1^{-1}\nabla\rho_1^2\tilde\partial_{\rho_1}\varphi_{\alpha+},
  \label{eq:grad2}
\end{equation}
the first equation of \eqref{eq:il1-} follows as
\begin{equation}
  \partial_t\del{\L}{\u} + \lie_{\u}\del{\L}{\u} =
  \rho_{n_\alpha}\nabla\del{\L}{\rho_{n_\alpha}},
  \label{eq:il1-ep}
\end{equation}
upon some cancellation.

The above suggests that the \IL PE model equations \eqref{eq:il1-}
admit an Euler--Poincare variational formulation. In fact, interpreting
$\L$ as a \emph{Lagrangian}, with $K$ and $\varphi_{\alpha+}$ thus
corresponding to \emph{kinetic} and \emph{potential energy density},
respectively, \eqref{eq:il1-ep} extremizes the action
\begin{equation}
  \mathscr S := \int_{t_0}^{t_1}\d{t}\,\L[\u,\rho]
\end{equation}
under constrained variations
\begin{align}
  \delta\u &= \partial_t\v + [\u,\v],\label{eq:du}\\
  \delta\rho_{n_\alpha} &= -\nabla\cdot\rho_{n_\alpha}\v,
  \label{eq:drho}
\end{align}
where $\v(\x,t)$ is an arbitrary vector
field satisfying $\v(\x,t_0) = 0 = \v(\x,t_1)$ and
$\v\cdot\n\vert_{\partial D} = 0$, and
\begin{equation}
   [\a,\b] := (\a\cdot\nabla)\b - (\b\cdot\nabla)\a,
	\label{eq:comm}
\end{equation}
which is the \emph{commutator of vector fields}. Indeed,
\begin{align}
  \delta \mathscr  S &= \int_{t_0}^{t_1}\d{t}\int
  \del{\L}{\u}\cdot\delta\u +
  \del{\L}{\rho_{n_\alpha}}\delta\rho_{n_\alpha}\\
  &=-\int_{t_0}^{t_1}\d{t}\int
  \Big(\partial_t\del{\L}{\u} + \lie_{\u}\del{\L}{\u} -
  \rho_{n_\alpha}\nabla\del{\L}{\rho_{n_\alpha}}\Big)\cdot\v
\end{align}
upon integration by parts. This vanishes for arbitrary $\v$ when
\eqref{eq:il1-ep} holds. Thus \eqref{eq:il1-ep} can be interpreted
as Newton's second law for fluid motion: the rate of change of the
fluid's absolute momentum density, $\m = \L_{\u}$, equals the sum
of force densities. The second equation in \eqref{eq:il1-} is built
in the constraint \eqref{eq:drho} (cf.\ App.\@~\ref{app:constraints},
where the nature of the constraints is clarified).

\subsubsection{Kelvin circulation theorem}\label{sec:kelvin}

Dividing \eqref{eq:il1-ep} by $\rho_1$ and using continuity of
$\rho_1$ (the second equation in \eqref{eq:il1-} for $n_\alpha =
1$), one finds
\begin{equation}
  \Dt{}\frac{1}{\rho_1}\del{\L}{\u} +
  \frac{1}{\rho_1}\del{\L}{\bar u^j}\nabla\bar u^j-
  \nabla\del{\L}{\rho_1} -
  \tilde\rho_{\tilde n_\alpha}\nabla\del{\L}{\rho_{\tilde n_\alpha}} =
  0.
  \label{eq:il1-lpDt}
\end{equation}
Introducing the circulation
\begin{equation}
  \gamma :=  \oint_{\partial
  D_{\u}}\frac{1}{\rho_1}\del{\L}{\u}\cdot\d{\x} =  
  \oint_{\partial D_{\u}} (\u+\f)\cdot\d{\x}
  \label{eq:g}
\end{equation}
($D_{\u}\subset D$ is transported by the flow of $\u$ {as
is its closed material boundary $\partial D_{\u}$}), from
\eqref{eq:il1-lpDt} it follows that
\begin{align}
  \dt{\gamma} &=  \int_{D_{\u}}\d{}^2\x\,
  \left\{\tilde\rho_{\tilde n_\alpha},\del{\L}{\rho_{\tilde n_\alpha}}\right\}_{xy} 
  \nonumber\\
  &=
  \int_{D_{\u}}\d{}^2\x\,
  \left\{\partial_{\tilde\rho_{\tilde
  n_\alpha}}\varphi_{\alpha+},\tilde\rho_{\tilde n_\alpha}\right\}_{xy},
  \label{eq:KN}
\end{align}
upon using $\a\cdot\Dt{}\d{\x} = \a\cdot\d{\u} = \a\cdot\partial_j\u\d{x}^j
= \a_j\nabla \bar u^j\cdot\d{\x}$ along with Stokes theorem, where
\begin{equation}
  \{A,B\}_{xy} := \z\cdot\nabla A\times\nabla B
  \label{eq:PBxy}
\end{equation}
for all $A,B(\x,t)$ is the \emph{canonical Poisson bracket}, which
is used to denote Jacobian of functions. Equation \eqref{eq:KN} is
the statement of the \emph{Kelvin circulation theorem}.

In general, $\gamma$ is not conserved; it is created (or destroyed)
by the misalignment of the gradients of $\tilde\rho_{\tilde n_\alpha}$ and
$\partial_{\tilde\rho_{\tilde n_\alpha}}\varphi_{\alpha+}$.  One
exception is the case in which only one density, i.e., layer thickness
or $\rho_1$, is included in \eqref{eq:il1-}.  Another exception is
the situation in which \eqref{eq:il1-} is initialized from
$\tilde\rho_{\tilde n_\alpha} = \const$, since they are transported
by the flow of $\u$ \eqref{eq:rhotilde} and thus remain constant
at all times. Indeed, the set $\{\tilde\rho_{\tilde n_\alpha}(\x,t)
= \const\}$ is an invariant manifold of \eqref{eq:il1-}.  We will
get back to this below when discussing submodels of the \IL PE
model.

On the other hand, if $\partial D_{\u}$ is replaced by the solid
boundary $\partial D$ of the flow domain, then $\gamma$, as in
\eqref{eq:g} or simply in this case $\gamma = \smash{\oint_{\partial
D}} \u\cdot\d{\x}$, is conserved because $\smash{\oint_{\partial
D} (\partial_{\tilde\rho_{(\tilde n_\alpha)}}\varphi_{\alpha+})
\nabla\tilde\rho_{(\tilde n_\alpha)}\cdot\d{\x} \equiv 0}$ by the
boundary condition \eqref{eq:bc}, clarifying the nature of this
boundary condition.  This holds along each connected component of
$\partial D$ when $D$ has multiple islands.

\subsubsection{Potential vorticity evolution}

Finally, using Stokes theorem we have
\begin{equation}
  \gamma \equiv \int_{D_{\u}}\d{}^2\x\,\rho_1\bar q;
\end{equation}
then by continuity of $\rho_1$, one finds
\begin{equation}
  \Dt{\bar q} = \rho_1^{-1}
  \left\{\partial_{\tilde\rho_{\tilde
  n_\alpha}}\varphi_{\alpha+},\tilde\rho_{\tilde n_\alpha}\right\}_{xy}.
  \label{eq:DqDt}
\end{equation}
In general, the potential vorticity is not transported by the flow
of $\u$.  Exceptions are the aforementioned ones, namely, the
single-density case and initialization on the constant density
invariant subspace.

\subsection{Lie--Poisson Hamiltonian structure}

Consider 
\begin{equation}
  \E[\m,\rho] := \int \m\cdot\u - \rho_1\big(K-\varphi_{\alpha+}\big),
  \label{eq:E}
\end{equation}
which is nothing but the \emph{total energy} of system \eqref{eq:il1-},
viz.,
\begin{equation}
  \E \equiv \int \rho_1\big(\tfrac{1}{2}|\u|^2 +
  \varphi_{\alpha+}\big).
\end{equation}
The functional derivatives of $\eqref{eq:E}$ are:
\begin{align}
  \del{\E}{\m} &= \u,\\
  \del{\E}{\rho_{n_\alpha}} &\equiv - \del{\L}{\rho_{n_\alpha}}.
\end{align}
Then taking \eqref{eq:grad}--\eqref{eq:grad2} into account, set
\eqref{eq:il1-} follows from
\begin{equation}
  \left.
  \begin{aligned}
	 \partial_t\m + \lie_{\E_{\m}}\m +
	 \rho_{n_\alpha}\nabla\del{\E}{\rho_{n_\alpha}}
    &= 0,\\
	 \partial_t\rho_{n_\alpha} + \nabla \cdot \rho_{n_\alpha}\del{\E}{\m} &= 0,
  \end{aligned}
  \right\}
  \label{eq:il1-lp}
\end{equation}
after some cancellation.

The above suggests that the \IL PE model \eqref{eq:il1-} possesses a
generalized Hamiltonian structure, i.e., it can be cast in the form
\begin{equation}
  \partial_t\mu = \{\mu,\H\} = \J\del{\H}{\mu},
  \label{eq:JdH}
\end{equation}
where
\begin{equation}
  \{\U,\V\}[\mu] :=  \int\del{\U}{\mu}\bdot\J\del{\V}{\mu}
  \label{eq:PBdef}
\end{equation}
for any functionals $\U,\V$.  Here, $\mu$ represents the state of
the system as a ``point'' in an infinite-dimensional phase space;
$\H$ is the \emph{Hamiltonian}; and $\{\,,\hspace{.1em}\}$ and $\J$
are called \emph{Poisson bracket} and \emph{operator}, respectively.
The bracket is assumed to satisfy two properties that do not follow
from its definition, namely, $\{\U+\V,\W\} = \{\U,\W\} + \{\V,\W\}$
(bilinearity) and $\{\U\V,\W\} = \U\{\V,\W\} + \{\U,\W\}\V$ (Leibniz
rule) for all functionals of state $\U,\V,\W$.  The properties in
question are: $\{\U,\V\} = - \{\V,\U\}$ (antisymmetry) and
$\{\{\U,\V\},\W\} + \cp = 0$ where $\cp$ denotes the two other terms
obtained by cyclic permutation of the functionals (Jacobi identity).

Setting $\mu = (\m,\rho)$ and identifying $\H$ with $\E$ as given
by \eqref{eq:E}, from \eqref{eq:il1-lp} it readily follows that the
Poisson operator for \eqref{eq:il1-} is
\begin{equation}
  \J = -
  \begin{pmatrix}
  \lie_{(\cdot)}\m & \rho_1\nabla(\cdot) & \cdots & \rho_{\alpha+2}\nabla(\cdot)\\
  \nabla\cdot \rho_1(\cdot) & 0 & \cdots & 0 \\
  \vdots & \vdots & \ddots & \vdots \\
  \nabla\cdot \rho_{\alpha+2}(\cdot) & 0 & \cdots & 0
  \end{pmatrix}.
  \label{eq:J}
\end{equation}
This leads, upon integration by parts, to the Poisson bracket
\begin{subequations}
\begin{equation}
  \{\U,\V\}[\m,\rho] = \{\U,\V\}^{\m} +
  \sum\{\U,\V\}^{\rho_{n_\alpha}}  
\end{equation}
where
\begin{align}
  \{\U,\V\}^{\m}\!&:=\!-\!\int \m\cdot \left[\del{\U}{\m},\del{\V}{\m}\right],\\
  \{\U,\V\}^{\rho_{n_\alpha}} \!&:=\!-\!\int
  \!\rho_{(n_\alpha)}\!\left(\del{\U}{\m}\!\cdot\!\nabla\del{\V}{\rho_{(n_\alpha)}}\!-\!
  \del{\V}{\m}\!\cdot\!\nabla\del{\U}{\rho_{(n_\alpha)}}\right).
\end{align}
\label{eq:PB}%
\end{subequations}

\begin{remark}
  To guarantee cancellation of the boundary term $\oint_{\partial
  D} (\m\cdot\U_{\m} + \rho_{n_\alpha}\U_{\rho_{n_\alpha}})\,\z\times\V_{\m}
  \,\cdot\,\d{\x}$ when the flow domain has a solid boundary, we
  require, as is customary, \citep{McIntyre-Shepherd-87}
  $\U_{\m}\cdot\n\vert_{\partial D} = 0$ for any $\U$. The space
  of admissible functionals of course is assumed \citep{Morrison-98}
  to be closed: the bracket of two admissible functionals produces
  an admissible functional.  Alternatively, boundary terms may be
  added to the definition of the bracket, \citep{Lewis-etal-86} yet
  not without imposing a restriction on the class of admissible
  functionals.
\end{remark}

The bracket \eqref{eq:PB} is manifestly antisymmetric, and indeed
satisfies the Jacobi identity.  A proof is given in App.\@~\ref{app:jac}.
This leads to an extension of an earlier geometric mechanics result
by \citet{Marsden-etal-84}, outlined in App.\@~\ref{app:lie}, where
a much deeper geometric interpretation of the \eqref{eq:il1-lp}
is provided.

\begin{remark}
  The Euler--Poincare variational formulation \eqref{eq:il1-ep} and
  the equivalent Lie--Poisson Hamiltonian formulation \eqref{eq:il1-lp}
  of the \IL PE are equivalent. The connection between them is
  provided by the transformation $(\u,\rho) \mapsto (\m,\rho)$
  defined by \eqref{eq:E}, which represents a \emph{partial Legendre
  transformation} \citep{Holm-etal-98a} analogous to that connecting
  the Lagrangian and (canonical) Hamiltonian formulations of
  mechanical systems (cf.\@~further details in App.\@~\ref{app:lie}).
\end{remark}

\subsubsection{Conservation laws}

More generally than \eqref{eq:JdH}, one has $\dot\U = \{\U,\H\}$
for any (admissible) functional of state $\U$.  By the antisymmetry
of the bracket, $\H$ commutes with itself in the bracket. Thus $\H
(=\E)$ is conserved under the dynamics, which can be verified
directly from \eqref{eq:il1-} (after multiplying the first equation
by $\u$ and integrating by parts using continuity of $\rho_1$).
This conservation law is a consequence of \emph{Noether's theorem},
which relates it to an explicit symmetry as follows.

\paragraph{Noether's theorem.}

Consider the one-parameter family of infinitesimal variations
generated by a functional $\G$ defined by \citep{Shepherd-90}
\begin{equation}
  \delta_\G := -\varepsilon\{\G,\cdot\,\},
\end{equation}
where $\varepsilon > 0$ is small. The change generated by $\G$ on
any functional $\U$ is
\begin{equation}
  \Delta_\G\U := \U[\mu + \delta_\G\mu] - \U[\mu] \sim \varepsilon\{\U,\G\}.
  \label{eq:DGU}
\end{equation}
Consequently,
\begin{equation}
  \dot{\Delta_\G\U} - \Delta_\G \dot\U \sim \varepsilon\{\U,
  \dot\G\}.
  \label{eq:sym}
\end{equation}
The condition $\{\U,\dot\G\} = 0$ is an expression of Noether's
theorem.  It says that $\G$ induces a symmetry in the most general
sense that \emph{applying a transformation and ``letting the time
run'' is the same as performing these operations in reverse
order}.\citep{Ripa-RMF-92a} Now if $\dot\G = 0$, i.e., $\G$ is an
integral of motion, then $\G$ induces a symmetry. Furthermore, in
such a case $\G$ \emph{transforms solutions into solutions}: since
$\partial_t\Delta_{\G}\mu = \Delta_{\G}\partial_t\mu$, on one hand,
and $\Delta_{\G}\{\mu,\H\} = \{\Delta_{\G}\mu,\H\}$, on the other.
The reciprocal of this theorem is not strictly true.  To see this,
one first must note that nontrivial solutions $\C[\mu]$, called
\emph{Casimirs}, of
\begin{equation}
  \{\U,\C\} = 0\,\forall\U\Longleftarrow 
  \{\mu,\C\} = \J\del{\C}{\mu}
  = 0
  \label{eq:C}
\end{equation}
represent integrals of motion that are not related to explicit
symmetries.\footnote{{The Casimirs are related to
the particle relabelling symmetry of fluid dynamics, which permits
one to switch between the Lagrangian and Eulerian descriptions
\citep{Marsden-etal-84}.}} Indeed, $\C$ does not generate any
variation, i.e., $\delta_\C\mu = -\varepsilon\{\C,\mu\} =
\varepsilon\J\C_{\mu} \equiv 0$.  As for the reciprocal of the
Noether's theorem, if $\smash{\dot{\Delta_\G\U}} - \smash{\Delta_\G\dot\U}
= 0$, then $\dot\G$ is equal to a \emph{distinguished function}
$F(\C)$, for any $F$, since $\{\U,F(\C)\} \equiv 0$.  However, if
$\G$ is replaced by $\smash{\tilde\G = \G - \int^t F\d{t}}$, which
does not alter \eqref{eq:sym}, then $\smash{\tilde\G}$ is
conserved.\citep{Ripa-RMF-92a} Clearly, finding the generator $\G$
of a given symmetry is not in general a simple task.  It is easier
to start with $\G$ and check if this leaves $\H$ invariant.  In
such a case it will be conserved, by \eqref{eq:DGU}, and as a result
it will represent a symmetry.

\paragraph{Energy and momenta.}

Assume that $\G$ is the generator of an infinitesimal time shift
$t\to t-\varepsilon$, i.e., $\delta_\G\mu = -\varepsilon\partial_t\mu$.
Clearly, $\G = -\H$ (modulo a Casimir).  Now, $\Delta_{\H}\H \sim
\varepsilon \dot\H$ by \eqref{eq:DGU}.  So invariance of the $\H$
under time shifts, i.e., generated by $\H$ itself, is equivalent
to conservation of $\H$ (modulo a Casimir). This implies
$\smash{\dot{\Delta_\H\U}} \equiv \smash{\Delta_\H\dot\U}$ for any
$\U$, by \eqref{eq:sym}.

Now, let $\hat{\boldsymbol\chi}$ indicate a direction in $D\subset
\mathbb R^2$, and suppose that $\G$ is the generator of an infinitesimal
translation along $\chi = \x\cdot \hat{\boldsymbol\chi}$, $\chi \to
\chi-\varepsilon$.  Then $\delta_\G\mu = -\varepsilon\partial_\chi\mu$.
Furthermore, \emph{define} $\M^\chi$ such that $\partial_\chi\mu =
- \{\mu,\M^\chi\}$ and identify it with the $\chi$-\emph{momentum}.
Then $\G = \M^\chi$.  Consequently, $\Delta_{\M^\chi}\H \sim
\varepsilon \dot \M^\chi$ by \eqref{eq:DGU}.  In this case, invariance
of the $\H$ under $\chi$-translations, i.e., generated by $\M^\chi$,
is equivalent to conservation of $\M^\chi$ (modulo a Casimir).
Consequently, by\eqref{eq:sym}, $\smash{\dot{\Delta_{\M^\chi}\U}}
\equiv \smash{\Delta_{\M^\chi}\dot\U}$ for any $\U$.

Examples of conserved momenta are
\begin{equation}
  \M^x := \int \m\cdot\hat{\mathbf x} = \int \rho_1\big(\bar u + f_0
  y + \tfrac{1}{2}\beta y^2\big)
  \label{eq:M}
\end{equation}
when the flow domain $D$ is zonally symmetric, and
\begin{equation}
  \M^\phi :=  \int (r\hat{\mathbf r}\times\m) \cdot \z = \int
  \rho_1 \big(\bar u_\phi r + \tfrac{1}{2}f_0r^2\big)
  \label{eq:A}
\end{equation}
where $r$ (resp., $\phi$) is the radial (resp., azimuthal) coordinate,
and $\bar u_\phi$ is the azimuthal velocity component, which requires
$D$ to be an axisymmetric domain of the \emph{$f$-plane}.  

Indeed, when $D$ is $x$-symmetric, $\H$ is invariant under
$x$-translations.  Now, in order for $\{\rho_{n_\alpha},\M^x\} =
-\nabla\cdot\rho_{n_\alpha}\M^x_{\m} = -\partial_x\rho_{n_\alpha}$,
it must be $\M^x_{\m} = \x$.  On the other hand, $\{\m,\M^x\} = -
\lie_{\M^x_{\m}}\m - \rho_{n_\alpha}\nabla\M^x_{\rho_{n_\alpha}} =
- \partial_x\m$ iff $\M^x_{\rho_{n_\alpha}} = 0$ for all $n_\alpha$.
Hence, \eqref{eq:M} follows.

In turn, when $D$ is axisymmetric, $\H$ is invariant under rotations.
In  this case, $\smash{\{\rho_{n_\alpha},\M^\phi\}} =
-\smash{\nabla\cdot\rho_{n_\alpha}\M^\phi_{\m}} =
-\smash{\partial_\phi\rho_{n_\alpha}}$ requires $\smash{\M^\phi_{\m}
= r\hat{\boldsymbol\phi}}$.  On the other hand, $\smash{\{\m,\M^\phi\}}
= - \smash{\lie_{\M^\phi_{\m}}\m} -
\smash{\rho_{n_\alpha}\nabla\M^\phi_{\rho_{n_\alpha}} = -
\partial_\phi\m}$ iff $\smash{\M^\phi_{\rho_{n_\alpha}} = 0}$ for
all $n_\alpha$, and finally \eqref{eq:A} follows.

\begin{table}
  \centering \renewcommand*{\arraystretch}{1.5}
  \begin{tabular}{cccc}
	 \hline\hline
	 \multicolumn{1}{r}{$\alpha$} &&&
	 \multicolumn{1}{c}{$\C = \int $}	 
	 \\
	 \multicolumn{1}{r}{$-1$} &&& 
	 \multicolumn{1}{l}{$\rho_1F(\bar q)$} 	 
	 \\
	 \multicolumn{1}{r}{$0$} &&& 
	 \multicolumn{1}{l}{$\rho_1\bar q F(\tilde\rho_2) + \rho_1G(\tilde\rho_2)$}
	 \\
	 \multicolumn{1}{r}{$\ge1$} &&& 
	 \multicolumn{1}{l}{$\rho_1\bar q + \rho_1
	 F(\tilde\rho_2,\dotsc,\tilde\rho_{\alpha+2})$}
	 \\
	 \hline
  \end{tabular}
  \renewcommand*{\arraystretch}{0.5} \caption{Casimirs of the Poisson
  bracket \eqref{eq:PB}, where $F$ and $G$ are arbitrary.} 
  \label{tab:cas}
\end{table}

\begin{remark}
  The quantity $\rho_1^{-1}\m\cdot\hat{\mathbf x} = \bar u + f_0 y
  + \smash{\tfrac{1}{2}\beta y^2}$ actually is the $\beta$-plane
  representation of the \emph{angular} momentum per unit mass, with
  respect to the center of the Earth and in the direction of its
  axis of rotation.\citep{Ripa-JPO-97b, Beron-03} This should not
  be confused with $\rho_1^{-1}(r\hat{\mathbf r}\times\m)\cdot \z
  = \bar u_\phi r + \smash{\tfrac{1}{2}f_0r^2}$, which has the
  interpretation of angular momentum per unit mass, but as a
  symmetry-inducing generator of infinitesimal rotations on the
  $f$-plane.
\end{remark}

\paragraph{Casimirs.}

These are listed in Table \ref{tab:cas}.  The particular form taken
by each of them depends on the number of densities considered.
Conservation of these quantities can be verified directly; to
\emph{obtain} them from \eqref{eq:C}, as we do it below, instead
of the Poisson operator \eqref{eq:J}, it is more convenient to use
$\J =  -$\arraycolsep.05cm
\begin{equation} 
  \begin{pmatrix} 
	 \bar q \z \times (\cdot) & \nabla(\cdot)  &
	 -(\cdot)\rho_1^{-1}\nabla\tilde\rho_2 & \cdots &
	 -(\cdot)\rho_1^{-1}\nabla\tilde\rho_{\alpha+2}\\
    \nabla\cdot (\cdot) & 0 & 0 & \cdots & 0\\
    \rho_1^{-1}(\cdot)\cdot\nabla\tilde\rho_2 & 0 & 0 & \cdots & 0\\
	 \vdots & \vdots & \vdots & \ddots & \vdots\\
	 \rho_1^{-1}(\cdot)\cdot\nabla\tilde\rho_{\alpha+2} & 0 & 0 & \cdots & 0\\
  \end{pmatrix};
  \label{eq:Jpedro}
\end{equation}
\arraycolsep.1cm this gives a Poisson bracket, $\{\U,\V\}[\u, \rho_1,
\tilde\rho] = - \int$
\begin{multline}
  \bar q\z\cdot \del{\V}{\u}\times \del{\U}{\u}\\
  \hspace{-7em}+ \del{\U}{\rho_1}\nabla\cdot\del{\V}{\u}
  - \del{\V}{\rho_1}\nabla\cdot\del{\U}{\u}\\
  -\rho_1^{-1}\nabla\tilde\rho_{\tilde n_\alpha}\cdot
  \left(\del{\U}{\tilde\rho_{\tilde n_\alpha}}\del{\V}{\u} -
  \del{\V}{\tilde\rho_{\tilde n_\alpha}}\del{\U}{\u}\right),
  \label{eq:PBpedro}
\end{multline}
which reduces to \eqref{eq:PB} by applying the chain rule,
\begin{align}
  \del{}{\u} &= \rho_1\del{}{\m},\\
  \left.\del{}{\rho_1}\right\vert_{\u,\tilde\rho} &=
  \left.\del{}{\rho_1}\right\vert_{\m,\rho_{\tilde n_\alpha}} +
  \frac{\m}{\rho_1}\cdot\del{}{\m} +
  \frac{\rho_{n_\alpha}}{\rho_1}
  \del{}{\rho_{n_\alpha}},\\
  \del{}{\tilde\rho_{\tilde n_\alpha}} &=
  \rho_1\del{}{\rho_{\tilde n_\alpha}},
\end{align}
and substantial cancellation. The $\J$ in \eqref{eq:Jpedro} is
\emph{suggested} when the momentum equation of the \IL PE model is
written in the form $\partial_t\u + \rho_1\bar q \z\times\u +
\nabla\smash{\frac{1}{2}}|\u|^2+
\rho_1^{-1}\nabla\rho_1^2\tilde\partial_{\rho_1}\varphi = 0$ and
using \eqref{eq:E} as a Hamiltonian, but viewed as a functional of
$(\u, \rho_1, \tilde\rho)$. An explicit proof for the Jacobi identity
of the (Poisson) bracket \eqref{eq:PBpedro} for $(\u, \rho_1,
\tilde\rho_2)$ is given in the appendix of \citet{Ripa-GAFD-93}.

Consider the case $\alpha\ge 1$. Using the Poisson operator in
\eqref{eq:Jpedro} and the variables $(\u,\rho_1,\tilde\rho)$, we
have $\{\tilde\rho_{\tilde n_\alpha},\C\} =
-\rho_1^{-1}\C_{\u}\cdot\nabla\tilde\rho_{\tilde n_\alpha} = 0$,
from which it follows that $\C_{\u} = 0$ (since it must hold for
all $\alpha\ge 1$).  This gives $\{\rho_1,\C\} = \nabla\cdot\C_{\u}
= 0$, trivially.  Finally, $\{\u,\C\} = \nabla\C_{\rho_1} -
\rho_1^{-1}\C_{\tilde\rho_{n_\alpha}}\nabla \tilde\rho_{n_\alpha}
= 0$ is fulfilled by the Casimir in the bottom row of Table
\ref{tab:cas} (where $F$ is arbitrary) since
\begin{equation}
  \delta\bar q = \rho_1^{-1} \z\cdot\nabla\times\delta\u -
  \rho_1^{-1}\bar q\delta\rho_1,
  \label{eq:dq}
\end{equation}
which guarantees that $\C_{\u} \equiv 0$ and $\C_{\rho_1} =
F(\tilde\rho)$.

Consider now the case $\alpha=0$.  We have $\{\tilde\rho_2,\C\} =
- \rho_1^{-1}\C_{\u}\cdot\nabla\tilde\rho_2 = 0$, which is satisfied
by $\C_{\u} = -\z\times\nabla F(\tilde\rho_2)$ for arbitrary $F$
(the sign is arbitrary as is the sign of each of the terms in the
Casimir listed in the second-to-bottom row of Table \ref{tab:cas}).
From $\{\u,\C\} = - \bar q\z\times \C_{\u} + \nabla\C_{\rho_1} -
\rho_1^{-1}\C_{\tilde\rho_2}\nabla\tilde\rho_2 = 0$ it follows, on
one hand, that $\C_{\rho_1} = - G(\tilde\rho_2)$ for arbitrary $G$
since $\C_{\u} \perp \nabla\tilde\rho_2$.  On the other hand, we
have $\C_{\tilde\rho_2}\nabla\tilde\rho_2 = \rho_1(\bar q \nabla F
+ \nabla G)$, or, equivalently, $\C_{\tilde\rho_2} = \rho_1(\bar q
F' + G')$ (modulo a constant). The Casimir follows upon integrating
by parts with \eqref{eq:dq} in mind and by \eqref{eq:bc}, which
guarantees that $F(\tilde\rho_2)\vert_{\partial D} = \const$.

Finally, when $\alpha=-1$, from $\{\u,\C\} = -\bar q\z\times \C_{\u} -
\nabla \C_{\rho_1} = 0$ one has $\C_{\u} = \bar q^{-1} \z\times
\nabla \C_{\rho_1}$.  Then, $\{\rho_1,\C\} = - \nabla\cdot \C_{\u}
= 0$ iff $\C_{\rho_1} = A(\bar q) + qB(\bar q)$, for any $A,B$.
While this specifies $\C_{\u}$, obtaining $\C$ from its functional
derivates as given seems not possible.  So additional information
is needed.  This is given by $\partial_t \rho_1F(\bar q) + \nabla\cdot
\rho_1F(\bar q)\u = 0$, for any $F$, because $\bar q$ is transported
by the flow of $\u$ when $n = 1$ \eqref{eq:DqDt} and due to continuity
of $\rho_1$.  This already integrates to the Casimir in the
second-to-top row of Table \ref{tab:cas}, and thus suggests $A =
F$ and $B = F'$.  This is consistent with $\C_{\u} = \nabla F'
\times\z$.  The Casimir follows upon invoking the admissibility
condition, $\C_{\u}\cdot\n\vert_{\partial D} = 0$, which translates
to $F'(\bar q)\vert_{\partial D} = \const$, with \eqref{eq:dq} in
mind.

\begin{remark}
  The assumed iso-$\tilde\rho_{n_\alpha}$ nature of $\partial D$
  \eqref{eq:bc}, beyond guaranteeing conservation of Kelvin
  circulation(s) along (cf.\@~\ref{sec:kelvin}), is needed to warrant
  the existence of Casimirs and hence possession of generalized
  Hamiltonian structure by the \IL PE when $\alpha = 0$. Yet, this
  is no more than the admissibility condition a Casimir is assumed
  to satisfy, namely, $\C_{\u}\cdot\n\vert_{\partial D} = 0$ (or
  $\C_{\m}\cdot\n\vert_{\partial D} = 0$).
\end{remark}

\section{Thermodynamics}\label{sec:thermo}

Having covered the geometry of the \IL PE in detail, I proceed to
interpret the model as a rotating shallow-water model with generalized
thermodynamics. This is done by first showing that the rotating
shallow-water model itself is a special case, living on an invariant
subspace of the system.  Two thermodynamically active submodels are
then discussed, prior to introducing the most general class of
thermodynamically active model(s).  {For each model
family I present the PE form along with the corresponding QG version,
a basic paradigm for the study of geophysical flow
stability.\citep{Pedlosky-87}} The rotating shallow-water model is
in particular used to illustrate the procedure to derive in each
case a QG approximation, and also to introduce generic geometric
mechanics results that apply to the rest of the models.  Some
background physical setup is necessary, which I introduce first.

Let $h(\x,t)$ and $\Hr = \const$ be instantaneous and reference
(i.e., unperturbed) layer thickness, respectively. Let $\gb = \const$
be the buoyancy jump at the base of the layer in the reference state
(I will focus on the reduced gravity case; the free-surface versions
of the models to be discussed follow upon minimal reinterpretation
of the parameters, such as reinterpreting $\gb$ as $g$, the
acceleration of gravity). The \emph{instantaneous buoyancy field},
\begin{equation}
  \vartheta(\x,z,t) := - g\frac{\rho_\mathrm{top}(\x,z,t) -
  \rho_\mathrm{bot}}{\rho_0},
  \label{eq:theta}
\end{equation}
where $\rho_\mathrm{top}$ is the density in top (active) layer,
$\rho_\mathrm{bot} = \const$ is the density in the bottom (inactive)
layer, and $\rho_0 = \const$ is a reference density used in the
Boussinesq approximation.  Let $\Nr$ be the reference Brunt--Vaissalaa
frequency, namely, the vertical derivative of the unperturbed form
of \eqref{eq:theta}. We will assume $\Nr = \const$, meaning that
the reference state has \emph{uniform} vertical stratification,
i.e., the unperturbed form of \eqref{eq:theta} depends linearly in
$z$.  In such a case, the stratification within the (active) layer
is conveniently measured by the nondimensional parameter
\citep{Ripa-JFM-00}
\begin{equation}
  s := \frac{\Nr^2\Hr}{\gb} > 0.
  \label{eq:s}
\end{equation}
The quantities
\begin{equation}
  R^2 := \frac{\gb\Hr}{f_0^2},\quad L^2:=sR^2,
  \label{eq:RL}
\end{equation}
where $R$ is the external (equivalent barotropic) Rossby deformation
radius and $L$ is proportional to the internal Rossby deformation
radius (for normal mode perturbations on a reference state with
uniform stratification).  When $s$ is small, i.e., the stratification
is weak, the above scales are well separated.  Finally, consider
the rescaled vertical coordinate \citep{Ripa-JFM-95}
\begin{equation}
  \sigma := 1 + 2\frac{z}{h(\x,t)},  \label{eq:sigma}
\end{equation}
which allows one to compute a (vertical) average over $-h(\x,t) \le
z \le 0$ simply as one half the integral over $-1\le \sigma \le 1$.

\subsection{The HL$+$ family}

\subsubsection{The HL$+$PE}

As noted above, the set {$M_n := \{\tilde\rho_{\tilde
n_\alpha} = \const\}$} is an invariant manifold of the \IL PE system:
once initialized on $M_n$, the system remains on $M_n$, for all
$t$.  Identifying $\rho_1$ with layer thickness ($h$), on $M_n$,
the dynamics are controlled by
\begin{equation}
\left.  
\begin{aligned}
  \partial_t\m + \lie_{\u}\m - h\nabla K + \nabla
  h^2\varphi_{-1+}'(h)
  &= 0,\\
  \partial_t h + \nabla \cdot h\u &= 0.
\end{aligned}
\right\}
\label{eq:hl+}
\end{equation}
This is system \eqref{eq:il1-} for $\alpha = -1$, with the
identification $\rho_1 = h$ (so $\m = h(\u+\f)$ now, and on), and
noting that $\varphi_{\alpha+} = \varphi_{-1+}(h)$. For arbitrary
$\varphi_{-1+}$, the resulting system can be viewed as the equations
for a shallow layer of constant density fluid on the $\beta$-plane
with a generalized pressure ``gradient'' force, given by
$\smash{-h^{-1}\nabla h^2\varphi_{-1+}'(h)}$.  Making $\varphi_{-1+}
= \tfrac{1}{2}\gb h$, the standard rotating shallow-water equations
follow. We refer to these models respectively as HL$+$PE and HLPE
(instead of IL$^{(0,-1)+}$PE and IL$^{(0,-1)}$PE) where \emph{HL}
stands for \emph{homogeneous layer}, to make explicit that they
cannot accommodate thermodynamics as the fluid density is constant
or, equivalently, $\vartheta = \gb$.  Clearly, both model equations
admit Euler--Poincare variational formulations and form Lie--Poisson
systems, and thus the potential energy density choice $\varphi_{-1+}
= \tfrac{1}{2}\gb h + F(h)$ for some $F$ enables the investigation
of isothermal mixed-layer dynamics with forcing in a conservative
context.  {This justifies the + sign notation, used
to mean that the HL$+$PE adds the noted potentially additional
feature to the HLPE.  This notation is similarly adopted below.}

\subsubsection{The HL$+$QG}

Let $\varepsilon > 0$ be a small parameter taken to represent a
Rossby number, relating the ratio of inertial to Coriolis
forces, e.g., 
\begin{equation}
  \varepsilon = U/|f_0|R \ll 1, 
\end{equation}
where $U$ is a characteristic velocity magnitude.  In the QG scaling
\citep{Pedlosky-87} 
\begin{equation}
  (|\u|,h-\Hr,\partial_t,\beta y) = O(\varepsilon
  U, \varepsilon R, \varepsilon f_0,  \varepsilon f_0)
  \label{eq:QGscaling}
\end{equation}
Consistent with this, we write
\begin{align}
  \left.
  \begin{array}{ccccccc}
	 \u/U & = & & & \z\times\nabla\p/U & + & \dotsc,\\
	 h/\Hr & = & 1  & + & \frac{\p}{f_0R_+^2} & + & \dotsc,\\
	 O & : & 1 &  & \varepsilon &  & \varepsilon^2
  \end{array}
  \right\}
\end{align}
where $\p(\x,t)$ is a streamfunction and
\begin{equation}
  R_{-1+}^2 := \frac{2\Hr\varphi_{-1+}'(\Hr) + \Hr^2\varphi_{-1+}''(\Hr)}{f_0^2}.
  \label{eq:R+}
\end{equation}
The potential vorticity, $\bar q = \Hr^{-1}(\nabla^2\p - R_{-1+}^{-2}\p
+ f) + O(\varepsilon^2)$.  Since $\bar q$ is transported for the
HL class (the rhs of \eqref{eq:DqDt} vanishes), to lowest-order in
$\varepsilon$, i.e., $O(\varepsilon^2)$, one has that the dynamics
are controlled by
\begin{subequations}
\begin{equation}
  \partial_t\xx + \{\p,\xx\}_{xy} = 0
  \label{eq:hl+qg}
\end{equation}
where
\begin{equation}
  \nabla^2\p - R_{-1+}^{-2}\p = \xx - \beta y.
  \label{eq:hl+qg-pv}
\end{equation}
\label{eq:hl+qg-sys}%
\end{subequations}
Equation \eqref{eq:hl+qg} with the \emph{invertibility principle}
\eqref{eq:hl+qg-pv} forms the HL$+$QG model.  The standard HLQG is
recovered upon setting $\varphi_{-1+} = \smash{\frac{1}{2}}\gb\Hr$, which
gives $R_{-1+} = R$.

\begin{remark}
  A peculiarity (cf., e.g., \citet{Shepherd-90}) of \eqref{eq:hl+qg-sys}
  is that while $\dot\gamma = 0$ (namely, constancy of the circulation
  of $\u$ along $\partial D$) holds to lowest-order in $\varepsilon$,
  it cannot be deduced from \eqref{eq:hl+qg-sys}, \emph{unless}
  $R_{-1+}\uparrow\infty$ (i.e., the bottom of the layer is effectively
  rigid).  In the general case with a soft bottom, i.e., when $R_{-1+}$
  is finite, the evolution equation \eqref{eq:hl+qg} must be
  complemented with $\dot\gamma = 0$. More generally, when $D$
  includes islands, i.e., when $D$ is multiply connected, constancy
  of $\gamma$ along the boundary of each of these islands must be
  appended. If $D$ is a zonal channel, constancy of $\gamma$ along
  both coasts must be included.
\end{remark}

\paragraph{Hamiltonian structure.}

The HLQG system is well-known \citep{Morrison-98} to possess a
generalized Hamiltonian structure. Such a structure is conveyed to
\eqref{eq:hl+qg-sys} by the Hamiltonian (energy),
\begin{align}
  \H[\xx] 
  &:= \tfrac{1}{2} \int |\nabla\p|^2 + R_{-1+}^{-2}\p^2
  = - \tfrac{1}{2} \int \p(\xx-\beta y)
  \nonumber\\ 
  &\equiv - \tfrac{1}{2}\int (\xx-\beta y)(\nabla^2-R_{-1+}^{-2})^{-1}(\xx-\beta
  y),
  \label{eq:hl+qg-H}
\end{align}
by $\z\times\nabla\p\cdot\n\vert_{\partial D} = 0$ and where
$(\nabla^2-R_{-1+}^{-2})^{-1}(\xx-\beta y)$ represents a convolution of
$\xx-\beta y$ with the Green's function  of the elliptic problem
\eqref{eq:hl+qg-pv}, and the Poisson operator 
\begin{equation}
  \J = -\{\xx,\cdot\hspace{.05em}\}_{xy},
\end{equation}
so the Poisson bracket
\begin{equation}
  \{\U,\V\}[\xx] = \int \xx\left\{\del{\U}{\xx},\del{\V}{\xx}\right\}_{xy},
  \label{eq:hl+qg-PB}
\end{equation}
with the admissibility condition, $\nabla\U_{\xx}\cdot\n\vert_{\partial
D} = 0$ for all functionals $\U$ of the state variables, which above
have been taken to be composed of simply $\xx$ under the assumption
that $D$ does not include islands (or has the topology of a zonal
channel).  If this is not the case, the state variables should be
augmented to $\mu = (\xx,\gamma_1,\gamma_2,\dotsc)$ with as many
circulations as appropriate.\citep{Shepherd-90}  We will ignore
this technicality here and below, which does not have an immediate
consequence for our purposes (stability analyses \citep{Ripa-JFM-93}
do require one to account for it, though). Clearly, \eqref{eq:hl+qg}
follows from $\partial_t\xx = \{\xx,\H\} = \J\H_{\xx}$ since $\H_{\xx}
= -\p$.

For future reference, we note that \eqref{eq:hl+qg-PB} is a
special case of Lie--Poisson brackets of the general form
\citep{Thiffeault-Morrison-00}
\begin{equation}
  \{\U,\V\}[\mu] = W^{ab}_c\int\mu^c\left\{\del{\U}{\mu^a},
  \del{\V}{\mu^b}\right\}_{xy},
  \label{eq:PB-qg}
\end{equation}
where the constants $W^{ab}_c$ transform like the components of a
(2,1)-tensor under linear transformations of $\mu$. The tensor $W$
is symmetric in its upper indices, viz., 
\begin{subequations}
\begin{equation}
  W^{ab}_c = W^{ba}_c,
  \label{eq:W1}
\end{equation}
so the bracket \eqref{eq:PB-qg} is antisymmetric.  Furthermore,
$\smash{W^{ab}_cW^{a'b'}_a} = \smash{W^{ab'}_cW^{ba'}_a}$, which
guarantees it satisfies the Jacobi identity.  If $W$ is viewed as
a collection of matrices $\mathsf W^{(b)}$ with $(c,a)$-th entry
given by $\smash{W^{a(b)}_c}$, the latter property means that these
matrices commute, namely
\begin{equation}
  \mathsf W^{(a)}\mathsf W^{(b)} = \mathsf W^{(b)}\mathsf
  W^{(a)},\quad a\neq b.
  \label{eq:W2}
\end{equation}
\label{eq:W}%
\end{subequations}
In \eqref{eq:hl+qg-PB}, $W = 1$, simply.  (A deeper geometric
interpretation of the generalized Hamiltonian formulation of QG
systems, which is different than that of PE systems, is outlined
in App.\@~\ref{app:lie}.)

\begin{remark}
  Generalized Hamiltonian systems with Lie--Poisson brackets of the
  form seem \eqref{eq:PB-qg} do not seem possible to be obtained
  via a Legendre transformation as are systems with brackets of the
  form \eqref{eq:PB}.  Thus Euler--Poincare variational formulations
  may not exist or at least are difficult to be derived for QG-type
  systems.  However, ad-hoc variational formulations have been
  proposed in the literature.\cite{Virasoro-81, Holm-Zeitlin-98,
  Morrison-etal-14}
\end{remark}

\paragraph{Conservation laws.}

Symmetry-related integrals of motion of \eqref{eq:hl+qg-sys} are
the energy ($\H$), the zonal momentum 
\begin{equation}
  \M^x := \int y\xx
  \label{eq:Mx}
\end{equation}
(when $D$ is $x$-symmetric), and the angular momentum 
\begin{equation}
  \M^\phi := -\int r\xx
  \label{eq:Mphi}
\end{equation}
(on an $f$-plane when $D$ is an axisymmetric domain).\citep{Shepherd-90}
The Casimirs of the bracket \eqref{eq:PB-qg} are $\C = \int F(\xx)$
for any $F$ (e.g., \citet{Morrison-98}).

\subsection{The IL$^{0+}$ family}

\subsubsection{The IL$^{0+}$PE}

We are now ready to introduce the first family of shallow-water
models with thermodynamics.  This follows from the \IL PE model
\eqref{eq:il1-} upon setting $\alpha = 0$. Making $\rho_1 = h$, as
before, and $\tilde\rho_2 = \t$, viz., the vertical average of the
buoyancy field \eqref{eq:theta} across the layer extent, in
\eqref{eq:il1-} we obtain
\begin{equation}
\left.  
\begin{aligned}
  \partial_t\m + \lie_{\u}\m - h\nabla K + \nabla
  h^2\partial_h\varphi_{0+}(h,\t) &= 0,\\
  \partial_t h + \nabla \cdot h\u &= 0,\\
  \partial_t\t + \u\cdot\nabla\t &= 0.
\end{aligned}
\right\}
\label{eq:il0+}
\end{equation}
Choosing the potential energy density as
\begin{equation}
  \varphi_{0+} := \varphi_0 = \tfrac{1}{2}h\t
  \label{eq:phi0}
\end{equation}
reduces \eqref{eq:il0+} to the IL$^0$PE model,\citep{Ripa-GAFD-93}
which is the rotating shallow-water model with buoyancy varying in
horizontal position and time, i.e., thermodynamically active, that
we sought to extend. No variation in the vertical is allowed for
the dynamical fields, which justifies the superscript in IL$^0$PE.
The IL$^0$PE formally follows from the IL$^\infty$PE by replacing
the horizontal velocity and buoyancy in the model by their vertical
averages, $\u$ and $\t$, respectively, and further by vertically
averaging the resulting pressure gradient, namely,
\begin{equation}
  \nabla p = \nabla(z+h)\t,
\end{equation}
which gives 
\begin{equation}
  \overline{\nabla p} = \tfrac{1}{2}h^{-1}\nabla h^2\t \equiv
  h^{-1}\nabla h^2 \partial_h\varphi_0
  \label{eq:nablap-il0}
\end{equation}
with $\varphi_0$ as in \eqref{eq:phi0}.  It must be noted, however,
that while the horizontal velocity is set to $\u$, because
$\nabla\partial_z p = \nabla\t$, the velocity includes \citep{Ripa-JGR-96}
a linear vertical shear, implicitly, as it follows from the
thermal--wind balance, which dominates at low frequency; we will
return to this in the section that follows.  Clearly, \eqref{eq:nablap-il0}
is a special case of $\smash{h^{-1}\nabla h^2
\partial_h\varphi_{0+}(h,\t)}$ for arbitrary $\varphi_{0+}$, which
extends the IL$^0$PE to \eqref{eq:il0+}, referred here to as the
IL$^{0+}$PE.  With a generalized pressure ``gradient'' force, the
IL$^{0+}$PE model has the potential of expanding the realm of
applicability of the IL$^0$PE system. A suitable choice of
$\varphi_{0+}$ can allow one to study forced--dissipative mixed-layer
hydrodynamics with thermodynamics in a conservative setting, as the
IL$^{0+}$PE admits an Euler--Poincare formulation and has a generalized
Hamiltonian structure.

\paragraph{Connection with \citet{Morrison-Greene-80}.}

The nonrotating form of the IL$^{0+}$PE, i.e., with $f=0$, follows
from the magnetohydrodynamics (MHD) model considered by
\citet{Morrison-Greene-80} upon neglecting the magnetic field, i.e.,
$\smash{\vec{\mathrm B}} = 0$ in the notation of that paper,
particularizing the resulting system to two-space dimensions, and
reinterpreting the mass density ($\rho$, in the notation of
\citet{Morrison-Greene-80}) and entropy per unit mass ($s$, in their
notation) variables as layer thickness ($h$) and vertically averaged
buoyancy ($\t$), respectively, in the internal energy per unit mass,
$U(\rho,s)$.  That the (nonrotating) IL$^0$PE is thus equivalent
to Morrison and Greene's hydrodynamics system with $U(\rho,s) =
\smash{\frac{1}{2}\rho s}$ had remained elusive until
present.\footnote{\citet{Dellar-03} writes the Lie--Poisson bracket
for the IL$^0$PE, viz., \eqref{eq:PB} for $\alpha = 0$, referring
to \citet{Holm-etal-85}, where a planar compressible MHD system is
discussed in relation with that of \citet{Morrison-Greene-80}.}

\subsubsection{The IL$^{0+}$QG}

Consistent with the QG scaling \eqref{eq:QGscaling}, consider
\begin{align}
  \left.
  \begin{array}{ccccccc}
	 \u/U & = & & & \z\times\nabla\p/U & + & \dotsc,\\
	 h/\Hr & = & 1 & + & \frac{\p -
	 2\partial_{\Hr\gb}\Phi_0\ps}{f_0R_{0+}^2} & + & \dotsc,\\
	 \t/g_\mathrm{b} & = & 1 & + & \frac{2}{f_0R_{0+}^2}\ps & + &
	 \dotsc,\\
    O & : & 1 &  & \varepsilon &  & \varepsilon^2
  \end{array}
  \right\}
  \label{eq:il0exp}
\end{align}
where
\begin{equation}
  R_{0+}^2 := \frac{2\Hr\partial_{\Hr}\Phi_{0+} +
  \Hr^2\partial_{\Hr}^2\Phi_{0+}}{f_0^2},
  \label{eq:R0}
\end{equation}
and the shorthand notation $\Phi_{0+}$ for $\varphi_{0+}(\Hr,\gb)$ was
introduced.  The notation $\ps(\x,t)$ is clarified by noting that
\begin{equation}
  \partial_z\mathbf u = \frac{2\gb}{f_0^2R_{0+}^2} \z \times \nabla\ps
  + O(\varepsilon^2)
\end{equation}
is the vertical shear that the horizontal velocity \emph{implicitly}
has by the thermal-wind balance. When $\Phi_{0+} = \smash{\frac{1}{2}}\Hr\gb$,
the streamfunction would read (with \eqref{eq:RL} in mind) as
\begin{equation}
  \psi =\p + \big(1 + 2\tfrac{z}{\Hr}\big) \ps,
\end{equation}
which better justifies the notation (note that $\sigma = 1 +
2\smash{\frac{z}{\Hr}} + O(\varepsilon)$ by the QG scaling).  Plugging
the expansions \eqref{eq:il0exp} in the potential vorticity equation
\eqref{eq:DqDt} (with $\rho_1 = h$ and $\tilde\rho_2 = \t$, and
generic potential energy density $\varphi_{0+}$) and the equation
for $\tilde\rho_2 (= \t)$ in \eqref{eq:il0+}, we obtain, to
$O(\varepsilon^2)$, i.e., to lowest-order in $\varepsilon$, the
following set:
\begin{subequations}
\begin{equation}
\left.  
\begin{aligned}
  \partial_t\xx + \{\p,\xx\}_{xy} -
  \tilde R_{0+}^{-2}\{\p,\ps\}_{xy} &=0,\\
  \partial_t\ps + \{\p,\ps\}_{xy} &= 0,
\end{aligned}
\right\}  
\label{eq:il0+qg}
\end{equation}
where $\tilde R_{0+}^2 := (2\partial_{\Hr\gb}\Phi_{0+})^{-1} R_{0+}^{2}$ and
\begin{equation}
  \nabla^2\p - R_{0+}^{-2}\p = \xx - \tilde R_{0+}^{-2}\ps - \beta
  y =: H_{0+}(\xx,\ps).
  \label{eq:il0+qg-pv}
\end{equation}
\label{eq:il0+qg-sys}%
\end{subequations}
Equations \eqref{eq:il0+qg} with the invertibility principle
\eqref{eq:il0+qg-pv} form the IL$^{0+}$QG system.  The IL$^0$QG
model \citep{Ripa-RMF-96} follows as the special case $\Phi_{0+} =
\smash{\frac{1}{2}}\gb\Hr$, for which $\tilde R_{0+} = R_{0+} = R$.

\paragraph{Hamiltonian structure.}

The IL$^{0+}$QG system \eqref{eq:il0+qg-sys} possesses a
generalized Hamiltonian structure endowed by the Hamiltonian
\begin{align}
  \H[\xx,\ps] 
  &:= \tfrac{1}{2} \int |\nabla\p|^2 + R_{0+}^{-2}\p^2 
  = - \tfrac{1}{2} \int \p H_{0+}(\xx,\ps)
  \nonumber\\ 
  &\equiv - \tfrac{1}{2}\int H_{0+}(\xx,\ps)
  (\nabla^2-R_{0+}^{-2})^{-1}H_{0+}(\xx,\ps),
  \label{eq:il0+qg-H}
\end{align}
since $\z\times\nabla\p\cdot\n\vert_{\partial D} = 0$ and where
$(\nabla^2-R_{0+}^{-2})^{-1}H_{0+}(\xx,\ps)$ represents a convolution of
$H_{0+}(\xx,\ps)$ with the Green's function of the elliptic problem
\eqref{eq:il0+qg-pv}, and the Poisson operator
\begin{equation}
  \J = -
  \begin{pmatrix}
	 \phantom{-}\{\xx,\cdot\}_{xy} & \{\ps,\cdot\}_{xy}\\
    \{\ps,\cdot\}_{xy} & 0
   \end{pmatrix},
	\label{eq:J-il0+qg}
\end{equation}
which leads to the Lie--Poisson bracket \eqref{eq:PB-qg} on $\mu =
(\xx,\ps)$ with $W^{11}_1 = W^{12}_2 = W^{21}_2 = 1$ and zero
otherwise (note that $W$ satisfies the required symmetry property
\eqref{eq:W1} and further the resulting $\mathsf W$'s satisfy
\eqref{eq:W2} since $\mathsf W^{(1)} = \Id^{2\times2}$ and
\begin{equation}
  \mathsf W^{(2)} = 
  \begin{pmatrix}
	 0 & 0\\
	 1 & 0
  \end{pmatrix},
\end{equation}
which commute). The first and second equations in \eqref{eq:il0+qg}
follow from $\partial_t\xx = \{\xx,\H\}$ and $\partial_t\ps =
\{\ps,\H\}$, respectively, upon noting that.
\begin{equation}
  \del{\H}{\xx} = -\p,\quad
  \del{\H}{\ps} = \tilde R_{0+}^{-2}\p.
\end{equation}

\begin{remark}
  The Poisson operator \eqref{eq:J-il0+qg} and corresponding
  Lie--Poisson bracket turn out to be the same as those for
  ``low-$\beta$'' reduced magnetohydrodynamics
  \citep{Morrison-Hazeltine-84} and incompressible, nonhydrostatic,
  Boussinesq fluid dynamics on a vertical plane.\citep{Benjamin-84}
\end{remark}

\paragraph{Conservation laws.}

The Hamiltonian ($\H$) of the IL$^{0+}$QG \eqref{eq:il0+qg-H} is
invariant under time shifts.  As the generator of infinitesimal
such transformations,  by Noether's theorem $\H$ (i.e., the energy)
is preserved under dynamics of \eqref{eq:il0+qg-sys}.  In a zonally
symmetric domain $D$ of the $\beta$-plane, $\H$ is invariant under
$x$-translations.  The corresponding generator ($\M^x$) must be
conserved.  Since $\M^x$ must satisfy $\smash{\{\ps,\M^x_{\xx}\}_{xy}}
= \partial_x\ps$, it follows that $\M^x$ is given by \eqref{eq:Mx},
just as in the HL$+$QG model, sufficiently and necessarily: given
any two vectors $\a,\b$ on $D$, $\a\times\b = a_1\z$ iff $\b =
(1,0)$.  This immediately gives $\smash{\{\xx,\M^x_{\xx}\}_{xy}} +
\smash{\{\ps,\M^x_{\ps}\}_{xy}} = \partial_x\xx$. The
conservation of $\M^x$ can be verified directly from \eqref{eq:il0+qg-sys}
with a careful examination of the boundary terms.  On an $f$-plane
the Hamiltonian is invariant under rotations in an axisymmetric
domain.  The generator $\M^\phi$ of infinitesimal rotations is
conserved by Noether's theorem. This must satisfy
$\smash{\{\ps,\M^\phi_{\xx}\}_{r\phi}} = r\partial_\phi\ps$, which
holds iff $\M^\phi$ is given by \eqref{eq:Mphi}, as for
the HL$+$QG.  This immediately makes $\smash{\{\xx,\M^\phi_{\xx}\}_{r\phi}}
+ \smash{\{\ps,\M^\phi_{\ps}\}_{r\phi}} = r\partial_\phi\xx$.
Finally, the Casimirs of \eqref{eq:J-il0+qg} are given by $\C =
\int \xx F(\ps) + G(\ps)$ where $F,G$ are arbitrary, which have
been known for a long time. \citep{Morrison-Hazeltine-84, Benjamin-84,
Ripa-RMF-96}

\subsection{The IL$^{(0,1)+}$ family}\label{sec:IL01}

\subsubsection{The IL$^{(0,1)+}$PE}

One step above the IL$^0$ class in dynamical richness is the family
of shallow-water models with thermodynamics arise from \eqref{eq:il1-}
with $\alpha = 1$, assuming that the buoyancy field not only varies
in the horizontal and time, but also in the vertical, \emph{linearly}.
Namely,
\begin{equation}
  \vartheta(\x,\sigma,t) = \t(\x,t) + \sigma \ts(\x,t).
  \label{eq:theta-il05}
\end{equation}
Making $\rho_1 = h$ and $\tilde\rho_2 = \t$ as in the previous
section, and further setting $\tilde\rho_3 = \ts$, the equations
of the model read
\begin{equation}
\left.  
\begin{aligned}
  \partial_t\m + \lie_{\u}\m - h\nabla K + \nabla
  h^2\partial_h\varphi_{1+}(h,\t,\ts) &= 0,\\
  \partial_t h + \nabla \cdot h\u &= 0,\\
  \partial_t\t + \u\cdot\nabla\t &= 0,\\
  \partial_t\ts + \u\cdot\nabla\ts &= 0,
\end{aligned}
\right\}
\label{eq:il05+}
\end{equation}
where $\varphi_{1+}$ is arbitrary. 

The specific potential energy density choice 
\begin{equation}
  \varphi_{1+} = \varphi_{1} := \tfrac{1}{2}h\big(\t -
  \tfrac{1}{3}\ts\big) 
  \label{eq:phi05}
\end{equation}
leads to a model, which will be called the IL$^{(0,1)}$PE, that
offers a better representation of thermodynamics than the IL$^0$
class.  The first slot in the superscript in IL$^{(0,1)}$PE means
that velocity does not vary with the vertical coordinate, while the
second slot that buoyancy varies linearly with it. Important processes
such as \emph{mixed-layer restratification}, which can be be expected
\citep{Haine-Marshall-98} to result from baroclinic instability,
can now be represented. This follows from the freedom of the buoyancy
in the IL$^{(0,1)}$PE to vary, albeit linearly, in the vertical;
cf.\ \eqref{eq:theta-il05}.  Implicitly by the thermal--wind balance,
the velocity in the IL$^{(0,1)}$PE has quadratic vertical shear.

\begin{remark}
Ignoring forcing and fluid exchanges across interfaces, a model
similar to the IL$^{(0,1)}$PE was used by \citet{Schopf-Cane-83}
in an intermediate layer of a multilayer model to study the interplay
between hydrodynamics and thermodynamics in the equatorial near-surface
ocean.
\end{remark}

The IL$^{(0,1)}$PE can be obtained from the IL$^{\infty}$PE in a
series of steps as follows.

\noindent\textbf{1.} Replace the horizontal velocity with its vertical
average, $\u$, and the buoyancy field by \eqref{eq:theta-il05}.

\noindent\textbf{2.} Vertically average the ensuing horizontal pressure
gradient, viz.,
\begin{equation}
  \nabla p = (z+h)\t + z\left(1+\tfrac{z}{h}\right)\nabla\ts
  + \big(\t - \tfrac{z^2}{h^2}\ts\big)\nabla h.
\end{equation}
The result is 
\begin{equation}
  \overline{\nabla p} = \tfrac{1}{2}h^{-1}\nabla h^2(\t -
  \tfrac{1}{3}\ts) \equiv h^{-1}\nabla h^2
  \partial_h\varphi_{1};
  \label{eq:nablap-il05}
\end{equation}
cf.\@~\eqref{eq:phi05}. 

\noindent\textbf{3.} Note that consistent with the horizontal velocity being
independent of the vertical coordinate is the \emph{vertical}
velocity depending linearly in it:
\begin{equation}
  w_{\u} := \frac{\sigma - 1}{2} \Dt{h}.
  \label{eq:w}
\end{equation}
\noindent\textbf{4.} Realize that transportation under the flow of $\u$
of $\t$ and $\ts$ implies transportation of \eqref{eq:theta-il05}
under the flow of $(\u, w_{\u})$. 

To see the last step, it is best to write the three-dimensional
material derivative in the IL$^\infty$PE as \citep{Ripa-JFM-95}
$\smash{\partial_t\vert_\sigma} + \mathbf u\cdot\smash{\nabla\vert_\sigma}
+ w_\sigma\partial_\sigma$, where the $\sigma$-vertical velocity
\begin{equation}
  w_\sigma := \frac{2}{h}\left(\frac{1-\sigma}{h}(\partial_t h +
  \mathbf u\cdot\nabla h) + w\right)
\end{equation}
vanishes for vertically averaged horizontal velocity, $\mathbf u =
\u$, and $z$-vertical vertical $w =  w_{\u}$, i.e., as given in
\eqref{eq:w}.

Clearly, \eqref{eq:nablap-il05} represents a particular case of
$\smash{h^{-1}\nabla h^2\partial_h\varphi_{(0,1)+}(h,\t,\ts)}$ for
arbitrary $\varphi_{(0,1)+}$, which extends the IL$^{(0,1)}$PE to
\eqref{eq:il05+}.  We refer to \eqref{eq:il05+} as the IL$^{(0,1)+}$PE,
which has the power of expanding the domain of applicability of the
IL$^{(0,1)}$PE.  As noted for the IL$^0$ class, an appropriate
choice of $\varphi_{(0,1)+}$ enables investigation of mixed-layer
processes, such as restratification by baroclinic instability, with
forcing/dissipation, but in a Hamiltonian (i.e., conservative)
setting.

\subsubsection{The IL$^{(0,1)}$QG}

Next I will present a derivation of the QG approximation to the
IL$^{(0,1)}$PE.  Working with a generic energy potential density
$\varphi_{1+}$ complicates the algebra without shedding much light
on the problem.  If a specific choice of $\varphi_{1+}$ turns out
to be relevant, then steps similar to those below can be taken to
derive the corresponding QG set of equations.

With the above in mind, consistent with the QG scaling
\eqref{eq:QGscaling}, the following expansions are proposed:
\begin{align}
  \left.
  \begin{array}{ccccccc}
	 \u/U & = & & & \z\times\nabla\p/U & + & \dotsc,\\
	 h/H_\mathrm{r} & = & 1 & + & \frac{\p - \ps +
	 \frac{2}{3}\pss}{f_0R_1^2} & + & \dotsc,\\
    \t/g_\mathrm{b} & = & 1 & + & \frac{2}{f_0R^2}\ps & + & \dotsc,\\
	 \ts/\frac{1}{2}N^2_\mathrm{r}H_\mathrm{r} & = & 1 & + &
	 \frac{8}{f_0sR^2}\pss & + & \dotsc,\\
	 O & : & 1 &  & \varepsilon &  & \varepsilon^2
  \end{array}
  \right\}
  \label{eq:il05exp}
\end{align}
where  $R_1^2 := (1 - \frac{1}{6}s) R^{2}$. The origin of $\ps(\x,t)$
and $\pss(\x,t)$ is clarified upon realizing that
\begin{equation}
  \partial_z\mathbf u = \frac{2\gb}{f_0^2R^2} \z \times \nabla\ps +
  \frac{4\Nr^2\Hr}{f_0^2sR^2} \big(1 +
  2\tfrac{z}{\Hr}\big) \z \times \nabla\pss +
  O(\varepsilon^2)
\end{equation}
is the vertical shear that the horizontal velocity \emph{implicitly} has
by the thermal-wind balance. The streamfunction would then read
(with \eqref{eq:s}--\eqref{eq:RL} in mind) as 
\begin{equation}
  \psi = \p +  \big(1 + 2\tfrac{z}{\Hr}\big) \ps + \left(\big(1 +
  2\tfrac{z}{\Hr}\big)^2 - \tfrac{1}{3}\right) \pss.
\end{equation}

Plugging the expansions \eqref{eq:il05exp} in the potential vorticity
equation \eqref{eq:DqDt} (with $\rho_1 = h$, $\tilde\rho_2 = \t$,
and $\tilde\rho_3 = \ts$, the potential energy density \eqref{eq:phi05},
specifically) and the equations for $\tilde\rho_2 (= \t)$ and
$\tilde\rho_3 (= \ts)$ and in \eqref{eq:il0+}, we obtain, to
$O(\varepsilon^2)$, the following system:
\begin{subequations}
\begin{equation}
\left.  
\begin{aligned}
  \partial_t\xx + \{\p,\xi\}_{xy} - R_1^{-2}
  \{\p,\ps-\tfrac{2}{3}\pss\}_{xy} &= 0,\\
  \partial_t\ps + \{\p,\ps\}_{xy} &= 0,\\
  \partial_t\pss + \{\p,\pss\}_{xy} &= 0,
\end{aligned}
\right\}  
\label{eq:il05qg}
\end{equation}
where
\begin{align}
  \nabla^2\p - R_1^{-2}\p &= \xx - R_1^{-2} \big(\ps -
  \tfrac{2}{3}\pss\big) - \beta y \nonumber\\ &=: H_{1}(\xx,\ps,\pss).
  \label{eq:il05qg-pv}
\end{align}
\label{eq:il05qg-sys}%
\end{subequations}
Equations \eqref{eq:il05qg} with the invertibility principle
\eqref{eq:il05qg-pv} form the IL$^{(0,1)}$QG model.  The IL$^0$QG,
i.e., \eqref{eq:il0+qg} with $\tilde R_{0+} = R_{0+} = R$, is
recovered upon ignoring $\pss$ and making $s = 0$.

\paragraph{Hamiltonian structure.}

The IL$^{(0,1)}$QG system \eqref{eq:il05qg-sys} possesses a
generalized Hamiltonian structure conveyed by the Hamiltonian
\begin{align}
  \H[\xx,\ps,\pss] 
  &:= \tfrac{1}{2} \int |\nabla\p|^2 + R_1^{-2}\p^2
  \nonumber\\
  &= - \tfrac{1}{2} \int \p H_{1}(\xx,\ps,\pss)
  \nonumber\\ 
  &\equiv - \tfrac{1}{2}\int H_{1}
  (\nabla^2-R_1^{-2})^{-1}H_{1},
  \label{eq:il05qg-H}
\end{align}
where $\z\times\nabla\p\cdot\n\vert_{\partial D} = 0$ was used and
$\smash{(\nabla^2-R_1^{-2})^{-1}}H_{1}(\xx,\ps,\pss)$ is a convolution
of $H_{1}(\xx,\ps,\pss)$ with the Green's function of the elliptic
problem \eqref{eq:il0+qg-pv}, and the Poisson operator
\begin{equation}
  \J = -
  \begin{pmatrix}
	 \phantom{-}\{\xx,\cdot\}_{xy} & \{\ps,\cdot\}_{xy} &  \{\pss,\cdot\}_{xy}\\
    \{\ps,\cdot\}_{xy} & 0 & 0\\
	 \{\pss,\cdot\}_{xy} & 0 & 0
   \end{pmatrix},
	\label{eq:J-il05qg}
\end{equation}
which leads to the Lie--Poisson bracket \eqref{eq:PB-qg} on $\mu =
(\xx,\ps,\pss)$ with $W^{11}_1 = W^{12}_2 = W^{21}_2 =  W^{13}_3 =
W^{31}_3 = 1$ and zero otherwise. Note that $W$ indeed satisfies
the required symmetry property \eqref{eq:W1} and further
the $\mathsf W$'s satisfy \eqref{eq:W2} since $\mathsf W^{(1)} =
\Id^{3\times3}$, 
\begin{equation}
  \mathsf W^{(2)} = 
  \begin{pmatrix}
	 0 & 0 & 0\\
	 1 & 0 & 0\\
	 0 & 0 & 0
  \end{pmatrix},\quad
  \mathsf W^{(3)} = 
  \begin{pmatrix}
	 0 & 0 & 0\\
	 0 & 0 & 0\\
	 1 & 0 & 0
  \end{pmatrix},
\end{equation}
which commute.

Equations in \eqref{eq:il05qg} follow, in order, from $\partial_t\xx
= \{\xx,\H\}_{xy}$, $\partial_t\ps = \{\ps,\H\}_{xy}$, and
$\partial_t\pss = \{\pss,\H\}_{xy}$, upon noting that
\begin{equation}
  \del{\H}{\xx} = -\p,\quad
  \del{\H}{\ps} = R_1^{-2}\p,\quad
  \del{\H}{\pss} = - \tfrac{2}{3}R_1^{-2}\p.
\end{equation}

\paragraph{Conservation laws.}

The integrand of the Hamiltonian ($\H$) of the IL$^{(0,1)}$QG
\eqref{eq:il05qg-H} does not depend explicitly on $t$.  So $\H$ is
invariant under time shifts.  Being the generator of infinitesimal
such transformations, by Noether's theorem $\H$ (i.e., the energy)
is preserved under dynamics of \eqref{eq:il05qg-sys}.  In a zonally
symmetric domain $D$ of the $\beta$-plane, $\H$ is invariant under
$x$-translations.  The corresponding generator ($\M^x$) must be
conserved.  As for the HL$+$QG and IL$^{0+}$QG models, this turns
out be given by \eqref{eq:Mx} as shown next. Since $\M^x$ must
satisfy $\smash{\{\ps,\M^x_{\xx}\}_{xy}} = \partial_x\ps$ and
$\smash{\{\pss,\M^x_{\xx}\}_{xy}} = \partial_x\pss$, it follows
that $\M^x = \int y\xx$, sufficiently and necessarily by the same
argument given above: given any two vectors $\a,\b$ on $D$, $\a\times\b
= a_1\z$ iff $\b = (1,0)$.  This immediately gives
$\smash{\{\xx,\M^x_{\xx}\}_{xy}} + \smash{\{\ps,\M^x_{\ps}\}_{xy}}
+  \smash{\{\pss,\M^x_{\pss}\}_{xy}} = \partial_x\xx$.  On
an $f$-plane the Hamiltonian is invariant under rotations in an
axisymmetric domain.  The generator $\M^\phi$ of infinitesimal
rotations is conserved by Noether's theorem. This must satisfy
$\smash{\{\ps,\M^\phi_{\xx}\}_{r\phi}} = r\partial_\phi\ps$ and
$\smash{\{\pss,\M^\phi_{\xx}\}_{r\phi}} = r\partial_\phi\pss$, which
holds if and only if $\M^\phi$ is given by \eqref{eq:Mphi}, just
as for the HL$+$QG and IL$^{0+}$QG.  This immediately makes
$\smash{\{\xx,\M^\phi_{\xx}\}_{r\phi}} +
\smash{\{\ps,\M^\phi_{\ps}\}_{r\phi}} +
\smash{\{\pss,\M^\phi_{\pss}\}_{r\phi}} = r\partial_\phi\xx$.
Finally, the Casimirs of \eqref{eq:J-il05qg} are given by
\begin{equation}
  \C = \int \xx + F(\ps,\pss)
  \label{eq:il05qg-C}
\end{equation}
where $F$ is an arbitrary function.  Indeed, $\{\ps,\C_{\xx}\}_{xy}
= 0 = \smash{\{\pss,\C_{\xx}\}_{xy}}$ is satisfied iff $\smash{\C_{\xx}
= \const}$.  On the other hand, $\{\xx,\C_{\xx}\}_{xy} + \{\ps,\C_{\ps}\}_{xy}
+ \smash{\{\pss,\C_{\pss}\}_{xy}} = 0$ is satisfied iff
$\partial_{\pss}\C_{\ps} = \smash{\partial_{\ps}\C_{\pss}}$.  Thus
\eqref{eq:il05qg-C} follows. This Casimir should be possible
to be derived using the method developed by \citet{Thiffeault-Morrison-00}.

\begin{table*}
  \centering 
  \renewcommand*{\arraystretch}{1.5}
  \begin{tabular}{lllll}
	 \hline\hline
	 & & \multicolumn{2}{c}{Type} & \\
	 \cline{3-4}
	 \multicolumn{1}{c}{Class} & \multicolumn{1}{c}{$\varphi_{\alpha+}$} &
	 \multicolumn{1}{c}{PE} & \multicolumn{1}{c}{QG}
	 &\multicolumn{1}{c}{N.B.}\\
	 \hline
	 HL & $\tfrac{1}{2}\gb h$ & $hF(\bar q)$ & $F(\xx)$ & $\bar q
	 - f_0/\Hr \sim \xx$\\
	 IL$^0$ & $\tfrac{1}{2}\t h$ & $h\bar q F(\t) + hG(\t)$ & $\xx F(\ps) + G(\ps)$ &
	 $\t - \gb \sim \ps$\\
	 IL$^{(0,1)}$ & $\tfrac{1}{2}\big(\t - \tfrac{1}{3}\ts\big)h$ & $h\bar q + hF(\t,\ts)$ & 
	 $\xx + F(\ps,\pss)$ &
	 $\ts - \tfrac{1}{2}\Nr^2\Hr \sim \pss$\\
	 IL$^{(0,\alpha)}$ & $\tfrac{1}{2}\big(\t - \sum_1^\alpha
    \overline{\sigma^{n+1}} \vartheta_{\sigma^n}\big)h$ & $h\bar q +
	 hF(\t,\ts,\dotsc,\vartheta_{\sigma^\alpha})$ & $\xx +
	 F(\ps,\dotsc,\psi_{\sigma^{\alpha+1}})$ &
	 $\vartheta_{\sigma^n} \sim \psi_{\sigma^{n+1}}$\\
	 \hline
  \end{tabular}
  \renewcommand*{\arraystretch}{0.5}%
  \caption{Casimir densities depending on submodel class (as defined
  by the potential energy density choice) and type (relative to the
  Rossby number ($\varepsilon$) size, finite or infinitesimally
  small).  The functions $F$ and $G$ are arbitrary.  The symbol
  $\sim$ is used to mean proportional to, asymptotically as
  $\varepsilon\to 0$.} 
  \label{tab:cas-qg}
\end{table*}

\subsection{The IL$^{(0,\alpha)}$ family}

\subsubsection{The IL$^{(0,\alpha)}$PE}

Finally, a general class of rotating shallow-water models with
thermodynamics is the \IL PE itself with the identifications $\rho_1
= h$, $\tilde\rho_2 = \t$, $\tilde\rho_3 = \ts$, $\smash{\tilde\rho_4
= \vartheta_{\sigma^2}}$, etc., in \eqref{eq:il1-}, and the
interpretation of the buoyancy field as the polynomial expansion
(no claim on convergence is made or implied)
\begin{equation}
  \vartheta(\x,\sigma,t) = \t(\x,t) +
  \sum_1^\alpha\big(\sigma^n -
  \overline{\sigma^n})
  \vartheta_{\sigma^n}(\x,t),
  \label{eq:theta-il1-}
\end{equation}  
where 
\begin{equation}
  \overline{\sigma^n} = 
  \begin{cases}
	 0 & n : \text{odd},\\
	 \frac{1}{n+1} & n : \text{even}.
  \end{cases}
\end{equation}
The full (i.e., in the IL$^\infty$PE model) hydrostatic pressure
field produced by \eqref{eq:theta-il1-},
\begin{equation}
  p = \tfrac{1}{2}h\t + \tfrac{1}{2}\sum_1^\alpha
  \Big(\frac{\sigma^{n+1} - (-1)^{n+1}}{n+1} -
  \overline{\sigma^n}(\sigma+1)\Big)
  h\vartheta_{\sigma^n}.
\end{equation}
Noting that $\nabla p = \nabla\vert_\sigma p + \frac{1}{2}
\vartheta(1-\sigma)\nabla h$, one finds
\begin{equation}
  \overline{\nabla p} = \tfrac{1}{2}h\Big(\nabla\t -
  \sum_1^\alpha\overline{\sigma^{n+1}}
  \nabla\vartheta_{\sigma^n}\Big) + \Big(\t -
  \sum_1^\alpha\overline{\sigma^{n+1}}
  \vartheta_{\sigma^n}\Big)\nabla h.
\end{equation}
Under the same considerations as for the derivation of the
IL$^{(0,1)}$PE (steps 1--4 in Sec.\@~\ref{sec:IL01}), the \IL PE
with the potential energy density choice
\begin{equation}
  \varphi_{\alpha+} = \varphi_\alpha := \tfrac{1}{2}h\Big(\t - \sum_1^\alpha
  \overline{\sigma^{n+1}} \vartheta_{\sigma^n}\Big)
  \label{eq:phi-alpha}
\end{equation}
gives the dynamical equations consistent with the buoyancy field
representation \eqref{eq:theta-il1-}. We can refer to the resulting
model as the IL$^{(0,\alpha)}$PE.  The above clarifies the superscript
in IL$^{(0,\alpha)}$PE: no vertical variation is allowed for the
horizontal velocity, while polynomial vertical variation is permitted
for the buoyancy field up to an $\alpha$-th order degree. The model
velocity has implicit vertical shear, with dependence in the vertical
coordinate being polynomial up to a degree exceeding in one unit
that of the buoyancy field.  Finally, potential energy density
choices more general than $\varphi_\alpha$, allowed by the \IL PE,
can expand the realm of applicability of the model; also, an
appropriate choice can enable the study of the response to forcing
and/or dissipation in a Hamiltonian context.

\subsubsection{The IL$^{(0,\alpha)}$QG}

The IL$^{(0,\alpha)}$QG is obtained from the IL$^{(0,\alpha)}$PE
by considering the expansions in \eqref{eq:il05exp}, but with that
for $h$ replaced by
\begin{equation}
	 h/\Hr = 1 + \frac{\p - \ps + \sum_1^\alpha
	 (n+1)\overline{\sigma^{n+1}}\psi_{\sigma^{n+1}}}{f_0R_\alpha^2}
	 + \dotsc,
\end{equation}
where  $R_\alpha^2 := (1 - \frac{1}{2}\sum_1^\alpha
\overline{{\sigma}^{n+1}}s) R^{2}$, and further proposing 
\begin{align}
  \left.
  \begin{array}{cccccc}
    \vartheta_{\sigma^n}/\tfrac{1}{2}\Nr^2\Hr & = &
	 \frac{4(n+1)}{f_0sR^2} \psi_{\sigma^{n+1}} & + & \dotsc,\\
	 O & : & \varepsilon &  & \varepsilon^2
  \end{array}
  \right\}
\end{align}
$n = 2,\dotsc,\alpha$, consistent with a uniform \emph{reference}
density stratification (recall that the QG fields are $O(\varepsilon)$
perturbations off such a reference state). The IL$^{(0,\alpha)}$QG
is given by
\begin{subequations}
\begin{equation}
\left.  
\begin{aligned}
  \partial_t\xx + \{\p,\xi\}_{xy} - R_\alpha^{-2}
  \{\p,\nu(\ps,\dotsc,\psi_{\sigma^{\alpha+1}})\}_{xy} &= 0,\\
  \partial_t\psi_{\sigma^n} + \{\p,\psi_{\sigma^n}\}_{xy} &= 0,
\end{aligned}
\right\}  
\label{eq:il0alphaqg}
\end{equation}
$n = 1,\dotsc,\alpha$, where
\begin{equation}
  \nu(\ps,\dotsc,\psi_{\sigma^{\alpha+1}}) :=
  \ps-\sum_1^\alpha(n+1)\overline{\sigma^{n+1}}\psi_{\sigma^{n+1}},
\end{equation}
with invertibility principle
\begin{equation}
  \nabla^2\p - R_\alpha^{-2}\p = \xx - R_\alpha^{-2}
  \nu(\ps,\dotsc,\psi_{\sigma^{\alpha+1}}) - \beta y.
\end{equation}
\label{eq:il0alphaqg-sys}%
\end{subequations}
This system possesses a generalized Hamiltonian structure with
Hamiltonian given by
\begin{align}
  \H[\xx,\ps,\dotsc,\psi_{\sigma^\alpha}] 
  &:= \tfrac{1}{2} \int |\nabla\p|^2 + R_\alpha^{-2}\p^2
  \label{eq:il0alphaqg-H}
\end{align}
and Lie--Poisson bracket \eqref{eq:PB-qg} on $\mu =
(\xx,\ps,\dotsc,\psi_{\sigma^\alpha})$ with $W^{1n}_n = W^{n 1}_n
= 1$, $n = 1,\dotsc,\alpha+1$, and zero otherwise.  Conservation
of energy \eqref{eq:il0alphaqg-H}, zonal momentum \eqref{eq:Mx},
and angular momentum \eqref{eq:Mphi} are related by Noether's theorem
with symmetry of \eqref{eq:il0alphaqg-sys} under time shifts,
$x$-translations (in a zonal domain), and rotations (in an axisymmetric
domain of the $f$ plane), respectively. Finally, the corresponding
infinite of Casimirs, which do not generate any explicit symmetry,
is given by
\begin{equation}
  \C = \int\xx + F(\ps,\dotsc,\psi_{\sigma^{\alpha+1}}),
\end{equation}
where $F$ is arbitrary.  In Table \ref{tab:cas-qg} the Casimirs of
all submodels discussed are compared.

{\section{Stratification
effects}\label{sec:strat}}

\begin{figure}[t!]
  \centering%
  \includegraphics[width=\linewidth]{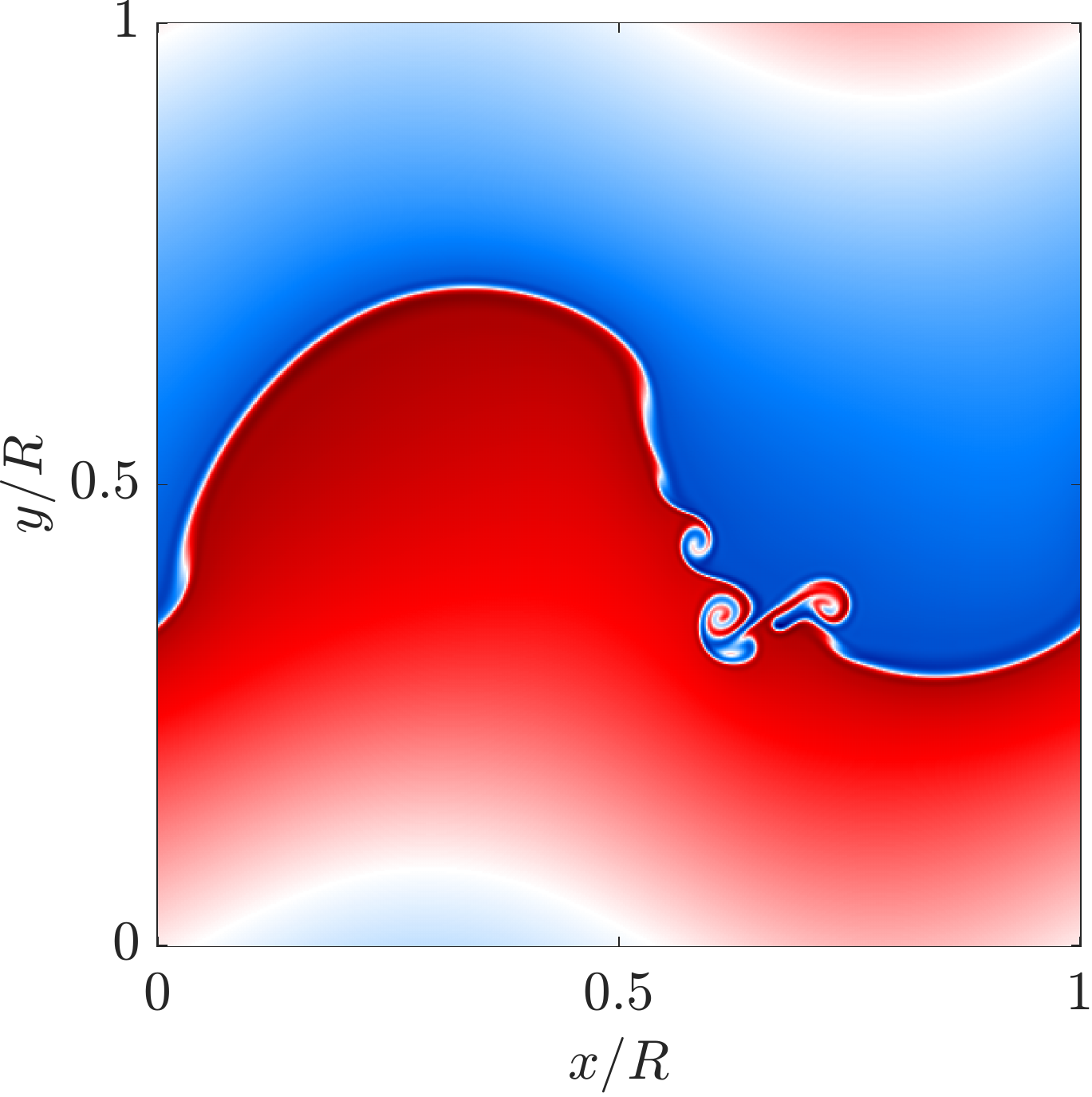}%
  \caption{As in Fig.\@~\ref{fig:il0}, but based on the
  IL$^{(0,1)}$QG derived in this paper.}
  \label{fig:il01}%
\end{figure}

{I had opened this paper showing in Fig.\@~\ref{fig:il0}
a snapshot at $tf_0 = 14$ (for additional snapshots,
cf.\@~Fig.\@~\ref{fig:il0-il01}, right panel, multimedia view) of
the buoyancy ($\bar\vartheta$) from a simulation of the IL$^0$QG
in a doubly periodic domain of the $f$-plane, and I close it by
showing in Fig.\@~\ref{fig:il01} the same but based on a simulation
of the IL$^{(0,1)}$QG, which has, in addition to lateral density
inhomogeneity, linear vertical stratification.  The initial conditions
are the same, viz., zero QG potential vorticity ($\xx$) and linear
$\bar\vartheta$.  As in \citet{Holm-etal-21}, topographic forcing
to the $\xx$-equation, of the form $R^{-2}\{\psi_\sigma,\psi_0\}_{xy}$
(which does not spoil the Hamiltonian structure of the system) where
$\psi_0(\x)$ is a prescribed small-amplitude sinusoid, was added
to speedup the development of small-scale (compared to $R$, the
equivalent-barotropic Rossby radius of deformation) Kelvin--Helmohltz-like
rollup filaments.  Note that the filamentation is much less intense
in the IL$^{(0,1)}$QG than in the IL$^0$QG.  It has been argued
\citep{Holm-etal-21} that the lack of Kelvin--Helmholtz circulation
conservation (more precisely, its creation and possibly ensuing
inverse-energy cascade suppression) in the IL$^0$ model plays a
role in the formation of small-scale rollup filaments.  The
IL$^{(0,1)}$ does not conserve circulation either, but the filamentation
is much weaker, as noted.  The inclusion of stratification seems
responsible for halting its development, possibly by introducing a
high-wavenumber instability cutoff, lacking in the
IL$^0$.\citep{Ripa-JGR-96, Beron-21-POFa} This might be expected
because the IL$^{(0,1)}$ can represent baroclinic instability at
long scales (i.e., of the order of $R$) as well as at short scales
(i.e., of the order of $sR$, when the stratification measure $s$,
defined in \eqref{eq:s}, is small).\citep{Beron-Ripa-97, Ripa-JFM-00}
This demands a thorough investigation of the stability properties
of the IL$^{(0,1)}$ (and more generally the IL$^{(0,\alpha)}$),
exploiting the symmetry-related conservation laws of the system(s),
and its consequences, which is reserved for future research.}

\section{Summary, discussion, and outlook}\label{sec:conclusions}

The renewed interest in thermodynamically-active rotating shallow-water
modeling \citep{Warneford-Dellar-13, Warneford-Dellar-14,
Warneford-Dellar-17, Gouzien-etal-17, Zeitlin-18, Eldred-etal-19,
Holm-etal-21, Beron-21-POFa, Beron-21-RMF, Moreles-etal-21} motivated
this work, which presented extended rotating shallow-water theories
with thermodynamics and geometry. With a focus on the ocean
mixed-layer, the topmost part of the ocean, a general model was
introduced and interpreted as such, considering in addition several
submodels, ranging from the shallow-water equations themself.  Unlike
the latter, all other models discussed have buoyancy varying
arbitrarily in horizontal position and time, and possibly also in
the vertical in polynomial form up to an arbitrary degree.  Unlike
the now classical ``thermal'' rotating shallow-water model, referred
to as Ripa's model by many authors due to Pedro Ripa's contribution
to its study,\citep{Ripa-GAFD-93, Ripa-JGR-96, Ripa-RMF-96,
Ripa-JFM-95, Ripa-DAO-99} the ``thermal'' models presented here are
capable of representing important mixed-layer ocean processes like
restratification by baroclinic instability.\citep{Haine-Marshall-98}
All models discussed were shown to admit Euler--Poincare variational
formulation \citep{Holm-etal-98a} and to possess generalized
Hamiltonian structure.\citep{Morrison-Greene-80}  This is important
because structure-preserving algorithms \citep{Morrison-17} can be
applied in simulations, preventing numerical artifacts.  Moreover,
flow-topology-preserving techniques to build parametrizations of
typically unresolved scales can be applied to investigate the effects
of these on transport at resolvable scales.\citep{Holm-15,
Cotter-etal-20}

Allowing variable density, the acronym IL, standing for ``inhomoegeneous
layer,'' was used to denote the models by appending to it superscripts
that indicate the amount of vertical variation allowed in the
horizontal velocity and buoyancy fields.  The most general model,
denoted IL$^{(0,\alpha)+}$, allows buoyancy to vary polynomially
up to degree $\alpha$ while velocity is kept constant in the vertical.
The $+$ sign indicates that the model enables, through an appropriate
choice of potential energy density, investigation of forced/dissipative
nonisothermal mixed-layer dynamics in a conservation (Hamiltonian)
setting.  Otherwise the IL$^{(0,\alpha)}$ has a potential energy
density strictly consistent with the buoyancy vertical structure
in the model. Quasigeostrophic (QG) versions of the corresponding
primitive equations (PE) were derived and their geometric structure
discussed in detail.  In particular, the IL$^{(0,0)}$PE, or simply
IL$^0$PE, which corresponds to Ripa's model, was explicitly connected
with earlier work by \citet{Morrison-Greene-80}: it is a special
case of the hydrodynamics form of the model discussed by these
authors.

{For future research I reserve a thorough investigation
of the stability of stratified steady-flow solutions of the model
derived here. This should lead to insight into the reported tendency
of stratification to halt the development of rollup filaments when
buoyancy is kept uniform in the vertical.}  Also left for future
research is including vertical shear in the velocity, as devised
in \citet{Ripa-JFM-95} but in a geometry-preserving manner.  This
is hoped to be achieved by performing truncations in the amount of
vertical variation allowed in the dynamical fields directly in the
Lagrangian,\citep{Salmon-83, Holm-Luesink-21} or the
Hamiltonian,\citep{Nore-Shepherd-97} of the IL$^\infty$PE, i.e.,
exact, model.  The mutilayer form \citep{Beron-21-RMF} of the IL$^1$
model of \citet{Ripa-JFM-95} was shown to behave in the ageostrophic
baroclinic instability problem quite accurately compared to the
IL$^\infty$PE model.  Yet issues with the numerical implementation
of the IL$^\infty$PE, and thus with its practical use, were
noted,\citep{Eldevik-02} possibly in connection with the seeming
lack of geometric structure of the model.

\begin{figure*}[t!]
  \centering%
  \animategraphics[width=\textwidth]{25}{il0plus-}{1}{201}
  \caption{Evolution of $\bar\vartheta$ (vertically averaged buoyancy)
  in the IL$^0$QG (left) and IL$^{(0,1)}$QG (right) models. The
  initial condition has uniform QG potential vorticity ($\xx$) and
  linear $\bar\vartheta$.  The domain of integration is doubly
  periodic, and lies on the $f$-plane.  Length is scaled by the
  external Rossby radius of deformation.  Time is scaled by the
  mean Coriolis parameter.  Topographic forcing to the $\xx$-equation,
  of the form $\{\bar\vartheta,\psi_0\}_{xy}$, where $\psi_0(\x)$
  is a prescribed small-amplitude sinusoid, is included.  A
  pseudospectral code with a small amount of hyperviscosity and a
  fourth-order Runge--Kutta time stepper is used.  The grid resolution
  is $512\times 512$. (Multimedia view.)}
  \label{fig:il0-il01}%
\end{figure*}

\section*{Author's contributions}

This paper is authored by a single individual who entirely carried
out the work.

\section*{Data availability}

This paper does not involve the use of data. 

\acknowledgements 

The comments by an anonymous reviewer led to improvements to this
paper.  I thank Prof.\ Philip J.\ Morrison for calling my attention
to \citet{Morrison-Greene-80} and for lively discussions on Poisson
brackets during the 2021 Aspen Center for Physics workshop ``Transport
and Mixing of Tracers in Geophysics and Astrophysics,'' where this
and additional ongoing work was initiated.  Further stimulating
discussions with Prof.\ Darryl D.\ Holm on ``thermal fluids'' are
acknowledged.  The Aspen Center for Physics is supported by NSF
grant PHY-1607611.

\appendix

\section{Proof for the Jacobi identity of \eqref{eq:PB}}\label{app:jac}

The commutator in \eqref{eq:comm} is antisymmetric (i.e., $[\a,\b]
= - [\b,\a]$ for any vectors $\a,\b$) and satisfies the Jacobi
identity (viz., $[[\a,\b],\c] + \cp = 0$ for all vectors $\a,\b,\c$).
The first property is obvious; the second one involves some algebra
but is otherwise quite straightforward to verify:
\begin{align}
  [[\a,\b],\c] = {} &
  + 
  \underbrace{\Big(\big((\a\cdot\nabla)\b\big)\cdot\nabla\Big)\c}_{\boxed{1}}
  - 
  \underbrace{\Big(\big((\b\cdot\nabla)\a\big)\cdot\nabla\Big)\c}_{\boxed{2}}\nonumber\\
  &
  - 
  \underbrace{(\c\cdot\nabla)\big((\a\cdot\nabla)\b\big)}_{\boxed{3} 
  + (\c\a:\nabla\nabla)\b}
  +
  \underbrace{(\c\cdot\nabla)\big((\b\cdot\nabla)\a\big)}_{\boxed{4}
  + (\c\b:\nabla\nabla)\a},\label{eq:com-jac-1}\\
  [[\c,\a],\b] = {} &
  + 
  \underbrace{\Big(\big((\c\cdot\nabla)\a\big)\cdot\nabla\Big)\b}_{\boxed{3}}
  - 
  \underbrace{\Big(\big((\a\cdot\nabla)\c\big)\cdot\nabla\Big)\b}_{\boxed{5}}\nonumber\\
  &
  - 
  \underbrace{(\b\cdot\nabla)\big((\c\cdot\nabla)\a\big)}_{\boxed{6}
  + (\b\c:\nabla\nabla)\a}
  +
  \underbrace{(\b\cdot\nabla)\big((\a\cdot\nabla)\c\big)}_{\boxed{2}
  + (\b\a:\nabla\nabla)\c},\label{eq:com-jac-2}\\
  [[\b,\c],\a] = {} &
  + 
  \underbrace{\Big(\big((\b\cdot\nabla)\c\big)\cdot\nabla\Big)\a}_{\boxed{6}}
  - 
  \underbrace{\Big(\big((\c\cdot\nabla)\b\big)\cdot\nabla\Big)\a}_{\boxed{4}}\nonumber\\
  &
  - 
 \underbrace{(\a\cdot\nabla)\big((\b\cdot\nabla)\c\big)}_{\boxed{1}
 + (\a\b:\nabla\nabla)\c}
  +
  \underbrace{(\a\cdot\nabla)\big((\c\cdot\nabla)\b\big)}_{\boxed{5}
  + (\a\c:\nabla\nabla)\b},\label{eq:com-jac-3}
\end{align}
where $\a\b : \c\mathbf d = a^ib^j c_id_j$; adding
\eqref{eq:com-jac-1}--\eqref{eq:com-jac-3} with $\a\b = \b\a$ in
mind proves the Jacobi identity for $[\,,\hspace{.05em}]$.

Now,
\begin{equation}
  \{\U,\V\}^{\m} = - \{\V,\U\}^{\m},\quad
  \{\U,\V\}^{\rho_{n_\alpha}} = - \{\V,\U\}^{\rho_{n_\alpha}},
\end{equation}
manifestly for any functionals $\U,\V$. Our goal is to demonstrate
that
\begin{equation}
  \{\U,\V\} := \{\U,\V\}^{\m} +
  \sum\{\U,\V\}^{\rho_{n_\alpha}}  
  \label{eq:PBalpha}
\end{equation}
satisfies $\{\{\U,\V\},\W\} + \cp = 0$ for all functionals
$\U,\V,\W$. More precisely, we seek to
show that
\begin{equation}
  \{\{\U,\V\},\W\} = \{\{\U,\V\},\W\}^{\m} + \sum
  \{\{\U,\V\},\W\}^{\rho_{n_\alpha}}
  \label{eq:mrho}
\end{equation}
vanishes upon $+\cp$. To do it, we consider the $\m$ bracket and
the sum of $\rho_{n_\alpha}$ brackets in \eqref{eq:mrho} separately,
with the following in mind:
\begin{align}
  \{\U,\V\}_{\m} &= - [\U_{\m},\V_{\m}],\label{eq:izq}\\
  \{\U,\V\}_{\rho_{n_\alpha}} &= - 
  \big(\U_{\m}\cdot\nabla\V_{\rho_{n_\alpha}} -
  \V_{\m}\cdot\nabla\U_{\rho_{n_\alpha}}\big),
  \label{eq:der}
\end{align}
where terms involving second-order functional derivatives of $\U$
and $\V$ have been omitted.  This can be done with no harm because
the (manifestly) skew-adjointness of the Poisson operator $\J$,
defined in \eqref{eq:J}, accounts \citep{Morrison-82} for the lack
of contribution of these terms to $\{\{\U,\V\},\W\} + \cp$. For
completeness, I give a quick proof here.  First note that, for
$\U[\mu] = \int U(\x;\mu,\nabla\mu,\dotsc)$, the second variational
derivative is defined as the unique element
$\smash{\frac{\delta^2}{\delta\mu^2}}\U$ satisfying
\begin{equation}
  \left.\frac{\d{}^2}{\d{\varepsilon^2}}\right\vert_{\varepsilon=0}
  \U[\mu+\varepsilon\delta\mu] =
  \int\delta\mu\bdot\frac{\delta^2\U}{\delta\mu^2}\delta\mu.
\end{equation}
Since $\U_\mu = \partial_\mu U$, it is clear then that $\U_{\mu\mu}
:= \smash{\frac{\delta^2}{\delta\mu^2}}\U = \partial_\mu \U_\mu$.
Skew-adjointness of $\J$ means $\int \U_\mu\bdot\J\V_\mu = - \int
\V_\mu\bdot\J\U_\mu$, where $\mu = (\m,\rho)$. The ignored terms
in $\{\U,\V\}_\mu$ are $\U_{\mu\mu}\bdot \J\V_\mu - \V_{\mu\mu}
\bdot \J\U_\mu$, where skew-adjointness of $\J$ was used and
$\U_{\mu\mu}\bdot\J\V_\mu = (\J\V_\mu)^a\U_{\mu\mu^a}$). Then we
have, including these terms only, $\{\{\U,\V\},\W\} = (\U_{\mu\mu}\bdot
\J\V_\mu) \bdot \J\W_\mu - (\V_{\mu\mu} \bdot \J\U_\mu) \bdot
\J\W_\mu$. To see that $\{\{\U,\V\},\W\} + \cp = 0$ (when the terms
in question are included only), it is enough to realize that
$(\U_{\mu\mu}\bdot \J\V_\mu) \bdot \J\W_\mu \equiv (\U_{\mu\mu}
\bdot \J\W_\mu) \bdot \J\V_\mu$, noting that $\U_{\mu^a\mu^b} =
\U_{\mu^b\mu^a}$.

Let us start with the $\m$ bracket, which, using \eqref{eq:izq}, reads
\begin{align}
  \{\{\U,\V\},\W\}^{\m} &= - \int\m\cdot
  \left[\{\U,\V\}_{\m},\W_{\m}\right]\nonumber\\ &= + \int\m\cdot
  \left[[\U_{\m},\V_{\m}],\W_{\m}\right].
\end{align}
Since $[\,,\hspace{.05em}]$ satisfies the Jacobi identity, we readily find
\begin{equation}
  \{\{\U,\V\},\W\}^{\m} + \cp = 0.
  \label{eq:jac_m}
\end{equation}

We now turn to the sum of $\rho_{n_\alpha}$ brackets in \eqref{eq:mrho},
which require more elaboration. It is enough to consider one term
only, though. More precisely, $\{\{\U,\V\},\W\}^{\rho_{n_\alpha}}$
\begin{align}
  = {} & 
  - \int
  \rho_{(n_\alpha)}\left(\{\U,\V\}_{\m}\cdot\nabla\W_{\rho_{(n_\alpha)}} -
  \W_{\m}\cdot\nabla\{\U,\V\}_{\rho_{(n_\alpha)}}\right)
  \nonumber\\ 
  = {} & 
  + \int \rho_{(n_\alpha)}\Big([\U_{\m},\V_{\m}]\cdot \nabla\W_{\rho_{(n_\alpha)}}
  \nonumber\\ 
  &-
  \W_{\m}\cdot\nabla \big(\U_{\m}\cdot\nabla\V_{\rho_{(n_\alpha)}} -
  \V_{\m}\cdot\nabla\U_{\rho_{(n_\alpha)}}\big)\Big)
  \nonumber\\ 
  = {} & 
  + \int
  \rho_{(n_\alpha)}\Big([\U_{\m},\V_{\m}]\cdot \nabla\W_{\rho_{(n_\alpha)}}
  \nonumber\\ 
  & -
  ((\W_{\m}\cdot\nabla)\U_{\m}\big)\cdot\nabla\V_{\rho_{(n_\alpha)}}
  -
  (\W_{\m}\cdot\nabla)\V_{\m}\big)\cdot\nabla\U_{\rho_{(n_\alpha)}}\Big)
  \label{eq:rho}
\end{align}
where, in order, we took into account \eqref{eq:izq}--\eqref{eq:der} and
\begin{equation}
  (\a\cdot\nabla)\b\cdot\mathbf c = ((\a\cdot\nabla)\b)\cdot \mathbf
  c + \a\b:\nabla\mathbf c
  \label{eq:rho-iden}
\end{equation}
(recalling that $\a\b = \b\a$). More explicitly,  we have:
$\{\{\U,\V\},\W\}^{\rho_{n_\alpha}}= \int\rho_{(n_\alpha)}$
\begin{align}
  &
  \Big( 
  + 
  \underbrace{\big((\Fm\cdot\nabla)\Gm\big)\cdot\nabla\Hs}_{\boxed{1}}
  - 
  \underbrace{\big((\Gm\cdot\nabla)\Fm\big)\cdot\nabla\Hs}_{\boxed{2}}\nonumber\\
  &
  - 
  \underbrace{\big((\Hm\cdot\nabla)\Fm\big)\cdot\nabla\Gs}_{\boxed{3}}
  +
  \underbrace{\big((\Hm\cdot\nabla)\Gm\big)\cdot\nabla\Fs}_{\boxed{4}}
  \Big).\label{eq:rho-jac-1}
\end{align}
Similarly, $\{\{\W,\U\},\V\}^{\rho_{n_\alpha}} = \int\rho_{(n_\alpha)}$
\begin{align}
  &
  \Big(
  + 
  \underbrace{\big((\Hm\cdot\nabla)\Fm\big)\cdot\nabla\Gs}_{\boxed{3}}
  - 
  \underbrace{\big((\Fm\cdot\nabla)\Hm\big)\cdot\nabla\Gs}_{\boxed{5}}\nonumber\\
  &
  - 
  \underbrace{\big((\Gm\cdot\nabla)\Hm\big)\cdot\nabla\Fs}_{\boxed{6}}
  +
  \underbrace{\big((\Gm\cdot\nabla)\Fm\big)\cdot\nabla\Hs}_{\boxed{2}}
  \Big)\label{eq:rho-jac-2}
\end{align}
and $\{\{\V,\W\},\U\}^{\rho_{n_\alpha}} = \int\rho_{(n_\alpha)}$
\begin{align}  
  &
  \Big(
  + 
  \underbrace{\big((\Gm\cdot\nabla)\Hm\big)\cdot\nabla\Fs}_{\boxed{6}}
  - 
  \underbrace{\big((\Hm\cdot\nabla)\Gm\big)\cdot\nabla\Fs}_{\boxed{4}}\nonumber\\
  &
  - 
 \underbrace{\big((\Fm\cdot\nabla)\Gm\big)\cdot\nabla\Hs}_{\boxed{1}}
  +
 \underbrace{\big((\Fm\cdot\nabla)\Hm\big)\cdot\nabla\Gs}_{\boxed{5}}
 \Big).\label{eq:rho-jac-3}
\end{align}
Adding \eqref{eq:rho-jac-1}--\eqref{eq:rho-jac-3}, one obtains
\begin{equation}
  \{\{\U,\V\},\W\}^{\rho_{n_\alpha}} + \cp = 0.
  \label{eq:jac_rho}
\end{equation}

Thus, \eqref{eq:jac_m} and \eqref{eq:jac_rho} together
produce
\begin{equation}
  \{\{\U,\V\},\W\} + \cp = 0,
\end{equation}
and the proof is completed.$\square$

\section{Clarification of the constraints
\eqref{eq:du}--\eqref{eq:drho}}\label{app:constraints}

The constraints \eqref{eq:du}--\eqref{eq:drho} allow one to ``skip
one step'' and obtain the equations of motion directly in Eulerian
(spatial) variables, i.e., without having to transform back to these
variables after extremizing an action under variations of particle
paths at fixed Lagrangian (material) labels and time. Indeed, the
constraints represent Eulerian variable variations induced by such
path variations.\citep{Newcomb-62}

Explicitly, $\v(\x,t) := \delta \q_t\circ \q_{-t}(\x)$, where
$\q_t(\l) := \q(\l,t)$ is the path in $D\subseteq \mathbb R^2$ of
the fluid particle marked with label $\l$, taken as position in $D$
at $t=0$, which is viewed as the \emph{reference configuration of
the fluid}.  The \emph{Lagrangian-to-Eulerian coordinates map} $\q_t
: \l\mapsto \x$ is assumed to be smooth as well as its inverse
$\q_t^{-1} = \q_{-t}$.  The relationship between $\q_t$ and $\mathbf
F_{t_0}^t$, i.e., the $({t_0,t})$-flow map of $\u(\x,t)$, obtained
by solving $\dot\x = \u(\x,t)$ with initial condition $\x_0 =
\x(t_0)$, is $\smash{\mathbf F_{t_0}^t} = \smash{\q_t\circ \q_{-t_0}}$.

Noting that $\u(\x,t) = \partial_t\vert_{\mathbf l}\q_t\circ
\q_{-t}(\x) = \dot\x$, upon taking its variation ($\delta\vert_{\mathbf
l,t}\u + (\v\cdot\nabla)\u = \delta\dot\x$) and the time derivative
of $\v$ ($\partial_t\v + (\u\cdot\nabla)\v = \delta\dot\x$),
constraint \eqref{eq:du} follows.

Constraint \eqref{eq:drho} follows upon noting first that
\eqref{eq:intrho} is equivalent to $\rho_{n_\alpha}(\x,t) = \rho_{n_\alpha}^0 J_t^{-1}
\circ \q_{-t}(\x)$, where $\rho_{n_\alpha}^0(\l)$ is the amount of ``density
mass'' carried by fluid particle $\l$, which is preserved in time,
and $J_t(\l):= \partial(\q_t(\l))/\partial(\l)$ is the Jacobian of
the map $\q_t$.  The variation of $\rho_{n_\alpha}$, $\delta\vert_{\mathbf
l,t}\rho_{n_\alpha} + \v\cdot\nabla\rho_{n_\alpha} = -\rho_{n_\alpha} \nabla\cdot\v$, by
invertibility of $\q_t$, which leads to \eqref{eq:drho}.

\section{Base Lie algebras, extensions, and realization
envelopings}\label{app:lie}

We recall a few abstract results \citep{Marsden-Ratiu-99} that are
needed to make formal statements on the geometry of results presented
in this paper. The set $\{\q_t\}$ forms a one-parameter group under
composition. As $\q_t$ is a diffeomorphism, i.e., it is smoothly
invertible, $\{\q_t\}$ has the structure of a differentiable manifold,
and, hence, $(\{\q_t\},\circ)$ represents a \emph{Lie group}, denoted
$\Diff(D)$, which provides a representation for the \emph{fluid
configuration space}: knowing $\q_t$ tells one where a fluid particle
goes.

The velocity $\u(\x,t)$ is an element of the \emph{tangent space}
to $\Diff(D)$ at $\x$, $T_{\x}\Diff(D)$. Since $\x = \q_t(\l)$, the
tangent space to $\Diff(D)$ at the identity $\q_0(\l) = \l$, namely,
$T_{\l}\Diff(D)$, uniquely determines all other tangent spaces to
$\Diff(D)$.  The tangent space has the structure of a vector space.
The elements of $T_{\l}\Diff(D)$ are invariant under composition
of $\q_t$ by the \emph{particle relabelling} $\r(\l)$ on the
\emph{right}.  Indeed, under $\l \mapsto \r(\l)$ it follows that
$\u \mapsto \smash{\partial_t\vert_{\r(\l)}(\q_t\circ\r^{-1})\circ
(\q_t\circ\r^{-1})^{-1}} = \smash{\partial_t\vert_{\l}\q_t\circ
\q_{-t} \circ (\r\circ\r^{-1})} \equiv \u$. The vector space of
right-invariant vectors of a Lie group forms a \emph{Lie algebra}.
For $\Diff(D)$, this is denoted $\mathfrak X$, which is isomorphic
to $T_{\l}\Diff(D)$.

In addition to its vector space structure, $\mathfrak X$ is equipped
with a bilinear, right-invariant \emph{product}
$[\,,\hspace{.05em}]_{\mathfrak X} : \mathfrak X \times \mathfrak
X \to \mathfrak X$, known as \emph{Lie bracket}, which satisfies
$[\a,\b]_{\mathfrak X} = - [\b,\a]_{\mathfrak X}$ (antisymmetry)
and $[[\a,\b]_{\mathfrak X},\c]_{\mathfrak X} + \cp = 0$ (Jacobi
identity) for all $\a,\b,\c\in\mathfrak X$.  The bracket is given
by minus the commutator of vector fields \eqref{eq:comm}, namely,
the Lie derivative of a vector field in $\mathfrak X$ along the
flow of another one in $\mathfrak X$, which expresses how the
elements of $\mathfrak X$ act on themselves.

When $J_t(\l) = 1$, i.e., $\q_t$ is an \emph{area preserving}
diffeomorphism, then $\u$ is divergence free. Namely, it is determined
by a streamfunction $\p(\x,t)$, i.e., $\u = \z\times\nabla\p$, where
$\p$ is viewed as an element of the vector space $\mathcal F$ of
smooth ($C^\infty$) time-dependent functions (in $D$).  The Lie
bracket of the corresponding Lie algebra, $\mathfrak X_\mathrm{area}
\simeq T_{\l}\Diffa(D)$, is $[\u_1,\u_2]_{\mathfrak X_\mathrm{area}}
= -[\u_1,\u_2] = -[\z\times\nabla\p_1,\z\times\nabla\p_2] \equiv
\z\times\nabla\{\p_1,\p_2\}_{xy}$, where $\{\,,\hspace{.05em}\}_{xy}$
is the \emph{canonical Poisson bracket} \eqref{eq:PBxy}.  This
bracket is manifestly antisymmetric and satisfies the Jacobi identity.
Furthermore, it is a derivation in each of its arguments; thus it
satisfies $\{UV,W\}_{xy} = U\{V,W\}_{xy} + \{U,W\}_{xy}V$ (Leibniz
rule).  The vector space $\mathcal F$ together with the canonical
bracket forms a \emph{Lie enveloping algebra},\citep{Morrison-06}
which we denote $\mathfrak F$.

\subsection{PE systems}

The proof in App.\ \ref{app:jac} expands Example 5.B of
\citet{Marsden-etal-84} as follows.  Let $\mathcal F^n$ represent
the $n$-dimensional vector space given by $n$ copies of $\mathcal
F$. The vector space $\mathfrak X \times \mathcal F^{\alpha+2}$
with the bracket given by $[\,,\hspace{.05em}]_{\mathfrak S_\alpha}
: \mathfrak S_\alpha \times \mathfrak S_\alpha \to \mathfrak
S_\alpha$; $((\a,A), (\b,B)) \mapsto ([\b,\a],\, \b\cdot \nabla A_1
- \a\cdot\nabla B_1, \dotsc, \, \b\cdot \nabla A_{\alpha+2} -
\a\cdot\nabla B_{\alpha+2})$ extends the Lie algebra $\mathfrak
X$ to the \emph{semidirect product} Lie algebra $\mathfrak S_\alpha$
of $\mathfrak X$ with the $(\alpha+2)$-dimensional vector space
$\mathcal F^{\alpha+2}$, denoted $\mathfrak X \rtimes \mathcal
F^{\alpha+2}$. The type of extension is determined by the particular
way $\mathfrak X$ acts on each component of $\mathcal F^{\alpha+2}$.
This action is provided by (minus) the Lie (i.e., directional)
derivative of a function in $\mathcal F$ along the flow of a vector
field in $\mathfrak X$.  The dual $\mathfrak S_\alpha^*$ of $\mathfrak
S_\alpha$, identified using $\smash{\int} \bdot$ to represent the
pairing $\mathfrak S_\alpha^*\times \mathfrak S_\alpha \to \mathbb
R$, carries the \emph{Lie--Poisson bracket} \eqref{eq:PB} for $\mu
= (\m,\rho)\in \mathfrak S_\alpha^*$.  Namely, $\{\U,\V\}_{\mathfrak
S_\alpha} = \smash{\int} \mu\bdot [\U_\mu,\V_\mu]_{\mathfrak
S_\alpha}$, where the Lie algebra is indicated in the ``outer''
bracket to make explicit its correspondence with the ``inner''
bracket. The (outer) bracket $\{\,,\hspace{.05em}\}_{\mathfrak
S_\alpha}$ makes $\mathfrak S_\alpha^*$ a \emph{Poisson manifold}
and conveys to $C^\infty(\mathfrak S_\alpha^*)$ a Lie algebra
structure that is a derivation in each of its arguments, i.e., a
\emph{realization of a Lie enveloping algebra on functionals}. The
equation of motion, $\partial_t\mu = \{\mu,\H\} = \J_{\mathfrak
S_\alpha}\H_\mu$ where $\J_{\mathfrak S_\alpha} =
\smash{[\mu,\cdot]_{\mathfrak S_\alpha}^\dagger}$.  Here, $\dagger$
stands for adjoint, defined by $\int
\mu\bdot\smash{[\U_\mu,\V_\mu]_{\mathfrak S_\alpha}} =: \int
\smash{[\mu,\V_\mu]_{\mathfrak S_\alpha}^\dagger\bdot \U_\mu}$ for
$\mu\in \mathfrak S_\alpha^*$ and  $\U_\mu,\V_\mu\in \mathfrak
S_\alpha$. Note that $\smash{[\,,\hspace{.05em}]_{\mathfrak
S_\alpha}^\dagger} : \mathfrak S_\alpha^* \times \mathfrak S_\alpha
\to \mathfrak S_\alpha^*$, which defines the \emph{coadjoint orbit}.

\begin{remark}
  While the Lie--Poisson system \eqref{eq:il1-lp} is defined on
  $\mathfrak X^* \times (\mathcal F^{\alpha+2})^*$, its Euler--Poincare
  counterpart, \eqref{eq:il1-ep} with the second equation of
  \eqref{eq:il1-}, is defined on $\mathfrak X \times (\mathcal
  F^{\alpha+2})^*$.\citep{Holm-etal-98a}  The connection between
  the two, equivalent formulations is provided by the \emph{partial
  Legendre transformation} \citep{Holm-etal-98a} $(\u,\rho) \mapsto
  (\m,\rho)$ defined by \eqref{eq:E}, viz., $\H[\m,\rho] = \int
  \m\cdot\u - \L[\u,\rho]$.
\end{remark}

\subsection{QG systems}

The bracket $[\,,\hspace{.05em}]_{\mathfrak F_W} : \mathfrak F_W
\times \mathfrak F_W \to \mathfrak F_W$; $(U,V) \mapsto
(\smash{W^{ab}_1\{U_a,V_b\}_{xy}}, \smash{W^{ab}_2\{U_a,V_b\}_{xy}},
\dotsc)$ for $U_a,V_a\in \mathcal F$ defines a Lie algebra $\mathfrak
F_W$ as an extension \citep{Thiffeault-Morrison-00} of the Lie
enveloping algebra $\mathfrak F$ of $\Diffa(D)$.  The dual $\mathfrak
F_W^*$ of $\mathfrak F_W$, identified using $\int\bdot$ to represent
the pairing $\mathfrak F_W^* \times \mathfrak F_W \to \mathbb R$,
carries the Lie--Poisson bracket \eqref{eq:PB-qg} for $\mu\in
\mathfrak F^*$. Namely, $\{\U,\V\}_{\mathfrak F_W} =
\int\mu\bdot[\U_\mu,\V_\mu]_{\mathfrak F_W}$, which makes $\mathfrak
F_W^*$ a Poisson manifold and is a product for a realization of a
Lie enveloping algebra on functionals in $C^\infty(\mathfrak F_W^*)$.
The motion equation, $\partial_t\mu = \J_{\mathfrak F_W}\H_\mu$
where $\J_{\mathfrak F_W} = \{\mu,\cdot\}_{\mathfrak F_W}$.

\begin{remark}
  For $\H[\mu]$ given, the motion equation $\partial_t\mu =
  \smash{[\mu,\H_\mu]_{\mathfrak F_W}^\dagger}$ does not seem
  possible to be obtained through Legendre transformation as is
  $\partial_t\mu = \smash{[\mu,\H_\mu]_{\mathfrak S}^\dagger}$.
  Thus an Euler--Poincare variational formulation may not exist or
  is quite difficult to be derived for QG-type systems.  Instead,
  ad-hoc variational formulations have been proposed in the
  literature.\cite{Virasoro-81, Holm-Zeitlin-98, Morrison-etal-14}
\end{remark}

With the product $[U,V]_{\mathfrak F_W} = (\{U_1,V_1\}_{xy},
\{U_1,V_2\}_{xy} - \{V_1,U_2\}_{xy}, \dotsc, \{U_1,V_n\}_{xy} -
\{V_1,U_n\}_{xy})$, the Lie enveloping algebra extension $\mathfrak
F_W$ represents a \emph{semidirect sum} \citep{Thiffeault-Morrison-00}
of $\mathfrak F$ and $\mathcal F^{n-1}$, where the representation
of $\mathfrak F$ on $\mathcal F^{n-1}$ is given by the canonical
Poisson bracket.  The ``inner'' brackets of all QG systems considered
here represent particular cases.

\section*{Author declarations}

\subsection*{Conflict of interest}

The author has no conflicts to disclose.

\bibliography{fot}

\begin{thebibliography}{72}%
\makeatletter
\providecommand \@ifxundefined [1]{%
 \@ifx{#1\undefined}
}%
\providecommand \@ifnum [1]{%
 \ifnum #1\expandafter \@firstoftwo
 \else \expandafter \@secondoftwo
 \fi
}%
\providecommand \@ifx [1]{%
 \ifx #1\expandafter \@firstoftwo
 \else \expandafter \@secondoftwo
 \fi
}%
\providecommand \natexlab [1]{#1}%
\providecommand \enquote  [1]{``#1''}%
\providecommand \bibnamefont  [1]{#1}%
\providecommand \bibfnamefont [1]{#1}%
\providecommand \citenamefont [1]{#1}%
\providecommand \href@noop [0]{\@secondoftwo}%
\providecommand \href [0]{\begingroup \@sanitize@url \@href}%
\providecommand \@href[1]{\@@startlink{#1}\@@href}%
\providecommand \@@href[1]{\endgroup#1\@@endlink}%
\providecommand \@sanitize@url [0]{\catcode `\\12\catcode `\$12\catcode
  `\&12\catcode `\#12\catcode `\^12\catcode `\_12\catcode `\%12\relax}%
\providecommand \@@startlink[1]{}%
\providecommand \@@endlink[0]{}%
\providecommand \url  [0]{\begingroup\@sanitize@url \@url }%
\providecommand \@url [1]{\endgroup\@href {#1}{\urlprefix }}%
\providecommand \urlprefix  [0]{URL }%
\providecommand \Eprint [0]{\href }%
\providecommand \doibase [0]{https://doi.org/}%
\providecommand \selectlanguage [0]{\@gobble}%
\providecommand \bibinfo  [0]{\@secondoftwo}%
\providecommand \bibfield  [0]{\@secondoftwo}%
\providecommand \translation [1]{[#1]}%
\providecommand \BibitemOpen [0]{}%
\providecommand \bibitemStop [0]{}%
\providecommand \bibitemNoStop [0]{.\EOS\space}%
\providecommand \EOS [0]{\spacefactor3000\relax}%
\providecommand \BibitemShut  [1]{\csname bibitem#1\endcsname}%
\let\auto@bib@innerbib\@empty
\bibitem [{\citenamefont {Arnold}(1966)}]{Arnold-66b}%
  \BibitemOpen
  \bibfield  {author} {\bibinfo {author} {\bibnamefont {Arnold}, \bibfnamefont
  {V.}},\ }\bibfield  {title} {\enquote {\bibinfo {title} {Sur la
  g{\'e}om{\'e}trie diff{\'e}rentielle des groupes de lie de dimension infinie
  et ses applications {\`a} l'hydrodynamique des fluides parfaits},}\
  }\href@noop {} {\bibfield  {journal} {\bibinfo  {journal} {Annales de
  l'Institut Fourier}\ }\textbf {\bibinfo {volume} {16}},\ \bibinfo {pages}
  {319--361} (\bibinfo {year} {1966})}\BibitemShut {NoStop}%
\bibitem [{\citenamefont {Beier}(1997)}]{Beier-97}%
  \BibitemOpen
  \bibfield  {author} {\bibinfo {author} {\bibnamefont {Beier}, \bibfnamefont
  {E.}},\ }\bibfield  {title} {\enquote {\bibinfo {title} {A numerical
  investigation of the annual variability in the {G}ulf of {C}alifornia},}\
  }\href@noop {} {\bibfield  {journal} {\bibinfo  {journal} {J. Phys.
  Oceanogr.}\ }\textbf {\bibinfo {volume} {27}},\ \bibinfo {pages} {615--632}
  (\bibinfo {year} {1997})}\BibitemShut {NoStop}%
\bibitem [{\citenamefont {Benjamin}(1984)}]{Benjamin-84}%
  \BibitemOpen
  \bibfield  {author} {\bibinfo {author} {\bibnamefont {Benjamin},
  \bibfnamefont {T.}},\ }\bibfield  {title} {\enquote {\bibinfo {title}
  {Impulse, flow force and variational principles},}\ }\href@noop {} {\bibfield
   {journal} {\bibinfo  {journal} {IMA J. Appl. Math.}\ }\textbf {\bibinfo
  {volume} {32}},\ \bibinfo {pages} {3--68} (\bibinfo {year}
  {1984})}\BibitemShut {NoStop}%
\bibitem [{\citenamefont {Beron-Vera}(2003)}]{Beron-03}%
  \BibitemOpen
  \bibfield  {author} {\bibinfo {author} {\bibnamefont {Beron-Vera},
  \bibfnamefont {F.~J.}},\ }\bibfield  {title} {\enquote {\bibinfo {title}
  {Constrained-{H}amiltonian shallow-water dynamics on the sphere},}\ }in\
  \href@noop {} {\emph {\bibinfo {booktitle} {{Nonlinear Processes in
  Geophysical Fluid Dynamics: A Tribute to the Scientific Work of Pedro
  Ripa}}}},\ \bibinfo {editor} {edited by\ \bibinfo {editor} {\bibfnamefont
  {O.~U.}\ \bibnamefont {Velasco-{F}uentes}}, \bibinfo {editor} {\bibfnamefont
  {J.}~\bibnamefont {Sheinbuam}}, \ and\ \bibinfo {editor} {\bibfnamefont
  {J.}~\bibnamefont {Ochoa}}}\ (\bibinfo  {publisher} {Kluwer},\ \bibinfo
  {year} {2003})\ pp.\ \bibinfo {pages} {29--51}\BibitemShut {NoStop}%
\bibitem [{\citenamefont {Beron-Vera}(2021{\natexlab{a}})}]{Beron-21-RMF}%
  \BibitemOpen
  \bibfield  {author} {\bibinfo {author} {\bibnamefont {Beron-Vera},
  \bibfnamefont {F.~J.}},\ }\bibfield  {title} {\enquote {\bibinfo {title}
  {Multilayer shallow-water model with stratification and shear},}\ }\href@noop
  {} {\bibfield  {journal} {\bibinfo  {journal} {Rev. Mex. Fis.}\ }\textbf
  {\bibinfo {volume} {67}},\ \bibinfo {pages} {351--364} (\bibinfo {year}
  {2021}{\natexlab{a}})}\BibitemShut {NoStop}%
\bibitem [{\citenamefont {Beron-Vera}(2021{\natexlab{b}})}]{Beron-21-POFa}%
  \BibitemOpen
  \bibfield  {author} {\bibinfo {author} {\bibnamefont {Beron-Vera},
  \bibfnamefont {F.~J.}},\ }\bibfield  {title} {\enquote {\bibinfo {title}
  {Nonlinear saturation of thermal instabilities},}\ }\href@noop {} {\bibfield
  {journal} {\bibinfo  {journal} {Phys. Fluid}\ }\textbf {\bibinfo {volume}
  {33}},\ \bibinfo {pages} {036608} (\bibinfo {year}
  {2021}{\natexlab{b}})}\BibitemShut {NoStop}%
\bibitem [{\citenamefont {Beron-Vera}\ and\ \citenamefont
  {Ripa}(1997)}]{Beron-Ripa-97}%
  \BibitemOpen
  \bibfield  {author} {\bibinfo {author} {\bibnamefont {Beron-Vera},
  \bibfnamefont {F.~J.}}and\ \bibinfo {author} {\bibnamefont {Ripa},
  \bibfnamefont {P.}},\ }\bibfield  {title} {\enquote {\bibinfo {title} {Free
  boundary effects on baroclinic instability},}\ }\href@noop {} {\bibfield
  {journal} {\bibinfo  {journal} {J. Fluid Mech.}\ }\textbf {\bibinfo {volume}
  {352}},\ \bibinfo {pages} {245--264} (\bibinfo {year} {1997})}\BibitemShut
  {NoStop}%
\bibitem [{\citenamefont {Cotter}\ \emph {et~al.}(2020)\citenamefont {Cotter},
  \citenamefont {Crisan}, \citenamefont {Holm}, \citenamefont {Pan},\ and\
  \citenamefont {Shevchenko}}]{Cotter-etal-20}%
  \BibitemOpen
  \bibfield  {author} {\bibinfo {author} {\bibnamefont {Cotter}, \bibfnamefont
  {C.}}, \bibinfo {author} {\bibnamefont {Crisan}, \bibfnamefont {D.}},
  \bibinfo {author} {\bibnamefont {Holm}, \bibfnamefont {D.}}, \bibinfo
  {author} {\bibnamefont {Pan}, \bibfnamefont {W.}}, and\ \bibinfo {author}
  {\bibnamefont {Shevchenko}, \bibfnamefont {I.}},\ }\bibfield  {title}
  {\enquote {\bibinfo {title} {Data assimilation for a quasi-geostrophic model
  with circulation-preserving stochastic transport noise},}\ }\href@noop {}
  {\bibfield  {journal} {\bibinfo  {journal} {J. Stat. Phys.}\ }\textbf
  {\bibinfo {volume} {179}},\ \bibinfo {pages} {1186 -- 1221} (\bibinfo {year}
  {2020})}\BibitemShut {NoStop}%
\bibitem [{\citenamefont {Dellar}(2003)}]{Dellar-03}%
  \BibitemOpen
  \bibfield  {author} {\bibinfo {author} {\bibnamefont {Dellar}, \bibfnamefont
  {P.~J.}},\ }\bibfield  {title} {\enquote {\bibinfo {title} {Common
  {H}amiltonian structure of the shallow water equations with horizontal
  temperature gradients and magnetic fields},}\ }\href@noop {} {\bibfield
  {journal} {\bibinfo  {journal} {Phys. Fluids}\ }\textbf {\bibinfo {volume}
  {15}},\ \bibinfo {pages} {292--297} (\bibinfo {year} {2003})}\BibitemShut
  {NoStop}%
\bibitem [{\citenamefont {Dronkers}(1969)}]{Dronkers-69}%
  \BibitemOpen
  \bibfield  {author} {\bibinfo {author} {\bibnamefont {Dronkers},
  \bibfnamefont {J.}},\ }\bibfield  {title} {\enquote {\bibinfo {title} {Tidal
  computations in rivers, coastal areas and seas},}\ }\href@noop {} {\bibfield
  {journal} {\bibinfo  {journal} {J. Hydrau. Div.}\ }\textbf {\bibinfo {volume}
  {95}},\ \bibinfo {pages} {44--77} (\bibinfo {year} {1969})}\BibitemShut
  {NoStop}%
\bibitem [{\citenamefont {Eldevik}(2002)}]{Eldevik-02}%
  \BibitemOpen
  \bibfield  {author} {\bibinfo {author} {\bibnamefont {Eldevik}, \bibfnamefont
  {T.}},\ }\bibfield  {title} {\enquote {\bibinfo {title} {On frontal dynamics
  in two model oceans},}\ }\href@noop {} {\bibfield  {journal} {\bibinfo
  {journal} {J. Phys. Oceanogr.}\ }\textbf {\bibinfo {volume} {32}},\ \bibinfo
  {pages} {2,915--2,925} (\bibinfo {year} {2002})}\BibitemShut {NoStop}%
\bibitem [{\citenamefont {Eldred}, \citenamefont {Dubos},\ and\ \citenamefont
  {Kritsikis}(2019)}]{Eldred-etal-19}%
  \BibitemOpen
  \bibfield  {author} {\bibinfo {author} {\bibnamefont {Eldred}, \bibfnamefont
  {C.}}, \bibinfo {author} {\bibnamefont {Dubos}, \bibfnamefont {T.}}, and\
  \bibinfo {author} {\bibnamefont {Kritsikis}, \bibfnamefont {E.}},\ }\bibfield
   {title} {\enquote {\bibinfo {title} {A quasi-hamiltonian discretization of
  the thermal shallow water equations},}\ }\href@noop {} {\bibfield  {journal}
  {\bibinfo  {journal} {Journal of Computational Physics}\ }\textbf {\bibinfo
  {volume} {379}},\ \bibinfo {pages} {1--31} (\bibinfo {year}
  {2019})}\BibitemShut {NoStop}%
\bibitem [{\citenamefont {Gawlik}\ and\ \citenamefont
  {{Gay-Balmaz}}(2021)}]{Gawlik-Gay-21}%
  \BibitemOpen
  \bibfield  {author} {\bibinfo {author} {\bibnamefont {Gawlik}, \bibfnamefont
  {E.~S.}}and\ \bibinfo {author} {\bibnamefont {{Gay-Balmaz}}, \bibfnamefont
  {F.}},\ }\bibfield  {title} {\enquote {\bibinfo {title} {{A
  structure-preserving finite element Mmethod for ccompressible ideal and
  resistive MHD}},}\ }\href@noop {} {\bibfield  {journal} {\bibinfo  {journal}
  {Journal of Plasma Physics}\ ,\ \bibinfo {pages} {in press}} (\bibinfo {year}
  {2021})}\BibitemShut {NoStop}%
\bibitem [{\citenamefont {Gouzien}\ \emph {et~al.}(2017)\citenamefont
  {Gouzien}, \citenamefont {Lahaye}, \citenamefont {Zeitlin},\ and\
  \citenamefont {Dubos}}]{Gouzien-etal-17}%
  \BibitemOpen
  \bibfield  {author} {\bibinfo {author} {\bibnamefont {Gouzien}, \bibfnamefont
  {E.}}, \bibinfo {author} {\bibnamefont {Lahaye}, \bibfnamefont {N.}},
  \bibinfo {author} {\bibnamefont {Zeitlin}, \bibfnamefont {V.}}, and\ \bibinfo
  {author} {\bibnamefont {Dubos}, \bibfnamefont {T.}},\ }\bibfield  {title}
  {\enquote {\bibinfo {title} {Thermal instability in rotating shallow water
  with horizontal temperature/density gradients},}\ }\href@noop {} {\bibfield
  {journal} {\bibinfo  {journal} {Physics of Fluids}\ }\textbf {\bibinfo
  {volume} {29}},\ \bibinfo {pages} {101702} (\bibinfo {year}
  {2017})}\BibitemShut {NoStop}%
\bibitem [{\citenamefont {Haine}\ and\ \citenamefont
  {Marshall}(1998)}]{Haine-Marshall-98}%
  \BibitemOpen
  \bibfield  {author} {\bibinfo {author} {\bibnamefont {Haine}, \bibfnamefont
  {T.~W.}}and\ \bibinfo {author} {\bibnamefont {Marshall}, \bibfnamefont
  {J.}},\ }\bibfield  {title} {\enquote {\bibinfo {title} {Gravitational,
  symmetric and baroclinic instability of the ocean mixed layer},}\ }\href@noop
  {} {\bibfield  {journal} {\bibinfo  {journal} {J. Phys. Oceanogr.}\ }\textbf
  {\bibinfo {volume} {28}},\ \bibinfo {pages} {634--658} (\bibinfo {year}
  {1998})}\BibitemShut {NoStop}%
\bibitem [{\citenamefont {Holm}\ \emph {et~al.}(1983)\citenamefont {Holm},
  \citenamefont {Marsden}, \citenamefont {Ratiu},\ and\ \citenamefont
  {Weinstein}}]{Holm-etal-83}%
  \BibitemOpen
  \bibfield  {author} {\bibinfo {author} {\bibnamefont {Holm}, \bibfnamefont
  {D.}}, \bibinfo {author} {\bibnamefont {Marsden}, \bibfnamefont {J.}},
  \bibinfo {author} {\bibnamefont {Ratiu}, \bibfnamefont {T.}}, and\ \bibinfo
  {author} {\bibnamefont {Weinstein}, \bibfnamefont {A.}},\ }\bibfield  {title}
  {\enquote {\bibinfo {title} {Nonlinear stability conditions and a priori
  estimates for barotropic hydrodynamics},}\ }\href@noop {} {\bibfield
  {journal} {\bibinfo  {journal} {Phys. Lett. A}\ }\textbf {\bibinfo {volume}
  {98}},\ \bibinfo {pages} {15--21} (\bibinfo {year} {1983})}\BibitemShut
  {NoStop}%
\bibitem [{\citenamefont {Holm}\ and\ \citenamefont
  {Zeitlin}(1998)}]{Holm-Zeitlin-98}%
  \BibitemOpen
  \bibfield  {author} {\bibinfo {author} {\bibnamefont {Holm}, \bibfnamefont
  {D.}}and\ \bibinfo {author} {\bibnamefont {Zeitlin}, \bibfnamefont {V.}},\
  }\bibfield  {title} {\enquote {\bibinfo {title} {Hamilton's principle for
  quasigeostrophic motion},}\ }\href@noop {} {\bibfield  {journal} {\bibinfo
  {journal} {Phys. Fluids}\ }\textbf {\bibinfo {volume} {10}},\ \bibinfo
  {pages} {800--806} (\bibinfo {year} {1998})}\BibitemShut {NoStop}%
\bibitem [{\citenamefont {Holm}(2015)}]{Holm-15}%
  \BibitemOpen
  \bibfield  {author} {\bibinfo {author} {\bibnamefont {Holm}, \bibfnamefont
  {D.~D.}},\ }\bibfield  {title} {\enquote {\bibinfo {title} {Variational
  principles for stochastic fluid dynamics},}\ }\href@noop {} {\bibfield
  {journal} {\bibinfo  {journal} {Proceedings of the Royal Society A:
  Mathematical, Physical and Engineering Sciences}\ }\textbf {\bibinfo {volume}
  {471}},\ \bibinfo {pages} {20140963} (\bibinfo {year} {2015})}\BibitemShut
  {NoStop}%
\bibitem [{\citenamefont {Holm}\ and\ \citenamefont
  {Luesink}(2021)}]{Holm-Luesink-21}%
  \BibitemOpen
  \bibfield  {author} {\bibinfo {author} {\bibnamefont {Holm}, \bibfnamefont
  {D.~D.}}and\ \bibinfo {author} {\bibnamefont {Luesink}, \bibfnamefont {E.}},\
  }\bibfield  {title} {\enquote {\bibinfo {title} {Stochastic wave--current
  interaction in thermal shallow water dynamics},}\ }\href@noop {} {\bibfield
  {journal} {\bibinfo  {journal} {J. Nonlinear Sci.}\ }\textbf {\bibinfo
  {volume} {31}},\ \bibinfo {pages} {29} (\bibinfo {year} {2021})}\BibitemShut
  {NoStop}%
\bibitem [{\citenamefont {Holm}, \citenamefont {Luesink},\ and\ \citenamefont
  {Pan}(2020)}]{Holm-etal-21}%
  \BibitemOpen
  \bibfield  {author} {\bibinfo {author} {\bibnamefont {Holm}, \bibfnamefont
  {D.~D.}}, \bibinfo {author} {\bibnamefont {Luesink}, \bibfnamefont {E.}},
  and\ \bibinfo {author} {\bibnamefont {Pan}, \bibfnamefont {W.}},\ }\bibfield
  {title} {\enquote {\bibinfo {title} {Stochastic mesoscale circulation
  dynamics in the thermal ocean},}\ }\href@noop {} {\bibfield  {journal}
  {\bibinfo  {journal} {Phys. Fluids}\ }\textbf {\bibinfo {volume} {33}},\
  \bibinfo {pages} {046603} (\bibinfo {year} {2020})}\BibitemShut {NoStop}%
\bibitem [{\citenamefont {Holm}, \citenamefont {Marsden},\ and\ \citenamefont
  {Ratiu}(1998)}]{Holm-etal-98a}%
  \BibitemOpen
  \bibfield  {author} {\bibinfo {author} {\bibnamefont {Holm}, \bibfnamefont
  {D.~D.}}, \bibinfo {author} {\bibnamefont {Marsden}, \bibfnamefont {J.~E.}},
  and\ \bibinfo {author} {\bibnamefont {Ratiu}, \bibfnamefont {T.}},\
  }\bibfield  {title} {\enquote {\bibinfo {title} {The {E}uler-{P}oincar\'e
  equations and semidirect products with applications to continuum theories},}\
  }\href@noop {} {\bibfield  {journal} {\bibinfo  {journal} {Adv. in Math.}\
  }\textbf {\bibinfo {volume} {137}},\ \bibinfo {pages} {1--81} (\bibinfo
  {year} {1998})}\BibitemShut {NoStop}%
\bibitem [{\citenamefont {Holm}, \citenamefont {Marsden},\ and\ \citenamefont
  {Ratiu}(2002)}]{Holm-etal-02}%
  \BibitemOpen
  \bibfield  {author} {\bibinfo {author} {\bibnamefont {Holm}, \bibfnamefont
  {D.~D.}}, \bibinfo {author} {\bibnamefont {Marsden}, \bibfnamefont {J.~E.}},
  and\ \bibinfo {author} {\bibnamefont {Ratiu}, \bibfnamefont {T.~S.}},\
  }\bibfield  {title} {\enquote {\bibinfo {title} {The {E}uler-{P}oincar\'e
  equations in geophysical fluid dynamics},}\ }in\ \href@noop {} {\emph
  {\bibinfo {booktitle} {Large-Scale Atmosphere-Ocean Dynamics {II}: Geometric
  Methods and Models}}},\ \bibinfo {editor} {edited by\ \bibinfo {editor}
  {\bibfnamefont {J.}~\bibnamefont {Norbury}}\ and\ \bibinfo {editor}
  {\bibfnamefont {I.}~\bibnamefont {Roulstone}}}\ (\bibinfo  {publisher}
  {Cambridge University},\ \bibinfo {year} {2002})\ Chap.~\bibinfo {chapter}
  {7}, pp.\ \bibinfo {pages} {251--299}\BibitemShut {NoStop}%
\bibitem [{\citenamefont {Kraus}, \citenamefont {Tassi},\ and\ \citenamefont
  {Grasso}(2016)}]{Kraus-etal-16}%
  \BibitemOpen
  \bibfield  {author} {\bibinfo {author} {\bibnamefont {Kraus}, \bibfnamefont
  {M.}}, \bibinfo {author} {\bibnamefont {Tassi}, \bibfnamefont {E.}}, and\
  \bibinfo {author} {\bibnamefont {Grasso}, \bibfnamefont {D.}},\ }\bibfield
  {title} {\enquote {\bibinfo {title} {Variational integrators for reduced
  magnetohydrodynamics},}\ }\href@noop {} {\bibfield  {journal} {\bibinfo
  {journal} {Journal of Computational Physics}\ }\textbf {\bibinfo {volume}
  {321}},\ \bibinfo {pages} {435--458} (\bibinfo {year} {2016})}\BibitemShut
  {NoStop}%
\bibitem [{\citenamefont {Lahaye}, \citenamefont {Zeitlin},\ and\ \citenamefont
  {Dubos}(2020)}]{Lahaye-etal-20}%
  \BibitemOpen
  \bibfield  {author} {\bibinfo {author} {\bibnamefont {Lahaye}, \bibfnamefont
  {N.}}, \bibinfo {author} {\bibnamefont {Zeitlin}, \bibfnamefont {V.}}, and\
  \bibinfo {author} {\bibnamefont {Dubos}, \bibfnamefont {T.}},\ }\bibfield
  {title} {\enquote {\bibinfo {title} {Coherent dipoles in a mixed layer with
  variable buoyancy: Theory compared to observations},}\ }\href@noop {}
  {\bibfield  {journal} {\bibinfo  {journal} {Ocean Modelling}\ }\textbf
  {\bibinfo {volume} {153}},\ \bibinfo {pages} {101673} (\bibinfo {year}
  {2020})}\BibitemShut {NoStop}%
\bibitem [{\citenamefont {Lavoie}(1972)}]{Lavoie-72}%
  \BibitemOpen
  \bibfield  {author} {\bibinfo {author} {\bibnamefont {Lavoie}, \bibfnamefont
  {R.}},\ }\bibfield  {title} {\enquote {\bibinfo {title} {A mesoscale
  numerical model of lake-effect storms},}\ }\href@noop {} {\bibfield
  {journal} {\bibinfo  {journal} {J. Atmos. Sci.}\ }\textbf {\bibinfo {volume}
  {29}},\ \bibinfo {pages} {1025 -- 1040} (\bibinfo {year} {1972})}\BibitemShut
  {NoStop}%
\bibitem [{\citenamefont {Lewis}, \citenamefont {Marsden},\ and\ \citenamefont
  {Montgomery}(1986)}]{Lewis-etal-86}%
  \BibitemOpen
  \bibfield  {author} {\bibinfo {author} {\bibnamefont {Lewis}, \bibfnamefont
  {D.}}, \bibinfo {author} {\bibnamefont {Marsden}, \bibfnamefont {J.}}, and\
  \bibinfo {author} {\bibnamefont {Montgomery}, \bibfnamefont {R.}},\
  }\bibfield  {title} {\enquote {\bibinfo {title} {{The Hamiltonian structure
  for dynamic free boundary problem}},}\ }\href@noop {} {\bibfield  {journal}
  {\bibinfo  {journal} {Physica D}\ }\textbf {\bibinfo {volume} {18}},\
  \bibinfo {pages} {391--404} (\bibinfo {year} {1986})}\BibitemShut {NoStop}%
\bibitem [{\citenamefont {Marle}(2013)}]{Marle-13}%
  \BibitemOpen
  \bibfield  {author} {\bibinfo {author} {\bibnamefont {Marle}, \bibfnamefont
  {C.-M.}},\ }\bibfield  {title} {\enquote {\bibinfo {title} {{On Henri
  Poincar\'e's Note ``Sur une forme nouvelle des \'equations de la
  M\'ecanique''}},}\ }\href@noop {} {\bibfield  {journal} {\bibinfo  {journal}
  {Journal of Geometry and Symmetry in Physics}\ }\textbf {\bibinfo {volume}
  {29}},\ \bibinfo {pages} {1 --38} (\bibinfo {year} {2013})}\BibitemShut
  {NoStop}%
\bibitem [{\citenamefont {Marsden}\ and\ \citenamefont
  {Weinstein}(1983)}]{Marsden-Weinstein-83}%
  \BibitemOpen
  \bibfield  {author} {\bibinfo {author} {\bibnamefont {Marsden}, \bibfnamefont
  {J.}}and\ \bibinfo {author} {\bibnamefont {Weinstein}, \bibfnamefont {A.}},\
  }\bibfield  {title} {\enquote {\bibinfo {title} {Coadjoint orbits, vortices
  and clebsch variables for incompressible flows},}\ }\href@noop {} {\bibfield
  {journal} {\bibinfo  {journal} {Physica D}\ }\textbf {\bibinfo {volume}
  {7}},\ \bibinfo {pages} {305--323} (\bibinfo {year} {1983})}\BibitemShut
  {NoStop}%
\bibitem [{\citenamefont {Marsden}\ and\ \citenamefont
  {Ratiu}(1999)}]{Marsden-Ratiu-99}%
  \BibitemOpen
  \bibfield  {author} {\bibinfo {author} {\bibnamefont {Marsden}, \bibfnamefont
  {J.~E.}}and\ \bibinfo {author} {\bibnamefont {Ratiu}, \bibfnamefont {T.}},\
  }\href@noop {} {\emph {\bibinfo {title} {Introduction to {M}echanics and
  {S}ymmetry}}},\ \bibinfo {edition} {2nd}\ ed.,\ \bibinfo {series} {Texts in
  {A}pplied {M}athematics}, Vol.~\bibinfo {volume} {17}\ (\bibinfo  {publisher}
  {Spinger},\ \bibinfo {year} {1999})\BibitemShut {NoStop}%
\bibitem [{\citenamefont {Marsden}, \citenamefont {Ratiu},\ and\ \citenamefont
  {Weinstein}(1984)}]{Marsden-etal-84}%
  \BibitemOpen
  \bibfield  {author} {\bibinfo {author} {\bibnamefont {Marsden}, \bibfnamefont
  {J.~E.}}, \bibinfo {author} {\bibnamefont {Ratiu}, \bibfnamefont {T.}}, and\
  \bibinfo {author} {\bibnamefont {Weinstein}, \bibfnamefont {A.}},\ }\bibfield
   {title} {\enquote {\bibinfo {title} {Semidirect products and reduction in
  mechanics},}\ }\href@noop {} {\bibfield  {journal} {\bibinfo  {journal}
  {Transactions of the American Mathematical Society}\ }\textbf {\bibinfo
  {volume} {281}},\ \bibinfo {pages} {147--177} (\bibinfo {year}
  {1984})}\BibitemShut {NoStop}%
\bibitem [{\citenamefont {Marsden}\ and\ \citenamefont
  {Weinstein}(1982)}]{Marsden-Weinstein-82}%
  \BibitemOpen
  \bibfield  {author} {\bibinfo {author} {\bibnamefont {Marsden}, \bibfnamefont
  {J.~E.}}and\ \bibinfo {author} {\bibnamefont {Weinstein}, \bibfnamefont
  {A.}},\ }\bibfield  {title} {\enquote {\bibinfo {title} {The {H}amiltonian
  structure of the {M}axwell--{V}lasov equations},}\ }\href@noop {} {\bibfield
  {journal} {\bibinfo  {journal} {Physica D}\ }\textbf {\bibinfo {volume}
  {4}},\ \bibinfo {pages} {349--406} (\bibinfo {year} {1982})}\BibitemShut
  {NoStop}%
\bibitem [{\citenamefont {McCreary}, \citenamefont {Kundu},\ and\ \citenamefont
  {Molinari}(1993)}]{McCreary-etal-93}%
  \BibitemOpen
  \bibfield  {author} {\bibinfo {author} {\bibnamefont {McCreary},
  \bibfnamefont {J.~P.}}, \bibinfo {author} {\bibnamefont {Kundu},
  \bibfnamefont {P.}}, and\ \bibinfo {author} {\bibnamefont {Molinari},
  \bibfnamefont {R.}},\ }\bibfield  {title} {\enquote {\bibinfo {title} {A
  numerical investigation of dynamics, thermodynamics and mixed-layer processes
  in the {I}ndian {O}cean},}\ }\href@noop {} {\bibfield  {journal} {\bibinfo
  {journal} {Prog. Oceanog.}\ }\textbf {\bibinfo {volume} {31}},\ \bibinfo
  {pages} {181--244} (\bibinfo {year} {1993})}\BibitemShut {NoStop}%
\bibitem [{\citenamefont {McIntyre}\ and\ \citenamefont
  {Shepherd}(1987)}]{McIntyre-Shepherd-87}%
  \BibitemOpen
  \bibfield  {author} {\bibinfo {author} {\bibnamefont {McIntyre},
  \bibfnamefont {M.~E.}}and\ \bibinfo {author} {\bibnamefont {Shepherd},
  \bibfnamefont {T.~G.}},\ }\bibfield  {title} {\enquote {\bibinfo {title} {An
  exact local conservation theorem for finite-amplitude disturbances to
  non-parallel shear flows, with remarks on {H}amiltonian structure and on
  {A}rno{l'd's} stability theorems},}\ }\href@noop {} {\bibfield  {journal}
  {\bibinfo  {journal} {J. Fluid Mech.}\ }\textbf {\bibinfo {volume} {181}},\
  \bibinfo {pages} {527--565} (\bibinfo {year} {1987})}\BibitemShut {NoStop}%
\bibitem [{\citenamefont {McWilliams}(2016)}]{McWilliams-16}%
  \BibitemOpen
  \bibfield  {author} {\bibinfo {author} {\bibnamefont {McWilliams},
  \bibfnamefont {J.~C.}},\ }\bibfield  {title} {\enquote {\bibinfo {title}
  {Submesoscale currents in the ocean},}\ }\href@noop {} {\bibfield  {journal}
  {\bibinfo  {journal} {Proc R Soc A}\ }\textbf {\bibinfo {volume} {472}},\
  \bibinfo {pages} {20160117} (\bibinfo {year} {2016})}\BibitemShut {NoStop}%
\bibitem [{\citenamefont {Moreles}\ \emph {et~al.}(2021)\citenamefont
  {Moreles}, \citenamefont {Zavala-Hidalgo}, \citenamefont {Martinez-Lopez},\
  and\ \citenamefont {Ruiz-Angulo}}]{Moreles-etal-21}%
  \BibitemOpen
  \bibfield  {author} {\bibinfo {author} {\bibnamefont {Moreles}, \bibfnamefont
  {E.}}, \bibinfo {author} {\bibnamefont {Zavala-Hidalgo}, \bibfnamefont {J.}},
  \bibinfo {author} {\bibnamefont {Martinez-Lopez}, \bibfnamefont {B.}}, and\
  \bibinfo {author} {\bibnamefont {Ruiz-Angulo}, \bibfnamefont {A.}},\
  }\bibfield  {title} {\enquote {\bibinfo {title} {Influence of stratification
  and {Yucatan Current} transport on the {Loop Current Eddy} shedding
  process},}\ }\href@noop {} {\bibfield  {journal} {\bibinfo  {journal}
  {Journal of Geophysical Research: Oceans}\ }\textbf {\bibinfo {volume}
  {126}},\ \bibinfo {pages} {e2020JC016315} (\bibinfo {year}
  {2021})}\BibitemShut {NoStop}%
\bibitem [{\citenamefont {Morrison}(1982)}]{Morrison-82}%
  \BibitemOpen
  \bibfield  {author} {\bibinfo {author} {\bibnamefont {Morrison},
  \bibfnamefont {P.~J.}},\ }\bibfield  {title} {\enquote {\bibinfo {title}
  {Poisson brackets for fluids and plasmas},}\ }in\ \href@noop {} {\emph
  {\bibinfo {booktitle} {Mathematical Methods in Hydrodynamics and
  Integrability in Dynamical Systems}}},\ \bibinfo {editor} {edited by\
  \bibinfo {editor} {\bibfnamefont {M.}~\bibnamefont {Tabor}}\ and\ \bibinfo
  {editor} {\bibfnamefont {Y.}~\bibnamefont {Treve}}}\ (\bibinfo {organization}
  {Institute of Physics Conference Proceedings 88},\ \bibinfo {year} {1982})\
  pp.\ \bibinfo {pages} {13--46}\BibitemShut {NoStop}%
\bibitem [{\citenamefont {Morrison}(1998)}]{Morrison-98}%
  \BibitemOpen
  \bibfield  {author} {\bibinfo {author} {\bibnamefont {Morrison},
  \bibfnamefont {P.~J.}},\ }\bibfield  {title} {\enquote {\bibinfo {title}
  {Hamiltonian description of the ideal fluid},}\ }\href@noop {} {\bibfield
  {journal} {\bibinfo  {journal} {Rev. Mod. Phys.}\ }\textbf {\bibinfo {volume}
  {70}},\ \bibinfo {pages} {467--521} (\bibinfo {year} {1998})}\BibitemShut
  {NoStop}%
\bibitem [{\citenamefont {Morrison}(2006)}]{Morrison-06}%
  \BibitemOpen
  \bibfield  {author} {\bibinfo {author} {\bibnamefont {Morrison},
  \bibfnamefont {P.~J.}},\ }\bibfield  {title} {\enquote {\bibinfo {title}
  {Hamiltonian fluid dynamics},}\ }in\ \href@noop {} {\emph {\bibinfo
  {booktitle} {Encyclopedia of Mathematical Physics}}},\ \bibinfo {editor}
  {edited by\ \bibinfo {editor} {\bibfnamefont {J.-P.}\ \bibnamefont
  {Francoise}}, \bibinfo {editor} {\bibfnamefont {G.~L.}\ \bibnamefont
  {Naber}}, \ and\ \bibinfo {editor} {\bibfnamefont {T.~S.}\ \bibnamefont
  {Tsun}}}\ (\bibinfo  {publisher} {Academic Press},\ \bibinfo {address}
  {Oxford},\ \bibinfo {year} {2006})\ pp.\ \bibinfo {pages}
  {593--600}\BibitemShut {NoStop}%
\bibitem [{\citenamefont {Morrison}(2017)}]{Morrison-17}%
  \BibitemOpen
  \bibfield  {author} {\bibinfo {author} {\bibnamefont {Morrison},
  \bibfnamefont {P.~J.}},\ }\bibfield  {title} {\enquote {\bibinfo {title}
  {Structure and structure-preserving algorithms for plasma physics},}\
  }\href@noop {} {\bibfield  {journal} {\bibinfo  {journal} {Physics of
  Plasmas}\ }\textbf {\bibinfo {volume} {24}},\ \bibinfo {pages} {055502}
  (\bibinfo {year} {2017})}\BibitemShut {NoStop}%
\bibitem [{\citenamefont {Morrison}\ and\ \citenamefont
  {Greene}(1980)}]{Morrison-Greene-80}%
  \BibitemOpen
  \bibfield  {author} {\bibinfo {author} {\bibnamefont {Morrison},
  \bibfnamefont {P.~J.}}and\ \bibinfo {author} {\bibnamefont {Greene},
  \bibfnamefont {J.~M.}},\ }\bibfield  {title} {\enquote {\bibinfo {title}
  {Noncanonical {H}amiltonian density formulation of hydrodynamics and ideal
  magnetohydrodynamics},}\ }\href@noop {} {\bibfield  {journal} {\bibinfo
  {journal} {Phys. Rev. Lett.}\ }\textbf {\bibinfo {volume} {45}},\ \bibinfo
  {pages} {790--794} (\bibinfo {year} {1980})}\BibitemShut {NoStop}%
\bibitem [{\citenamefont {Morrison}\ and\ \citenamefont
  {Hazeltine}(1984)}]{Morrison-Hazeltine-84}%
  \BibitemOpen
  \bibfield  {author} {\bibinfo {author} {\bibnamefont {Morrison},
  \bibfnamefont {P.~J.}}and\ \bibinfo {author} {\bibnamefont {Hazeltine},
  \bibfnamefont {R.~D.}},\ }\bibfield  {title} {\enquote {\bibinfo {title}
  {Hamiltonian formulation of reduced magnetohydrodynamics},}\ }\href@noop {}
  {\bibfield  {journal} {\bibinfo  {journal} {Phys. Fluids}\ }\textbf {\bibinfo
  {volume} {27}},\ \bibinfo {pages} {886--897} (\bibinfo {year}
  {1984})}\BibitemShut {NoStop}%
\bibitem [{\citenamefont {Morrison}, \citenamefont {Lingam},\ and\
  \citenamefont {Acevedo}(2014)}]{Morrison-etal-14}%
  \BibitemOpen
  \bibfield  {author} {\bibinfo {author} {\bibnamefont {Morrison},
  \bibfnamefont {P.~J.}}, \bibinfo {author} {\bibnamefont {Lingam},
  \bibfnamefont {M.}}, and\ \bibinfo {author} {\bibnamefont {Acevedo},
  \bibfnamefont {R.}},\ }\bibfield  {title} {\enquote {\bibinfo {title}
  {Hamiltonian and action formalisms for two-dimensional gyroviscous
  magnetohydrodynamics},}\ }\href@noop {} {\bibfield  {journal} {\bibinfo
  {journal} {Physics of Plasmas}\ }\textbf {\bibinfo {volume} {21}},\ \bibinfo
  {pages} {082102} (\bibinfo {year} {2014})}\BibitemShut {NoStop}%
\bibitem [{\citenamefont {Mungkasi}\ and\ \citenamefont
  {Roberts}(2016)}]{Mungkasi-Roberts-16}%
  \BibitemOpen
  \bibfield  {author} {\bibinfo {author} {\bibnamefont {Mungkasi},
  \bibfnamefont {S.}}and\ \bibinfo {author} {\bibnamefont {Roberts},
  \bibfnamefont {S.~G.}},\ }\bibfield  {title} {\enquote {\bibinfo {title} {{A
  smoothness indicator for numerical solutions tothe Ripa model}},}\
  }\href@noop {} {\bibfield  {journal} {\bibinfo  {journal} {J. Phys.: Conf.
  Ser.}\ }\textbf {\bibinfo {volume} {693}},\ \bibinfo {pages} {012011}
  (\bibinfo {year} {2016})}\BibitemShut {NoStop}%
\bibitem [{\citenamefont {Newcomb}(1962)}]{Newcomb-62}%
  \BibitemOpen
  \bibfield  {author} {\bibinfo {author} {\bibnamefont {Newcomb}, \bibfnamefont
  {W.~A.}},\ }\bibfield  {title} {\enquote {\bibinfo {title} {{Lagrangian and
  Hamiltonian methods in magnetohydrodynamics}},}\ }\href@noop {} {\bibfield
  {journal} {\bibinfo  {journal} {Nuclear Fusion: Supplement Part 2}\ }
  (\bibinfo {year} {1962})}\BibitemShut {NoStop}%
\bibitem [{\citenamefont {Nore}\ and\ \citenamefont
  {Shepherd}(1997)}]{Nore-Shepherd-97}%
  \BibitemOpen
  \bibfield  {author} {\bibinfo {author} {\bibnamefont {Nore}, \bibfnamefont
  {C.}}and\ \bibinfo {author} {\bibnamefont {Shepherd}, \bibfnamefont
  {T.~G.}},\ }\bibfield  {title} {\enquote {\bibinfo {title} {A hamiltonian
  weak-wave model for shallow-water flow},}\ }\href@noop {} {\bibfield
  {journal} {\bibinfo  {journal} {Proc. R. Soc. Lond. A}\ }\textbf {\bibinfo
  {volume} {453}},\ \bibinfo {pages} {563--580} (\bibinfo {year}
  {1997})}\BibitemShut {NoStop}%
\bibitem [{Note1()}]{Note1}%
  \BibitemOpen
  \bibinfo {note} {This object can be interpreted as a Lie derivative upon
  appropriate interpretation of $\protect \mathbf a$ and $\protect \mathbf b$,
  which I intentionally omit as the algebraic approach taken serves my
  purposes. Appendices \ref {app:constraints} and \ref {app:lie} do discuss
  some differential geometry aspects which are needed to provide a deeper
  geometric interpretation of some of the results of the paper, but these can
  be safely ignored by the reader if not interested.}\BibitemShut {Stop}%
\bibitem [{Note2()}]{Note2}%
  \BibitemOpen
  \bibinfo {note} {{The Casimirs are related to the particle relabelling
  symmetry of fluid dynamics, which permits one to switch between the
  Lagrangian and Eulerian descriptions \protect \citep
  {Marsden-etal-84}.}}\BibitemShut {Stop}%
\bibitem [{Note3()}]{Note3}%
  \BibitemOpen
  \bibinfo {note} {\protect \citet {Dellar-03} writes the Lie--Poisson bracket
  for the IL$^0$PE, viz., \protect \textup {\hbox {\mathsurround \z@ \protect
  \normalfont (\ignorespaces \ref {eq:PB}\unskip \@@italiccorr )}} for $\alpha
  = 0$, referring to \protect \citet {Holm-etal-85}, where a planar
  compressible MHD system is discussed in relation with that of \protect \citet
  {Morrison-Greene-80}.}\BibitemShut {Stop}%
\bibitem [{\citenamefont {O'{B}rien}\ and\ \citenamefont
  {Reid}(1967)}]{Obrien-Reid-67}%
  \BibitemOpen
  \bibfield  {author} {\bibinfo {author} {\bibnamefont {O'{B}rien},
  \bibfnamefont {J.~J.}}and\ \bibinfo {author} {\bibnamefont {Reid},
  \bibfnamefont {R.~O.}},\ }\bibfield  {title} {\enquote {\bibinfo {title} {The
  non-linear response of a two-layer, baroclinic ocean to a stationary,
  axially-symmetric hurricane: {P}art {I}: {U}pwelling induced by momentum
  transfer},}\ }\href@noop {} {\bibfield  {journal} {\bibinfo  {journal} {J.
  Atmos. Sci.}\ }\textbf {\bibinfo {volume} {24}},\ \bibinfo {pages} {197--207}
  (\bibinfo {year} {1967})}\BibitemShut {NoStop}%
\bibitem [{\citenamefont {Pedlosky}(1987)}]{Pedlosky-87}%
  \BibitemOpen
  \bibfield  {author} {\bibinfo {author} {\bibnamefont {Pedlosky},
  \bibfnamefont {J.}},\ }\href@noop {} {\emph {\bibinfo {title} {Geophysical
  {F}luid {D}ynamics}}},\ \bibinfo {edition} {2nd}\ ed.\ (\bibinfo  {publisher}
  {Springer},\ \bibinfo {year} {1987})\ p.\ \bibinfo {pages} {624
  pp.}\BibitemShut {Stop}%
\bibitem [{\citenamefont {Rehman}, \citenamefont {Ali},\ and\ \citenamefont
  {Qamar}(2018)}]{Rehman-etal-18}%
  \BibitemOpen
  \bibfield  {author} {\bibinfo {author} {\bibnamefont {Rehman}, \bibfnamefont
  {A.}}, \bibinfo {author} {\bibnamefont {Ali}, \bibfnamefont {I.}}, and\
  \bibinfo {author} {\bibnamefont {Qamar}, \bibfnamefont {S.}},\ }\bibfield
  {title} {\enquote {\bibinfo {title} {Exact riemann solutions of the ripa
  model for flat and non-flat bottom topographies},}\ }\href@noop {} {\bibfield
   {journal} {\bibinfo  {journal} {Results in Physics}\ }\textbf {\bibinfo
  {volume} {8}},\ \bibinfo {pages} {104 -- 113} (\bibinfo {year}
  {2018})}\BibitemShut {NoStop}%
\bibitem [{\citenamefont {Ripa}(1991)}]{Ripa-JFM-91}%
  \BibitemOpen
  \bibfield  {author} {\bibinfo {author} {\bibnamefont {Ripa}, \bibfnamefont
  {P.}},\ }\bibfield  {title} {\enquote {\bibinfo {title} {General stability
  conditions for a multi-layer model},}\ }\href@noop {} {\bibfield  {journal}
  {\bibinfo  {journal} {J. Fluid Mech.}\ }\textbf {\bibinfo {volume} {222}},\
  \bibinfo {pages} {119--137} (\bibinfo {year} {1991})}\BibitemShut {NoStop}%
\bibitem [{\citenamefont {Ripa}(1992)}]{Ripa-RMF-92a}%
  \BibitemOpen
  \bibfield  {author} {\bibinfo {author} {\bibnamefont {Ripa}, \bibfnamefont
  {P.}},\ }\bibfield  {title} {\enquote {\bibinfo {title} {Sistemas
  {H}amiltonianos singulares. {I}: Planteamiento del caso discreto, {T}eorema
  de {N}oether},}\ }\href@noop {} {\bibfield  {journal} {\bibinfo  {journal}
  {Rev. Mex. F\'{\i}s.}\ }\textbf {\bibinfo {volume} {38}},\ \bibinfo {pages}
  {984--1004} (\bibinfo {year} {1992})}\BibitemShut {NoStop}%
\bibitem [{\citenamefont {Ripa}(1993{\natexlab{a}})}]{Ripa-JFM-93}%
  \BibitemOpen
  \bibfield  {author} {\bibinfo {author} {\bibnamefont {Ripa}, \bibfnamefont
  {P.}},\ }\bibfield  {title} {\enquote {\bibinfo {title} {Arnol'd's second
  stability theorem for the equivalent barotropic model},}\ }\href@noop {}
  {\bibfield  {journal} {\bibinfo  {journal} {J. Fluid Mech.}\ }\textbf
  {\bibinfo {volume} {257}},\ \bibinfo {pages} {597--605} (\bibinfo {year}
  {1993}{\natexlab{a}})}\BibitemShut {NoStop}%
\bibitem [{\citenamefont {Ripa}(1993{\natexlab{b}})}]{Ripa-GAFD-93}%
  \BibitemOpen
  \bibfield  {author} {\bibinfo {author} {\bibnamefont {Ripa}, \bibfnamefont
  {P.}},\ }\bibfield  {title} {\enquote {\bibinfo {title} {Conservation laws
  for primitive equations models with inhomogeneous layers},}\ }\href@noop {}
  {\bibfield  {journal} {\bibinfo  {journal} {Geophys. Astrophys. Fluid Dyn.}\
  }\textbf {\bibinfo {volume} {70}},\ \bibinfo {pages} {85--111} (\bibinfo
  {year} {1993}{\natexlab{b}})}\BibitemShut {NoStop}%
\bibitem [{\citenamefont {Ripa}(1995)}]{Ripa-JFM-95}%
  \BibitemOpen
  \bibfield  {author} {\bibinfo {author} {\bibnamefont {Ripa}, \bibfnamefont
  {P.}},\ }\bibfield  {title} {\enquote {\bibinfo {title} {On improving a
  one-layer ocean model with thermodynamics},}\ }\href@noop {} {\bibfield
  {journal} {\bibinfo  {journal} {J. Fluid Mech.}\ }\textbf {\bibinfo {volume}
  {303}},\ \bibinfo {pages} {169--201} (\bibinfo {year} {1995})}\BibitemShut
  {NoStop}%
\bibitem [{\citenamefont {Ripa}(1996{\natexlab{a}})}]{Ripa-JGR-96}%
  \BibitemOpen
  \bibfield  {author} {\bibinfo {author} {\bibnamefont {Ripa}, \bibfnamefont
  {P.}},\ }\bibfield  {title} {\enquote {\bibinfo {title} {Linear waves in a
  one-layer ocean model with thermodynamics},}\ }\href@noop {} {\bibfield
  {journal} {\bibinfo  {journal} {J. Geophys. Res. C}\ }\textbf {\bibinfo
  {volume} {101}},\ \bibinfo {pages} {1233--1245} (\bibinfo {year}
  {1996}{\natexlab{a}})}\BibitemShut {NoStop}%
\bibitem [{\citenamefont {Ripa}(1996{\natexlab{b}})}]{Ripa-RMF-96}%
  \BibitemOpen
  \bibfield  {author} {\bibinfo {author} {\bibnamefont {Ripa}, \bibfnamefont
  {P.}},\ }\bibfield  {title} {\enquote {\bibinfo {title} {Low frequency
  approximation of a vertically integrated ocean model with thermodynamics},}\
  }\href@noop {} {\bibfield  {journal} {\bibinfo  {journal} {Rev. Mex.
  F\'{\i}s.}\ }\textbf {\bibinfo {volume} {42}},\ \bibinfo {pages} {117--135}
  (\bibinfo {year} {1996}{\natexlab{b}})}\BibitemShut {NoStop}%
\bibitem [{\citenamefont {Ripa}(1997)}]{Ripa-JPO-97b}%
  \BibitemOpen
  \bibfield  {author} {\bibinfo {author} {\bibnamefont {Ripa}, \bibfnamefont
  {P.}},\ }\bibfield  {title} {\enquote {\bibinfo {title} {``{I}nertial''
  oscillations and the $\beta $-plane approximation({s})},}\ }\href@noop {}
  {\bibfield  {journal} {\bibinfo  {journal} {J. Phys. Oceanogr.}\ }\textbf
  {\bibinfo {volume} {27}},\ \bibinfo {pages} {633--647} (\bibinfo {year}
  {1997})}\BibitemShut {NoStop}%
\bibitem [{\citenamefont {Ripa}(1999)}]{Ripa-DAO-99}%
  \BibitemOpen
  \bibfield  {author} {\bibinfo {author} {\bibnamefont {Ripa}, \bibfnamefont
  {P.}},\ }\bibfield  {title} {\enquote {\bibinfo {title} {On the validity of
  layered models of ocean dynamics and thermodynamics with reduced vertical
  resolution},}\ }\href@noop {} {\bibfield  {journal} {\bibinfo  {journal}
  {Dyn. Atmos. Oceans}\ }\textbf {\bibinfo {volume} {29}},\ \bibinfo {pages}
  {1--40} (\bibinfo {year} {1999})}\BibitemShut {NoStop}%
\bibitem [{\citenamefont {Ripa}(2000)}]{Ripa-JFM-00}%
  \BibitemOpen
  \bibfield  {author} {\bibinfo {author} {\bibnamefont {Ripa}, \bibfnamefont
  {P.}},\ }\bibfield  {title} {\enquote {\bibinfo {title} {Baroclinic
  instability in a reduced gravity, three-dimensional, quasi-geostrophic
  model},}\ }\href@noop {} {\bibfield  {journal} {\bibinfo  {journal} {J. Fluid
  Mech.}\ }\textbf {\bibinfo {volume} {403}},\ \bibinfo {pages} {1--22}
  (\bibinfo {year} {2000})}\BibitemShut {NoStop}%
\bibitem [{\citenamefont {Salmon}(1983)}]{Salmon-83}%
  \BibitemOpen
  \bibfield  {author} {\bibinfo {author} {\bibnamefont {Salmon}, \bibfnamefont
  {R.}},\ }\bibfield  {title} {\enquote {\bibinfo {title} {Practical use of
  {H}amilton's principle},}\ }\href@noop {} {\bibfield  {journal} {\bibinfo
  {journal} {J. Fluid Mech.}\ }\textbf {\bibinfo {volume} {132}},\ \bibinfo
  {pages} {431--444} (\bibinfo {year} {1983})}\BibitemShut {NoStop}%
\bibitem [{\citenamefont {Sanchez-Linares}, \citenamefont {{de Luna}},\ and\
  \citenamefont {{Castro Diaz}}(2016)}]{Sanchez-etal-16}%
  \BibitemOpen
  \bibfield  {author} {\bibinfo {author} {\bibnamefont {Sanchez-Linares},
  \bibfnamefont {C.}}, \bibinfo {author} {\bibnamefont {{de Luna}},
  \bibfnamefont {T.~M.}}, and\ \bibinfo {author} {\bibnamefont {{Castro Diaz}},
  \bibfnamefont {M.~J.}},\ }\bibfield  {title} {\enquote {\bibinfo {title} {{A
  HLLC scheme for Ripa model}},}\ }\href@noop {} {\bibfield  {journal}
  {\bibinfo  {journal} {Applied Mathematics and Computation}\ }\textbf
  {\bibinfo {volume} {272}},\ \bibinfo {pages} {369--384} (\bibinfo {year}
  {2016})}\BibitemShut {NoStop}%
\bibitem [{\citenamefont {Schopf}\ and\ \citenamefont
  {Cane}(1983)}]{Schopf-Cane-83}%
  \BibitemOpen
  \bibfield  {author} {\bibinfo {author} {\bibnamefont {Schopf}, \bibfnamefont
  {P.}}and\ \bibinfo {author} {\bibnamefont {Cane}, \bibfnamefont {M.}},\
  }\bibfield  {title} {\enquote {\bibinfo {title} {On equatorial dynamics,
  mixed layer physics and sea surface temperature},}\ }\href@noop {} {\bibfield
   {journal} {\bibinfo  {journal} {J. Phys. Oceanogr.}\ }\textbf {\bibinfo
  {volume} {13}},\ \bibinfo {pages} {917--935} (\bibinfo {year}
  {1983})}\BibitemShut {NoStop}%
\bibitem [{\citenamefont {Shepherd}(1988)}]{Shepherd-88a}%
  \BibitemOpen
  \bibfield  {author} {\bibinfo {author} {\bibnamefont {Shepherd},
  \bibfnamefont {T.}},\ }\bibfield  {title} {\enquote {\bibinfo {title}
  {Rigorous bounds on the nonlinear saturation of instabilities to parallel
  shear flows},}\ }\href@noop {} {\bibfield  {journal} {\bibinfo  {journal} {J.
  Fluid Mech.}\ }\textbf {\bibinfo {volume} {196}},\ \bibinfo {pages}
  {291--322} (\bibinfo {year} {1988})}\BibitemShut {NoStop}%
\bibitem [{\citenamefont {Shepherd}(1990)}]{Shepherd-90}%
  \BibitemOpen
  \bibfield  {author} {\bibinfo {author} {\bibnamefont {Shepherd},
  \bibfnamefont {T.~G.}},\ }\bibfield  {title} {\enquote {\bibinfo {title}
  {Symmetries, conservation laws and {H}amiltonian structure in geophysical
  fluid dynamics},}\ }\href@noop {} {\bibfield  {journal} {\bibinfo  {journal}
  {Adv. Geophys.}\ }\textbf {\bibinfo {volume} {32}},\ \bibinfo {pages}
  {287--338} (\bibinfo {year} {1990})}\BibitemShut {NoStop}%
\bibitem [{\citenamefont {Thiffeault}\ and\ \citenamefont
  {Morrison}(2000)}]{Thiffeault-Morrison-00}%
  \BibitemOpen
  \bibfield  {author} {\bibinfo {author} {\bibnamefont {Thiffeault},
  \bibfnamefont {J.-L.}}and\ \bibinfo {author} {\bibnamefont {Morrison},
  \bibfnamefont {P.~J.}},\ }\bibfield  {title} {\enquote {\bibinfo {title}
  {{Classification and Casimir invariants of Lie--Poisson brackets}},}\
  }\href@noop {} {\bibfield  {journal} {\bibinfo  {journal} {Physica D}\
  }\textbf {\bibinfo {volume} {136}},\ \bibinfo {pages} {205--244} (\bibinfo
  {year} {2000})}\BibitemShut {NoStop}%
\bibitem [{\citenamefont {Virasoro}(1981)}]{Virasoro-81}%
  \BibitemOpen
  \bibfield  {author} {\bibinfo {author} {\bibnamefont {Virasoro},
  \bibfnamefont {M.~A.}},\ }\bibfield  {title} {\enquote {\bibinfo {title}
  {Variational principle for two-dimensional incompressible hydrodynamics and
  quasigeostrophic flows},}\ }\href@noop {} {\bibfield  {journal} {\bibinfo
  {journal} {Phys. Rev. Lett.}\ }\textbf {\bibinfo {volume} {47}},\ \bibinfo
  {pages} {1,181--1,183} (\bibinfo {year} {1981})}\BibitemShut {NoStop}%
\bibitem [{\citenamefont {Warneford}\ and\ \citenamefont
  {Dellar}(2013)}]{Warneford-Dellar-13}%
  \BibitemOpen
  \bibfield  {author} {\bibinfo {author} {\bibnamefont {Warneford},
  \bibfnamefont {E.~S.}}and\ \bibinfo {author} {\bibnamefont {Dellar},
  \bibfnamefont {P.~J.}},\ }\bibfield  {title} {\enquote {\bibinfo {title} {The
  quasi-geostrophic theory of the thermal shallow water equations},}\
  }\href@noop {} {\bibfield  {journal} {\bibinfo  {journal} {J. Fluid Mech.}\
  }\textbf {\bibinfo {volume} {723}},\ \bibinfo {pages} {374--403} (\bibinfo
  {year} {2013})}\BibitemShut {NoStop}%
\bibitem [{\citenamefont {Warneford}\ and\ \citenamefont
  {Dellar}(2014)}]{Warneford-Dellar-14}%
  \BibitemOpen
  \bibfield  {author} {\bibinfo {author} {\bibnamefont {Warneford},
  \bibfnamefont {E.~S.}}and\ \bibinfo {author} {\bibnamefont {Dellar},
  \bibfnamefont {P.~J.}},\ }\bibfield  {title} {\enquote {\bibinfo {title}
  {{Thermal shallow water models of geostrophic turbulence in Jovian
  atmospheres}},}\ }\href@noop {} {\bibfield  {journal} {\bibinfo  {journal}
  {Physics of Fluids}\ }\textbf {\bibinfo {volume} {26}},\ \bibinfo {pages}
  {016603} (\bibinfo {year} {2014})}\BibitemShut {NoStop}%
\bibitem [{\citenamefont {Warneford}\ and\ \citenamefont
  {Dellar}(2017)}]{Warneford-Dellar-17}%
  \BibitemOpen
  \bibfield  {author} {\bibinfo {author} {\bibnamefont {Warneford},
  \bibfnamefont {E.~S.}}and\ \bibinfo {author} {\bibnamefont {Dellar},
  \bibfnamefont {P.~J.}},\ }\bibfield  {title} {\enquote {\bibinfo {title}
  {{Super- and sub-rotating equatorial jets in shallow water models of Jovian
  atmospheres: Newtonian cooling versus Rayleigh friction}},}\ }\href@noop {}
  {\bibfield  {journal} {\bibinfo  {journal} {Journal of Fluid Mechanics}\
  }\textbf {\bibinfo {volume} {822}},\ \bibinfo {pages} {484--511} (\bibinfo
  {year} {2017})}\BibitemShut {NoStop}%
\bibitem [{\citenamefont {Zeitlin}(2018)}]{Zeitlin-18}%
  \BibitemOpen
  \bibfield  {author} {\bibinfo {author} {\bibnamefont {Zeitlin}, \bibfnamefont
  {V.}},\ }\href@noop {} {\emph {\bibinfo {title} {Geophysical Fluid Dynamics:
  Understanding (Almost) Everything with Rotating Shallow Water Models}}}\
  (\bibinfo  {publisher} {Oxford University Press},\ \bibinfo {year}
  {2018})\BibitemShut {NoStop}%
\end{thebibliography}%

\end{document}